\documentclass[preprint2,twocolappendix]{aastex6}

\usepackage{longtable}

\shorttitle{SMG Structural Properties and NIR Morphologies}
\shortauthors{Chang et al.}

\begin{document}


\title{SCUBA-2 Ultra Deep Imaging EAO Survey (STUDIES) II: Structural Properties and Near-Infrared Morphologies of Faint Submillimeter Galaxies}



\author{Yu-Yen~Chang\altaffilmark{1}}
\email{yuyenchang.astro@gmail.com}
\author{Nicholas~Ferraro\altaffilmark{1,2}}
\author{Wei-Hao Wang\altaffilmark{1}}
\author{Chen-Fatt Lim\altaffilmark{1,3}}
\author{Yoshiki Toba\altaffilmark{1,4}}
\author{Fangxia An\altaffilmark{5,6,7}}
\author{Chian-Chou Chen\altaffilmark{8,9}}
\author{Ian Smail\altaffilmark{5}}
\author{Hyunjin Shim\altaffilmark{10}}
\author{Yiping Ao\altaffilmark{6,11}}
\author{Andy Bunker\altaffilmark{12,13}}
\author{Christopher J. Conselice\altaffilmark{14}}
\author{William Cowley\altaffilmark{15}}
\author{Elisabete~da~Cunha\altaffilmark{16}}
\author{Lulu Fan\altaffilmark{17}}
\author{Tomotsugu Goto\altaffilmark{18}}
\author{Kexin Guo\altaffilmark{19,20}}
\author{Luis C. Ho\altaffilmark{19,21}}
\author{Ho Seong Hwang\altaffilmark{22}}
\author{Chien-Hsiu Lee\altaffilmark{23}}
\author{Minju Lee\altaffilmark{11,24,25}}
\author{Micha{\l} J.~Micha{\l}owski\altaffilmark{26}}
\author{I. Oteo\altaffilmark{27,8}}
\author{Douglas Scott\altaffilmark{28}}
\author{Stephen Serjeant\altaffilmark{29}}
\author{Xinwen Shu\altaffilmark{30}}
\author{James Simpson\altaffilmark{1,31}}
\author{Sheona Urquhart\altaffilmark{29}}
\altaffiltext{1}{Academia Sinica Institute of Astronomy and Astrophysics, No.1 Section 4 Roosevelt Rd., 11F of Astro-Math Building, Taipei 10617, Taiwan}
\altaffiltext{2}{Department of Astronomy, University of Virginia, 530 McCormick Rd., Charlottesville, VA 22904, USA}
\altaffiltext{3}{Graduate Institute of Astrophysics, National Taiwan University, No.1 Section 4 Roosevelt Rd., Taipei 10617, Taiwan}
\altaffiltext{4}{Department of Astronomy, Kyoto University, Kitashirakawa-Oiwake-cho, Sakyo-ku, Kyoto 606-8502, Japan}
\altaffiltext{5}{Centre for Extragalactic Astronomy, Department of Physics, Durham University, Durham DH1 3LE, UK}
\altaffiltext{6}{Purple Mountain Observatory, Chinese Academy of Sciences, Nanjing 210008, China}
\altaffiltext{7}{University of Chinese Academy of Sciences, Beijing 100049, China}
\altaffiltext{8}{European Southern Observatory, Karl Schwarzschild Strasse 2, Garching, Germany}
\altaffiltext{9}{ESO Fellow}
\altaffiltext{10}{Department of Earth Science Education, Kyungpook National University, 80 Daehak-ro, Buk-gu, Daegu 41566, Republic of Korea}
\altaffiltext{11}{National Astronomical Observatory of Japan (NAOJ), Mitaka, Tokyo 181-8588, Japan}
\altaffiltext{12}{Department of Physics, University of Oxford, Denys Wilkinson Building, Keble Road, Oxford, OX1 3RH, U.K.}
\altaffiltext{13}{Affiliate Member, Kavli Institute for the Physics and Mathematics of the Universe, 5-1-5 Kashiwanoha, Kashiwa, 277-8583, Japan}
\altaffiltext{14}{School of Physics and Astronomy, The University of Nottingham, University Park, Nottingham NG7 2RD, UK}
\altaffiltext{15}{Kapteyn Astronomical Institute, University of Groningen, PO Box 800, NL-9700 AV Groningen, the Netherlands}
\altaffiltext{16}{The Australian National University, Mt Stromlo Observatory, Cotter Rd, Weston Creek, ACT 2611, Australia}
\altaffiltext{17}{Shandong Provincial Key Lab of Optical Astronomy and Solar-Terrestrial Environment, Institute of Space Science, Shandong University, Weihai, 264209, China}
\altaffiltext{18}{Institute of Astronomy, National Tsing Hua University, Hsinchu 30013, Taiwan}
\altaffiltext{19}{Kavli Institute for Astronomy and Astrophysics, Peking University, Beijing 100871, China}
\altaffiltext{20}{International Centre for Radio Astronomy Research, University of Western Australia M468, 35 Stirling Highway, Crawley, WA 6009, Australia}
\altaffiltext{21}{Department of Astronomy, School of Physics, Peking University, Beijing}
\altaffiltext{22}{Quantum Universe Center, Korea Institute for Advanced Study, 85 Hoegiro, Dongdaemun-gu, Seoul 02455, Korea}
\altaffiltext{23}{Subaru Telescope, NAOJ, Hilo, HI 96720, USA}
\altaffiltext{24}{Department of Astronomy, The University of Tokyo, 7-3-1 Hongo, Bunkyo-ku, Tokyo 133-0033, Japan}
\altaffiltext{25}{Department of Physics, Nagoya University, Furo-cho, Chikusa-ku, Nagoya 464-8602, Japan}
\altaffiltext{26}{Astronomical Observatory Institute, Faculty of Physics, Adam Mickiewicz University, 60-286 Pozna\'n, Poland}
\altaffiltext{27}{Institute for Astronomy, University of Edinburgh, Royal Observatory, Blackford Hill, Edinburgh EH9 3HJ, U.K.}
\altaffiltext{28}{Department of Physics and Astronomy, University of British Columbia, 6225 Agricultural Road, Vancouver, V6T 1Z1, Canada}
\altaffiltext{29}{School of Physical Sciences, The Open University, Milton Keynes, MK7 6AA, UK}
\altaffiltext{30}{Department of Physics, Anhui Normal University, Wuhu, Anhui, 241000, China}
\altaffiltext{31}{EACOA Fellow}

\begin{abstract}

We present structural parameters and morphological properties of faint 450-$\mu$m selected submillimeter galaxies (SMGs) from the JCMT Large Program, STUDIES, in the COSMOS-CANDELS region. Their properties are compared to an 850$\mu$m selected and a matched star-forming samples. We investigate stellar structures of 169 faint 450-$\mu$m sources ($S_{\rm 450}=2.8$--29.6~mJy; S/N$>4$) at $z<3$ using \emph{HST} near-infrared observations. Based on our spectral energy distribution fitting, half of such faint SMGs ($L_{\rm IR}=10^{11.65\pm0.98}~{\rm L_\odot}$) lie above the star-formation rate (SFR)/stellar mass plane. The size-mass relation shows that these SMGs are generally similar to less-luminous star-forming galaxies selected by ${\rm NUV}-r$ vs. $r-J$ colors. Because of the intrinsic luminosity of the sample, their rest-frame optical emission is less extended than the 850$\mu$m sources ($S_{\rm 850}>2$~mJy), and more extended than the star-forming galaxies in the same redshift range. For the stellar mass and SFR matched sample at $z\simeq1$ and $z\simeq2$, the size differences are marginal between faint SMGs and the matched galaxies. Moreover, faint SMGs have similar S\'ersic indices and projected axis ratios as star-forming galaxies with the same stellar mass and SFR. Both SMGs and the matched galaxies show high fractions ($\sim$70\%) of disturbed features at $z\simeq2$, and the fractions depend on the SFRs. These suggest that their star formation activity is related to galaxy merging, and the stellar structures of SMGs are similar to those of star-forming galaxies. We show that the depths of submillimeter surveys are approaching the lower luminosity end of star-forming galaxies, allowing us to detect galaxies on the main sequence. 
\end{abstract}




\section{Introduction}
\label{sec1}

The population known as ``Submillimeter galaxies'' (SMGs) was first discovered using the Submillimeter Common User Bolometer Array \citep[SCUBA]{1999MNRAS.303..659H} on the James Clerk Maxwell Telescope (JCMT) in the late 1990s in deep 850-$\mu$m images \citep{1997ApJ...490L...5S,1998Natur.394..248B,1998Natur.394..241H}. SMGs are understood to be a population of dusty starburst galaxies undergoing rapid stellar mass growth and thus they play an important role in our understanding of galaxy evolution and formation (see reviews by \citet{2002PhR...369..111B} and \citet{2014PhR...541...45C}). 

SMGs represents sources of the most luminous galaxies \citep[$L_{\rm IR}\gtrsim10^{12}{\rm L_\odot}$; e.g., ][]{2012A&A...539A.155M,2014MNRAS.438.1267S} at high redshifts \citep[$z\gtrsim2$; e.g., ][]{2005ApJ...622..772C,2014ApJ...788..125S}.
Their high luminosities are akin to local ultra-luminous infrared galaxies \citep[ULIRGs, see the review by][]{1996ARA&A..34..749S}, which are almost invariably mergers. All studies of local ULIRGS morphologies converge on a very high merger fraction \citep{1996MNRAS.279..477C,2000ApJ...529..170S,2001MNRAS.326.1333F,2002ApJS..143..315V}, according to their morphology in the optical and near-infrared (NIR).
However, theoretical models provide different formation routes for SMGs. They can be major mergers with significant starbursts, similar to local ULIRGs \citep[e.g.,][]{2010MNRAS.401.1613N}; a heterogeneous population of merger-driven starbursts and secularly evolving disk galaxies \citep[e.g.,][]{2011ApJ...743..159H}; or simply represent the most massive star-forming galaxy population at high redshift  \citep[e.g., ][]{2005MNRAS.363....2K,2010MNRAS.404.1355D,2015Natur.525..496N}. 
Moreover, \citet{2016MNRAS.462.3854L} suggested that SMGs are predominately disc-instability triggered starbursts.
Additionally, using large-scale simulations, \citet{2015MNRAS.446.1784C} found that SMGs detected in single-dish surveys can be chance superpositions of starbursting galaxies of very different redshifts along the same line of sight \citep[see also ][]{2013MNRAS.434.2572H,2015MNRAS.446.2291M}.
Therefore, it is important to investigate structures and morphologies of SMGs in large submillimeter surveys to verify these different possibilities.

At high redshift, morphologies of IR-luminous galaxies \citep[e.g.,][]{2009AJ....137.4854M,2010MNRAS.406..230R,2011ApJ...733...21B,2011ApJ...730..125Z,2012ApJ...757...23K,2012MNRAS.424.2232A,2012MNRAS.425.1320I,2013ApJ...768..164A,2016ApJ...827...57O,2017ApJ...844..106F} and massive galaxies \citep[e.g.,][]{2008ApJ...687L..61B} have been investigated.
Thanks to the high-resolution imaging available with the Hubble Space Telescope (\emph{HST}), the stellar structure of SMGs has been investigated.
\citet{2005MNRAS.358..149P} used \emph{HST}/Advanced Camera for Surveys (ACS) images to find larger sizes and a higher degree of asymmetry for 40 850 $\mu$m selected SMG. 
\citet{2010MNRAS.405..234S} analyzed the \emph{HST} $F160W$-band  images of 25 radio-identified SMGs ($S_{\rm 850}=3$--15 mJy) at $0.7<z<3.4$ from the \citet{2005ApJ...622..772C} survey, and found that the half-light radii of the SMGs and their asymmetries are not statistically distinct from a comparison sample of star-forming galaxies at similar redshifts.  However, the intermediate S\'ersic indices ($n\simeq2$) suggest that the stellar structure of SMGs is best described by a spheroid/elliptical galaxy light distribution. 
\citet{2011MNRAS.413...80C} used $F160W$-band images to study massive galaxies ($M_*>10^{11}{\rm M_\odot}$) at $1.7<z<2.9$, including galaxies detected in the submillimeter, finding that there is a gradual increase in size toward lower redshifts.  
\citet{2013MNRAS.432.2012T} used the Cosmic Assembly Near-infrared Deep Extragalactic Legacy Survey (CANDELS, \citealp{2011ApJS..197...35G,2011ApJS..197...36K}) $F160W$-band imaging to study 24 1.1mm and 870$\mu$m sources ($S_{\rm 870\,\mu m}=1.7$--9.1 mJy) at $1<z<3$. They found that almost all the (sub-) millimeter galaxies are well described by either a single exponential disk ($n\simeq1$), or a multiple-component system in which the dominant constituent is disk like. The extended structures are consistent with the sizes of other massive star-forming disks at $z\simeq2$.  \citet{2014ApJ...782...68T} showed that $3<z<6$ SMGs are consistent with being the progenitors of $z=2$ quiescent galaxies, based on their size distributions and other properties.

More recently, observations with the Atacama Large Millimeter/submillimeter Array (ALMA) help to refine the counterpart identification of single-dish samples.
\citet{2015ApJ...799..194C} analyzed \emph{HST} $F160W$-band imaging of 48 ALMA detected SMGs at $1<z<3$.
They found that 82\% of them appear to have disturbed morphologies, meaning that they are visually classified as either irregulars or interacting systems. They also found significant differences in the sizes and the S\'ersic indices between $2<z<3$ SMGs and $z\simeq2$ quiescent galaxies, and postulated that the majority of the $2<z<3$ SMGs with $S_{\rm 870}\gtrsim2$ mJy are early/mid-stage major mergers \citep[also see][]{2014ApJ...785..111W}.

Despite all the above studies, there does not seem to be a converging picture of whether SMGs are triggered by disc instability or mergers.
This might be caused by the differences in sample selections, redshift ranges, or methods of analysis.  
Furthermore, the previous studies focused on single-dish 850 $\mu$m or 1.1 mm selected SMGs, with typical fluxes of $S_{\rm 850} \gtrsim 2$ mJy, roughly corresponding to $L_{\rm IR}\gtrsim10^{12.3}{\rm L_\odot}$ (dust temperature $T_d\simeq30$K).   It is thus still difficult to study the variations as a function of star-formation rates (SFRs) from ULIRGs, luminous infrared galaxies (LIRGs), to normal star-forming galaxies that have $L_{\rm IR}<10^{12}{\rm L_\odot}$.  

The JCMT SCUBA-2 instrument \citep{2013MNRAS.430.2513H} enables 450-$\mu$m surveys that probe deeper (rms $\simeq$ 0.7 mJy; $L_{\rm IR} \simeq 5 \times 10^{11}{\rm L_\odot}$) than the 850-$\mu$m samples because of the roughly two times higher angular resolution (FWHM $\simeq$ 7$\arcsec$) and therefore lower confusion limit. 
Observations with SCUBA-2 at 450 $\mu$m can thus provide direct detections of fainter sources, and less ambiguous multi-wavelength counterpart identification. 
\citet{2013ApJ...762...81C,2013ApJ...776..131C} and \citet{2016ApJ...829...25H} carried out SCUBA-2 450 $\mu$m surveys in various blank fields and lensing cluster fields to detect 450 $\mu$m SMGs sources at $S_{\rm 450}=$1-10 mJy.
The SCUBA-2 Cosmology Legacy Survey \citep[S2CLS,][]{2013MNRAS.432...53G,2017MNRAS.465.1789G} and \citet{2013MNRAS.436.1919C} conducted
deep 450-$\mu$m imaging in the center of the Cosmic Evolution Survey \citep[COSMOS]{2007ApJS..172....1S} field and various other fields. 
\citet{2013MNRAS.436..430R} cross-identified 58 450-$\mu$m selected sources from the S2CLS sample \citep[$\sigma_{\rm 450}$= 1.5 mJy,][]{2013MNRAS.432...53G} with \emph{Spitzer} and \emph{HST}/WFC3 data. They showed a correlation between emissivity index $\beta$ and both stellar mass and effective radius. However, the depth was not sufficient to investigate faint SMGs, in the regime of more normal star-forming galaxies. 
\citet{2018MNRAS.475.5585Z} presented 64 sources ($\sigma_{\rm 450}\simeq$1.9 mJy; $ L_{\rm IR} \simeq 1.5 \times 10^{12} {\rm L_\odot}$ at $z<4$) of the S2CLS sample in the Extended Groth Strip field. They found that the dominant component for most of the galaxies at all redshifts is a disk-like structure (a median S\'ersic index $n\simeq1.4$ and half-light radius ${\rm r}_e\simeq4.8$kpc) by using the \emph{HST} $F160W$-band imaging.
They also showed a transition from irregular disks to disks with a spheroidal component at $z\simeq$1.4 and suggested that  SMGs are progenitors of massive elliptical galaxies. 

To further expand the 450 $\mu$m sample and to push to fainter depth ($\sigma_{\rm 450}\simeq$0.7 mJy) and lower luminosity ($L_{\rm IR} \simeq 5 \times 10^{11}{\rm L_\odot}$),  our team recently started a new program, the SCUBA-2 Ultra Deep Imaging EAO (East-Asian Observatory) Survey \citep[STUDIES,][]{2017ApJ...850...37W}. STUDIES targets the center of the COSMOS field where there are CANDELS NIR data ideal for a morphological study.
We combine all the SCUBA-2 data in the COSMOS-CANDELS region to reach a detection limit of $S_{\rm 450} \simeq 3$ mJy ($\sigma_{\rm 450}\simeq$0.7 mJy). 
Moreover,  the 450 $\mu$m selection does not just enable finding fainter samples. Both the 450 $\mu$m and the parallel deep 850 $\mu$m observations ($\sigma_{\rm 850}\simeq0.12$ mJy) help to constrain the shape of the spectral energy distribution (SED).
Our faint SMG sample therefore probes luminosities of approximately $L_{\rm IR} > 2$--$5 \times 10^{11} {\rm L_\odot}$ at $z=1$--2, corresponding to SFRs of 
$>40$--80 ${\rm {\rm M_\odot}}$ yr$^{-1}$, assuming the standard \citet{1998ApJ...498..541K} relation, overlapping with that of optically selected normal star-forming galaxies.
Therefore, we will be able to compare cool dusty galaxies to unobscured starbursts with similar redshifts, SFRs, and stellar masses.
The \emph{HST} NIR imaging across the STUDIES region enables us to investigate the stellar structures and morphological properties of these faint 450-$\mu$m sources. 

In this paper, we present morphological results based on structural analysis and visual classification for faint SMGs (450-$\mu$m sources) detected by STUDIES, as well as for a control sample matched to the STUDIES-SMGs.
The structure of this paper is as follows.
We describe the data, catalog matching, and SED fitting in Section \ref{sec2}.
We analyze the physical and structural properties in Section \ref{sec3}.
We discuss the implications in Section \ref{sec4} and summarize in Section \ref{sec5}.
We use AB magnitudes thrughout, adopt the cosmological parameters
($\Omega_{\rm M}$,$\Omega_\Lambda$,$h$)=(0.30,0.70,0.70), and assume the
 stellar initial mass function of \citet{2003PASP..115..763C}.
 

\begin{figure*}
\centering
\includegraphics[width=0.85\textwidth]{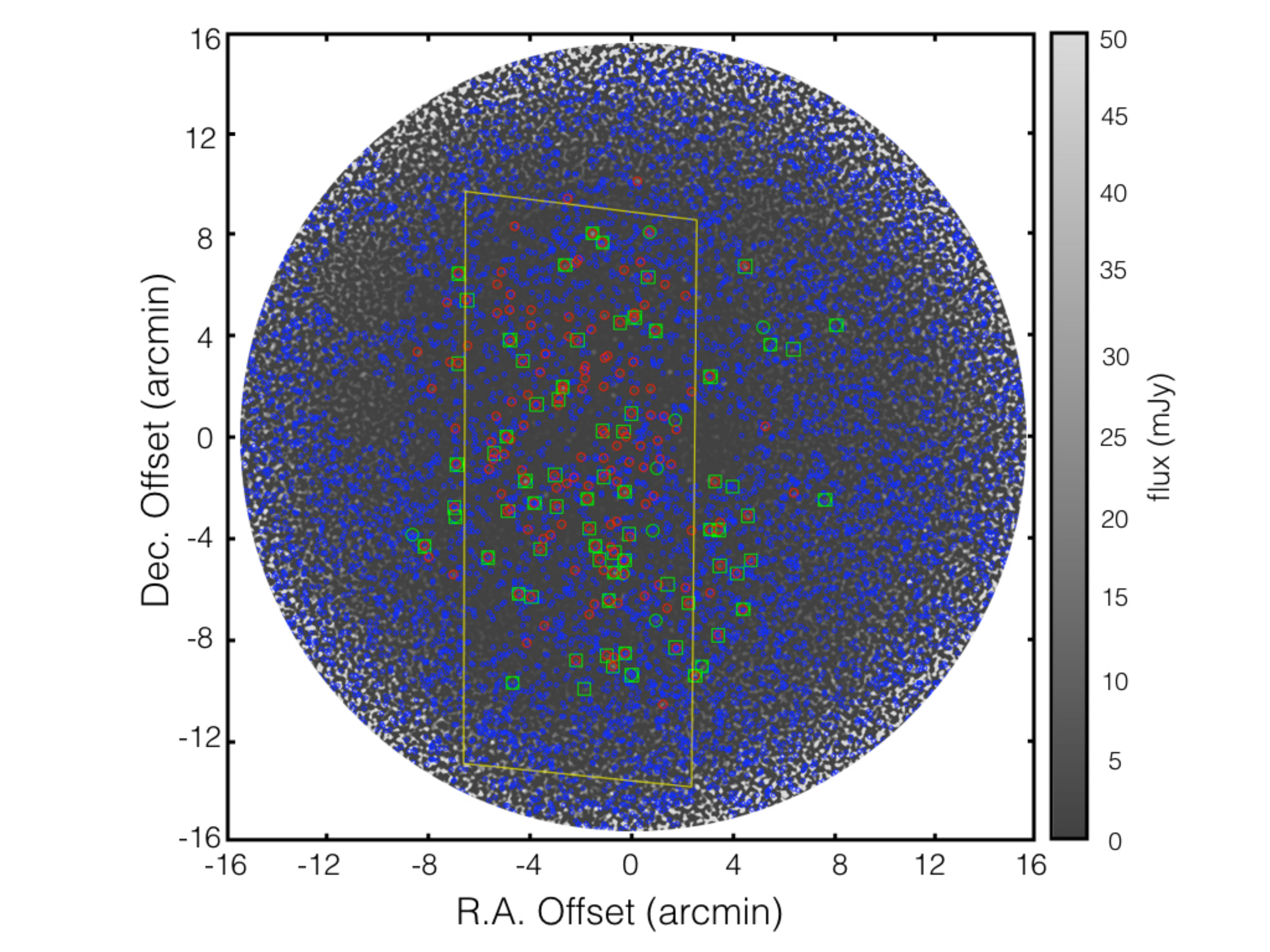} 
\caption{STUDIES 450-$\mu$m flux map which provides coverage over $\simeq$700 arcmin$^2$ centered at R.A.$=$10:00:22.26, decl.$=$+02:24:05.06. We show our sample selection of 450-$\mu$m sources (red circles with 10$\arcsec$ in radii, S/N $>$ 4, $S_{\rm 450}>2$mJy) and 850-$\mu$m sources (green circles with 15$\arcsec$ in radii, S/N $>$ 6, $S_{\rm 850}>2$mJy) from the machine-learning method, 850-$\mu$m sources (green box, S/N $>$ 6, $S_{\rm 850}>2$mJy) from the cross-matched method, along with the comparison sample (blue circles with 5$\arcsec$ in radii, $M_*>10^{10} {\rm M_\odot}$, ${\rm NUV}-r$ vs. $r-J$ selection). We consider star-forming galaxies inside the STUDIES coverage as the comparison sample. The yellow region shows the CANDELS footprint.}
\label{studies_region}
\end{figure*}

\begin{figure}
\centering
\includegraphics[width=0.95\columnwidth]{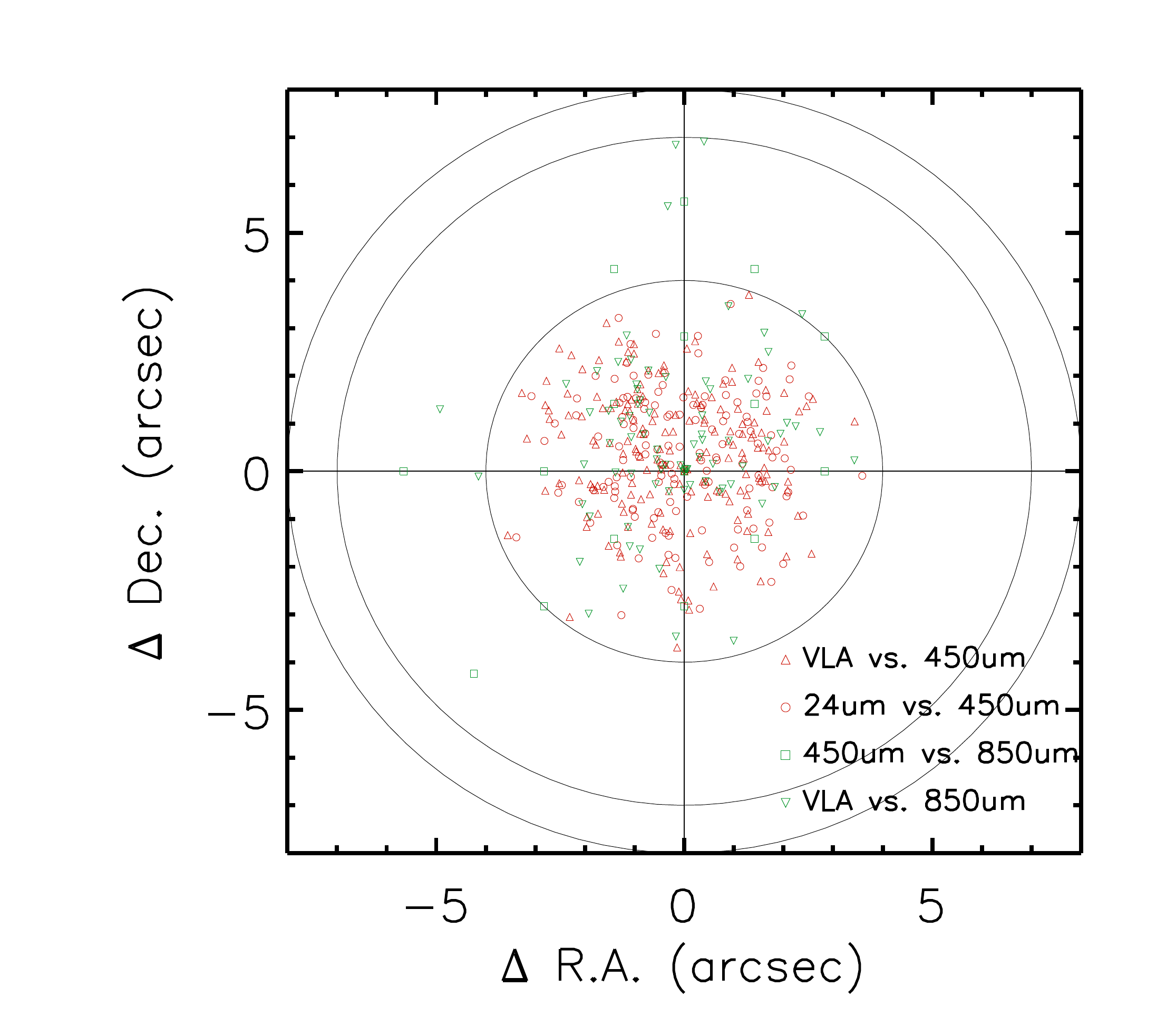} 
\caption{Offsets of the coordinates between the SMG sources and their counterparts: VLA vs. 450 $\mu$m, 24 $\mu$m vs. 450 $\mu$m, 450 $\mu$m vs. 850 $\mu$m, and VLA vs. 850 $\mu$m. The circles are search radii of $4\arcsec$, $7\arcsec$, and $8\arcsec$. }
\label{studies_id}
\end{figure}

\begin{figure}
\centering
\includegraphics[width=0.95\columnwidth]{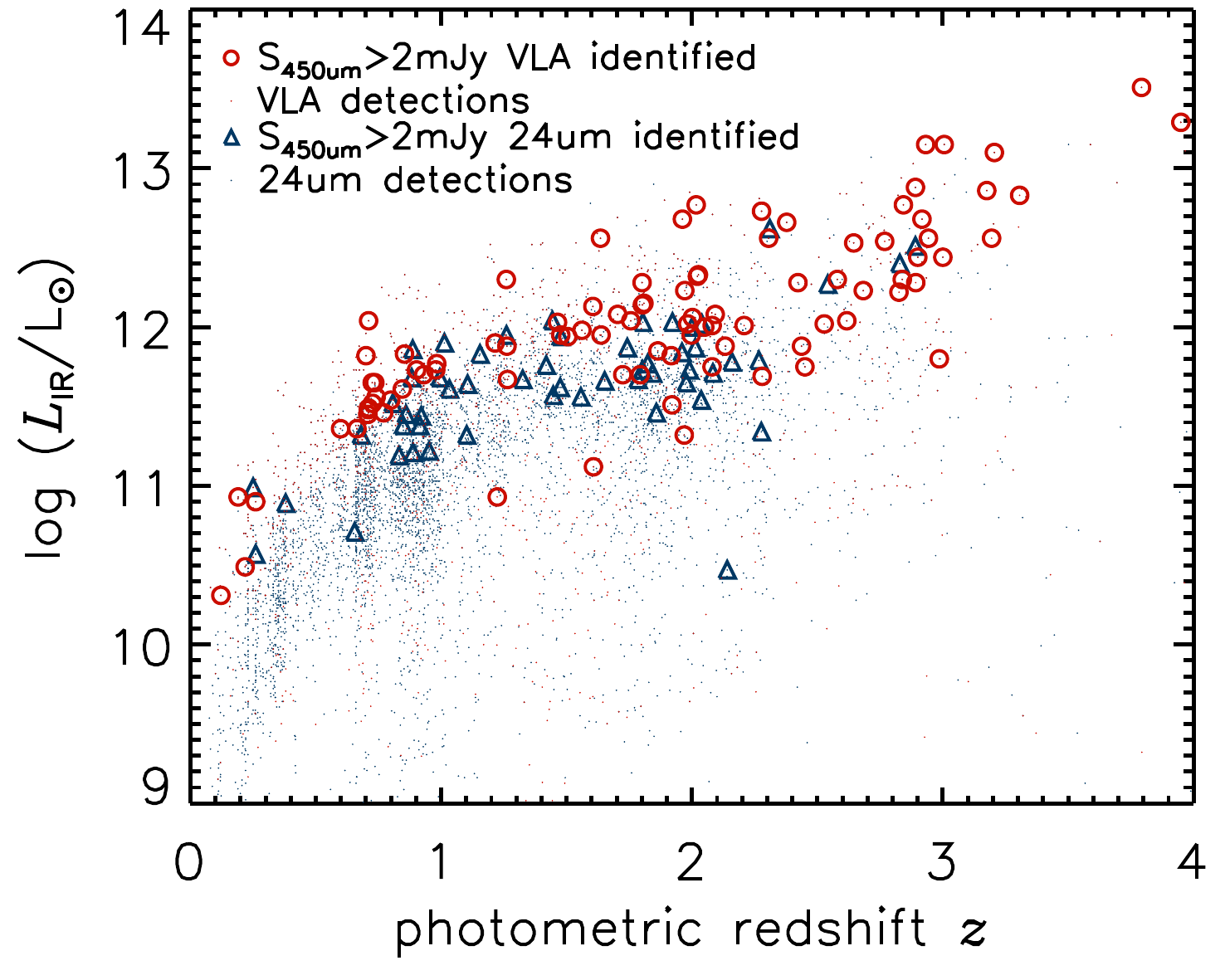} 
\caption{Photometric redshift to infrared luminosity plot for 450-$\mu$m sources identified by VLA and 24 $\mu$m position  (red circles and blue triangles), as well as VLA and 24 $\mu$m detections (red and blue points, which are not 450-$\mu$m detections). The infrared luminosity is derived by MAGPHYS (see \S~\ref{sec2_4} for more details).}
\label{studies_zlir}
\end{figure}

\begin{figure*}
\centering
\includegraphics[width=0.9\textwidth]{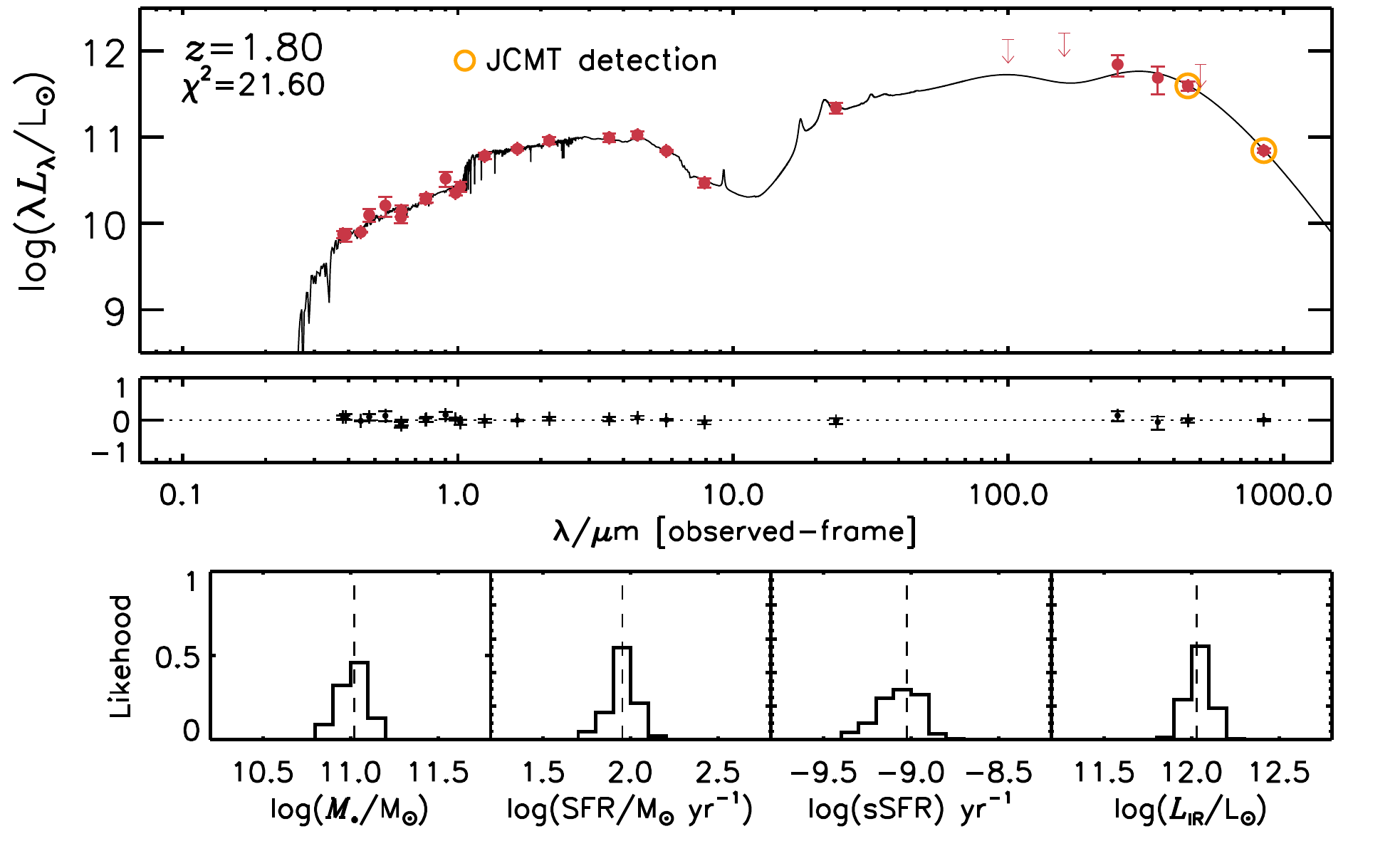} 
\includegraphics[width=0.9\textwidth]{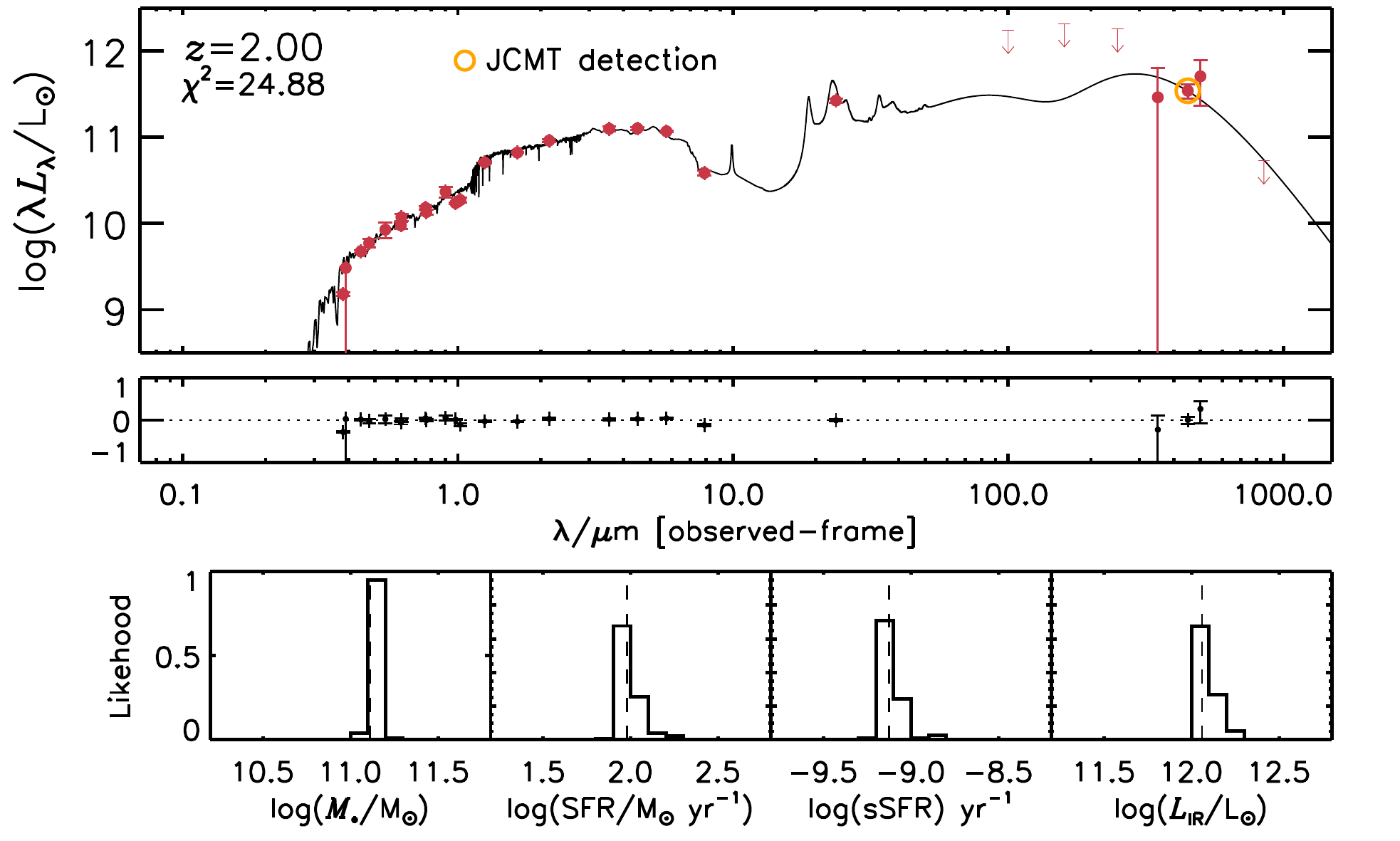}
\caption{Two typical SED fitting examples for 450-$\mu$m detected sources. The red points are the photometry and the red arrows are the upper limit of the photometry. The black lines show the best-fitting template. The orange circles label the JCMT detections. The upper example has both 450-$\mu$m and 850-$\mu$m detection, and the lower example has only 450-$\mu$m detection. The residuals and histograms of the physical parameters (stellar mass, SFR, sSFR, and infrared luminosity) are shown in the lower panels. In the histograms, the dashed lines are the median values.}
\label{studies_sed}
\end{figure*}

\section{Data}
\label{sec2}

\subsection{JCMT SCUBA-2 Data}
\label{sec2_1}

In this paper, we use our extremely deep 450 and 850-$\mu$m data obtained through the STUDIES program, as well as data from the JCMT archive.
A full description of STUDIES is given by \citet{2017ApJ...850...37W}, but we give a brief description here. STUDIES is a multi-year JCMT Large Program, aiming to reach the confusion limit and an rms noise below 0.6 mJy at 450-$\mu$m in the COSMOS-CANDELS region.
In this paper, we include the first two years of data from STUDIES (170 hours).  We also include the extremely deep archival data from the JCMT Legacy Program, S2CLS (\citealp{2013MNRAS.432...53G}, 150 hours, program code: MJLSC01), 
and the shallower but wider archival data of (\citealp{2013MNRAS.436.1919C}, 20 hours, program code: M11BH11A, M12AH11A and
M12BH21A).
The \citet{2013MNRAS.436.1919C} data do not substantially increase the depth in the area covered by STUDIES and S2CLS, 
but provide a wider area for us to expand the sample
size at the bright end.  The data reduction and source extraction are identical to those described in \citet{2017ApJ...850...37W}.
The final combined 450-$\mu$m map as shown in Figure\,~\ref{studies_region} achieves an rms sensitivity of 0.7 mJy in its center. 
The areas that have rms sensitivities better than 1 mJy and 5 mJy are $\sim10$ arcmin$^2$ and $\sim30$ arcmin$^2$, respectively.
For the 850-$\mu$m map, the areas that have rms sensitivities better than 0.12 mJy and 0.15 mJy are $\sim10$ arcmin$^2$ and $\sim30$ arcmin$^2$, respectively.

\subsection{Sample Selection and Catalog Matching}
\label{sec2_2}

In our combined map, there are 248 450-$\mu$m sources detected above 4$\sigma$ \footnote{According to Figure 9 in \citet{2017ApJ...850...37W}, the expected spurious fraction is around 1\% - 10\% for the sample at 20 mJy - 4 mJy. The confusion limit is around 1--2~mJy \citep[e.g., ][]{2013ApJ...762...81C,2013MNRAS.432...53G,2017ApJ...850...37W}.} ($\sigma_{\rm 450}<$ 5 mJy) with a wide range of flux densities (2.8 mJy $<S_{\rm 450}<$ 29.6 mJy with a median value of 6.4 mJy). In the same area, there are 128 850-$\mu$m sources ($\sigma_{\rm 850}<$1 mJy) above 2 mJy. The different cuts at 450$\sigma$m and 850$\sigma$m is due to their different confusion limits. \citet{2017ApJ...837..139C} determined that the confusion limit at 850 $\mu$m is $\simeq1.65$ mJy from their deep 850 $\mu$m map \citep[see also][]{2017MNRAS.465.1789G}. Although our 850-$\mu$m map is slightly deeper than that in \citet{2017ApJ...837..139C}, here we conservatively select sources above 2 mJy to be free from the confusion effect. Above this 2 mJy limit, all our 850-$\mu$m sources have S/N $>6$ and 2.0 mJy $<S_{\rm 850}<$ 16.7 mJy.

We used photometry from the optical to far-infrared (FIR) in the COSMOS2015 catalog \citep{2016ApJS..224...24L}.
First, we matched the 450-$\mu$m sources with sources in the VLA 3 GHz catalog  \citep{2017A&A...602A...6S,2017A&A...602A...3D} using a $4\arcsec$ search radii (expected false match rate, the expected number of objects which are mismatched over the total number of the sample, based on the number density of the population and the search area, is $\simeq$ 0.02). Among the  248 450-$\mu$m detected sources, 132 sources are matched to a VLA 3GHZ counterpart (see Figure\,~\ref{studies_id}). 
Then we used the VLA positions to find counterparts in the COSMOS2015 catalog. 
For the remaining 450-$\mu$m sources, we matched them with mid-infrared (MIR) 24-$\mu$m sources \citep{2009ApJ...703..222L} using $4\arcsec$ search radii (expected false match rate $\simeq$ 0.09), and 80 sources are found (Figure\,~\ref{studies_id}).  We then used $3\arcsec$ search radii to cross-match the 24-$\mu$m sources with \emph{Spitzer} IRAC sources.  
After that, we used the IRAC positions to find the counterparts in the COSMOS2015 catalog.  As a result, there are 198 450-$\mu$m sources (expected false match rate $\simeq$ 0.04) with COSMOS2015 counterparts.

For the 850-$\mu$m sources, we first matched them with the 450-$\mu$m sources with $8\arcsec$ search radii (expected false match rate $\simeq$ 0.09). We also matched 450-$\mu$m-undetected 850-$\mu$m sources with the VLA catalog using $7\arcsec$ search radii (expected false match rate $\simeq$ 0.07), and used the VLA positions to find their COSMOS2015 counterparts. 77 counterparts can be found with this cross-matching method.

We also employ the machine-learning technique to identify optical counterparts of 850-um single-dish sources (An et al., in preparation). The machine-learning method identifies the likely multi-wavelength counterparts to single-dish-detected submillimeter sources by utilizing a training set of precisely located SMGs from ALMA follow-up of the SCUBA-2 Cosmology Legacy Survey's UKIDSS-UDS field (AS2UDS). The precision of the machine-learning classification is 82 percent as shown in \citep{2018ApJ...862..101A}. In our work, the precision reaches to 88\% \citep[][private communication]{2018ApJ...862..101A} because we adopt a smaller search radius ($5\arcsec$) to match the machine-learning classified counterparts to 850-$\mu$m sources.

There are 44 near-infrared detected galaxies classified as the counterparts of 850-$\mu$m sources by the machine-learning method. 
Among them, 39 850-$\mu$m single-dish sources have a counterpart identified by the machine-learning and the cross-matched method mentioned above. Among the 39 sources, 36 of them ($>$92\%) lead to the same optical counterparts. 
The main results in this work are not changed no matter we use solely the cross-matched sample, solely the machine-learning sample, or both sample.
As a result, we adopt the 44 machine-learning classifications as 850-$\mu$m counterparts, and then include an additional 39 cross-matched 850-$\mu$m sources that do not have any counterparts in the machine-learning method (Figure\,~\ref{studies_region}).
Overall, there are 83 850-$\mu$m sources (expected false match rate $\simeq$ 0.07) with COSMOS2015 counterparts.

Among the 248 450-$\mu$m sources and 128 850-$\mu$m sources, 50 (20\%) and 44 (34\%), respectively, do not have any radio, MIR, and machine-learning counterparts.
A plausible explanation for the unidentified sources is that the radio and MIR observations are not deep enough at high redshift. 
The SMG population may start to drop outside the 3 GHz and 24-$\mu$m detection limits above $z\sim3$ as shown in Figure\,~\ref{studies_zlir}. The unidentified sources are likely to be at $z>3$, but we cannot confirm this until we have deeper radio and/or MIR observations or direct ALMA imaging. 

In our sample, we remove infrared and X-ray selected ($L_X$(2-10 keV)$>10^{42}$ ergs/s) AGNs  which are identified by previous work \citep{2016ApJ...819...62C,2016ApJ...817...34M,2017ApJS..233...19C}.
Among the sources with COSMOS2015 counterparts identified with the above procedure, there are 169 450-$\mu$m detected sources and 80 850-$\mu$m detected sources with reliable COSMOS2015 photometric redshifts \citep[a precision of $\sigma_{\Delta z/(1+z_s)}$=0.034 and a catastrophic failure fraction of $\eta$=10\% for $z\sim2$ sources according to][]{2016ApJS..224...24L}. 
We finally reach a sample of 188 sources that are detected at either 450 or 850 $\mu$m, or both, and have photometric redshifts. These are listed in Table~\ref{tab_1}.
We note that 64 of the sources in our sample are detected at both the 450 and 850 $\mu$m.  
There are 31 out of 188 SMGs with high confidence level spectroscopic redshifts \citep{2007ApJS..172...70L} in the COSMOS spectroscopic master catalog (Salvato et al., in preparation). These spectroscopic redshifts, the photometric redshifts of our SMGs are highly reliable (a precision of $\sigma_{\Delta z/(1+z_s)}$=0.024 and a catastrophic failure fraction of $\eta$=3\%). Therefore, we adopt the COSMOS2015 photometric redshifts in this paper. 

We identified star-forming galaxies across the STUDIES image using the COSMOS catalog and a ${\rm NUV}-r$ vs. $r-J$ selection (see \citealp{2013A&A...556A..55I} for more details). To perform a fair comparison, we only considered the 69,820 star-forming galaxies that are located in the same area (700 arcmin$^2$) as our JCMT sample (Figure\,~\ref{studies_region}) and are not classified as STUDIES SMGs. In this way, for all the star-forming galaxies undetected by SCUBA-2, we can set upper limits for their 450-$\mu$m and 850-$\mu$m flux densities for the SED fitting in \S~\ref{sec2_4}.  

To define our comparison sample of star-forming galaxies, we removed sources identified as SMGs.
However, at the high SFR end, some of these STUDIES-undetected star-forming galaxies might be still somewhat bright at 450 $\mu$m. They were not
detected simply because of the incompleteness of our source extraction and the shallower depth in the outer part of our 450 $\mu$m image.  To test if our normal star-forming galaxies are significantly contaminated by dusty SMGs that lie just below our 450 $\mu$m detection threshold, we conducted stacking analyses.  On star-forming galaxies with SFR $>100$ $M_{\odot}$~yr$^{-1}$ at $z=1$--3, we obtained a stacked flux of $0.96\pm0.23$ mJy, or approximately three times lower than the faintest 450 $\mu$m sources in our SMG sample.  We therefore conclude that there is not significant SMG contamination in our comparison sample. As a side note, what is interesting here is the obscured SFR in these galaxies.  The above stacked 450 $\mu$m flux corresponds to an infrared luminosity of $L_{\rm IR} = 1.6 \times 10^{11}~L_\odot$, and thus an obscured SFR of 26 $M_{\odot}$~yr$^{-1}$.  This is much smaller than their mean total SFR of  $166 M_{\odot}$~yr$^{-1}$ estimated by MAGPHYS, and implies that the majority of their star formation is unobscured and is seen in the rest-frame UV.

\subsection{CANDELS Imaging}
\label{sec2_3}

CANDELS \citep{2011ApJS..197...35G,2011ApJS..197...36K} is an \emph{HST} Multi-Cycle Treasury Program using the  Wide Field Camera 3 (WFC3) in the NIR $F125W$ and $F160W$ bands to target five legacy fields (COSMOS, GOODS-N, GOODS-S, UDS, and EGS).
In the COSMOS field \citep{2017ApJS..228....7N}, covers $9\arcmin\times24\arcmin$ to a limit of $H\simeq27$ mag ($F160W$, 5 $\sigma$).
The CANDELS imaging has been reduced and drizzled to a $0\farcs06$ pixel scale and high-resolution (FWHM $\simeq 0\farcs2$) mosaics.
In the CANDELS region (yellow polygon in Figure~\ref{studies_region}), 139 out of our 169 450-$\mu$m sources with redshifts and 58 out of our 80 850-$\mu$m sources with redshifts can be matched to CANDELS sources, 
among the 38,671 CANDELS sources in the COSMOS field \citep{2012ApJS..203...24V, 2017ApJS..228....7N} detected with \texttt{SExtractor} \citep{1996A&AS..117..393B}. For the comparison sample, 19,197 star-forming galaxies are matched to the CANDELS catalog.
We are considering optical light of dusty sources. 
At $z>3$, $F160W$ imaging traces rest-frame emission at $<$0.4 $\mu$m from galaxies.
Therefore we focus on $z<3$ sources in this paper. 
This leaves 128 450-$\mu$m sources, 46 850-$\mu$m sources, and 17,108 compared star-forming galaxies  for our stellar structural analysis.

\subsection{SED Fitting}
\label{sec2_4}

We model the observed photometry of our SMGs and the ${\rm NUV}-r$ vs. $r-J$ star-forming sample with the Multi-wavelength Analysis of Galaxy Physical Properties (MAGPHYS) code \citep{2008MNRAS.388.1595D}. MAGPHYS computes the emission from the stellar populations in galaxies from UV to NIR consistently with the emission from dust at MIR and FIR wavelengths using an energy balance technique. We use the version of the MAGPHYS code that has been modified for sources at high redshifts \citep[see ][]{2015ApJ...806..110D}.
In the fitting, we included photometry from COSMOS2015 \citep[optical: $u$, $B$, $V$, $i^+$, $z^{++}$; MIR: $Y$, $J$, $H$, $K_s$, 3.6 $\mu$m, 4.5 $\mu$m, 5.8 $\mu$m, 8.0 $\mu$m, 24 $\mu$m; FIR: 70 $\mu$m, 100 $\mu$m, 160 $\mu$m, 250 $\mu$m, 350 $\mu$m,  500 $\mu$m; the choice of aperture (3$\arcsec$) and corrections is identical to those used in][]{2017ApJS..233...19C} as well as  450 and 850-$\mu$m flux densities from the SCUBA-2 images. 

For sources that are undetected by SCUBA-2, we adopt 4$\sigma$ upper limits, and the higher value between 2mJy (confusion limit) and 6$\sigma$ as the upper limit at 850 $\mu$m.
Figure\,~\ref{studies_sed} shows two examples of the SMG photometry and SED fitting. We derive the stellar mass ($M_\star$), SFR, specific SFR (sSFR = SFR/$M_\star$) and infrared luminosity (dust luminosity at 3-2000 $\mu$m defined by MAGPHYS) from the SED fitting.  
The typical infrared luminosity of our faint SMGs (450-$\mu$m selected sources) is $10^{11.7}{\rm L_\odot}$ (mean $L_{\rm IR}=10^{11.65\pm0.98}{\rm L_\odot}$; median $L_{\rm IR}=10^{11.77}{\rm L_\odot}$), which places these sources between the LIRG ($L_{\rm IR}>10^{11}{\rm L_\odot}$) and ULIRG ($L_{\rm IR}>10^{12}{\rm L_\odot}$) limits as shown in  Figure\,~\ref{studies_zlir}. 

The SED fitting significantly underestimates the infrared luminosities of some of the sources (see Figure\,~\ref{studies_zlir}), so we visually inspected their SEDs.
We found that this is a result at both low significant photometry and catastrophic failure by MAGPHYS, which fits the optical and IR photometry simultaneously.
The latter is probably because the dominant optical and IR emission of these galaxies do not come from the same physical regions \citep[see][ for a recent example]{2017ApJ...844L..10S} and therefore the energy balance for the UV and FIR in MAGPHYS breaks down.
For significantly underestimated ($>4\sigma$ at 450-$\mu$m or 850-$\mu$m) sources ($\simeq$10\%), we fitted the optical and infrared parts of the SEDs separately. Their SFRs are replaced by the sum of the infrared SFR and uncorrected UV SFR according to \citet{1998ApJ...498..541K,2012ARA&A..50..531K}. In rare cases (4\%) where both the MAGPHYS fitting and our infrared fitting fail to reproduce the observed photometries (at 450-$\mu$m and 850-$\mu$m) within $4\sigma$, the problem seem to lie in the photometries themselves.  We thus remove those sources from our sample.

In this paper, we randomly select five star-forming galaxies with stellar mass and SFR (or sSFR) within 0.1 dex of every 450-$\mu$m SMG, in the same redshift bin as their comparison sample. 
However, we note that for $z\simeq2$, 26$\pm$2\% of the matched star-forming galaxies are included two to four times because of the limited sample size at the high stellar mass end. Therefore, we include these uncertainties by bootstrapping in our analyses.


\begin{figure*}
\centering
\includegraphics[width=0.99\textwidth]{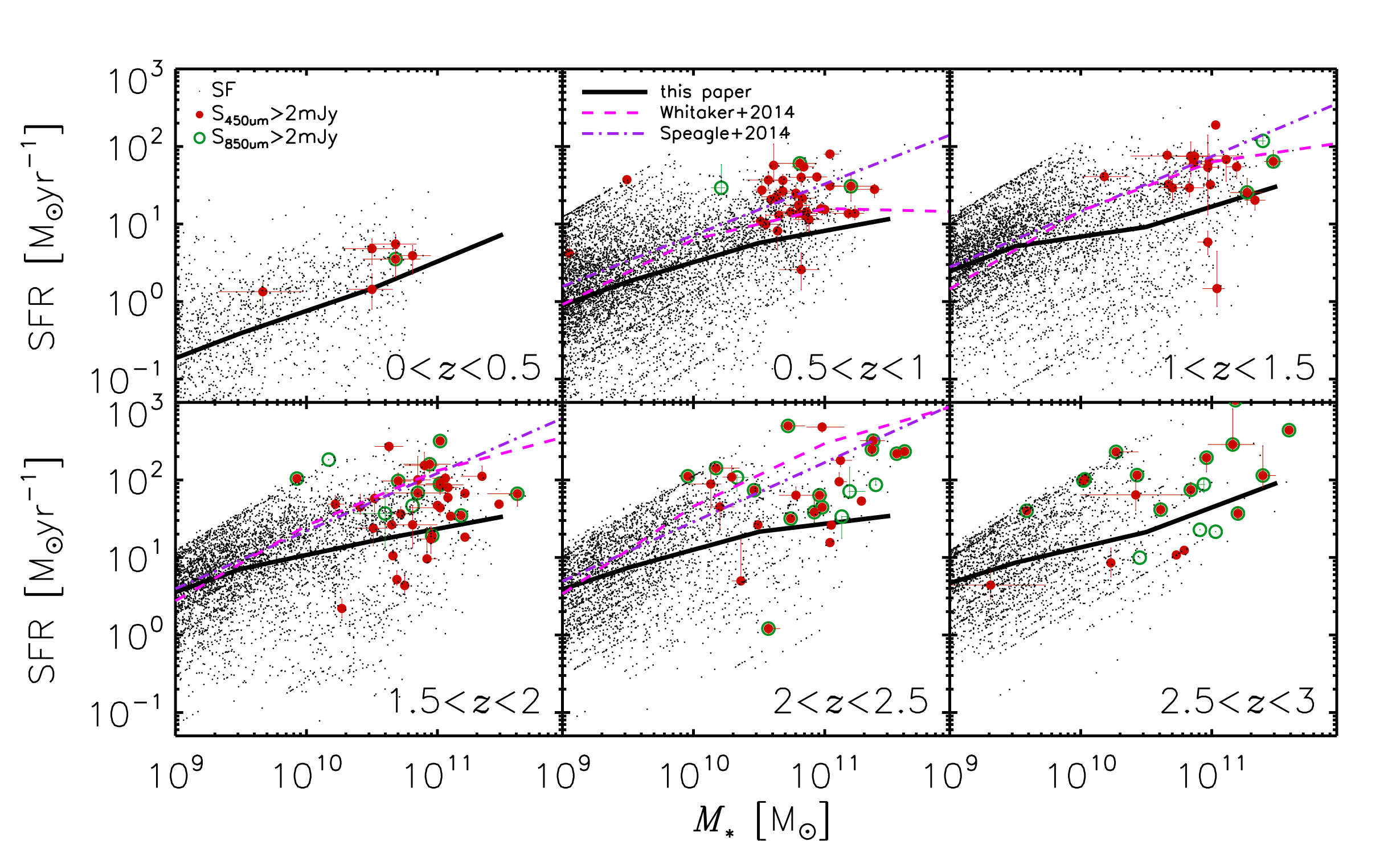} 
\caption{Stellar mass versus SFR relation at $z<3$ for 450-$\mu$m-detected sources (red), 850-$\mu$m sources (green), and a sample of star-forming galaxies (black). The values are derived from our SED fitting.  Most of the SMGs lie on or above the star-forming sequence. The black solid lines are median SFRs of star-forming galaxies in bins of 0.5 dex of stellar mass. We also show the sequences derived by \citet{2014ApJ...795..104W} (magenta dash lines) and \citet{2014ApJS..214...15S} (purple dash-dot lines) .} 
\label{studies_ms}
\end{figure*}

\begin{figure*}
\centering
\includegraphics[width=0.7\textwidth]{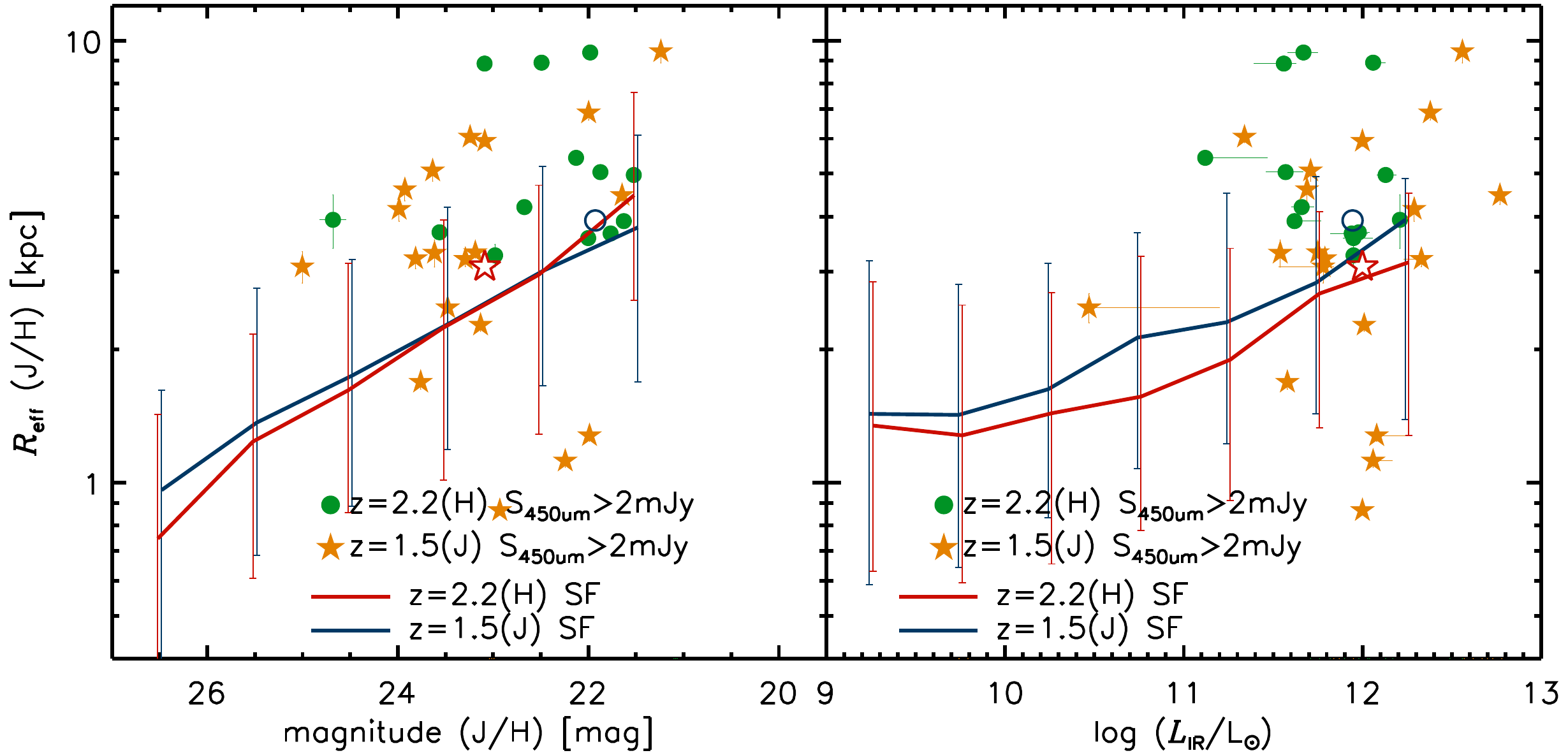} 
\caption{Size to magnitude and infrared luminosity plot. Here we show the effective radius and magnitude at $F160W$ for $z\simeq2.2$ sample (orange star: 450-$\mu$m- sources) and at $F125W$ for $z\simeq1.5$ sample (green circle: 450-$\mu$m- sources. The near-infrared observations correspond to a rest-frame wavelength of 5000{\AA}. The 16th, 50th, and 84th percentiles of size for the star-forming galaxies are shown at different magnitude and infrared luminosity bins. The median values of 450-$\mu$m- detected sources are labeled as red stars ($z\simeq2.2$) and blue circles ($z\simeq1.5$).} 
\label{studies_size}
\end{figure*}

\begin{figure*}
\centering
\includegraphics[width=0.99\textwidth]{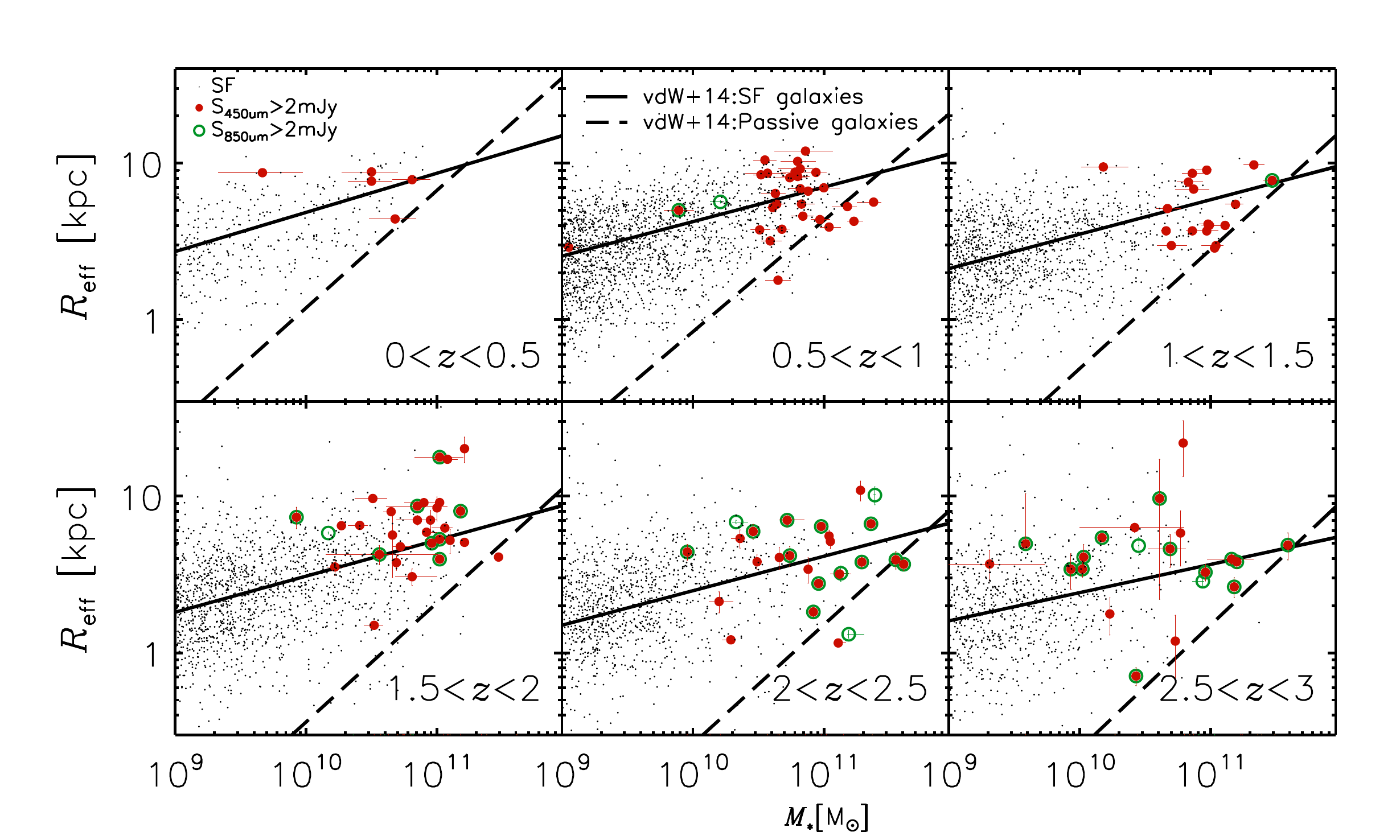} 
\caption{Stellar mass versus size relation at $z<3$ for 450-$\mu$m- detected sources, 850-$\mu$m- detected sources, and a comparison sample of star-forming galaxies. The black lines are fitted to normal star-forming and passive galaxies in \citet{2014ApJ...788...28V}. Most of the SMGs are similar to star-forming galaxies, rather than passive galaxies.}
\label{studies_mr}
\end{figure*}

\begin{figure*}
\centering
\includegraphics[width=0.49\textwidth]{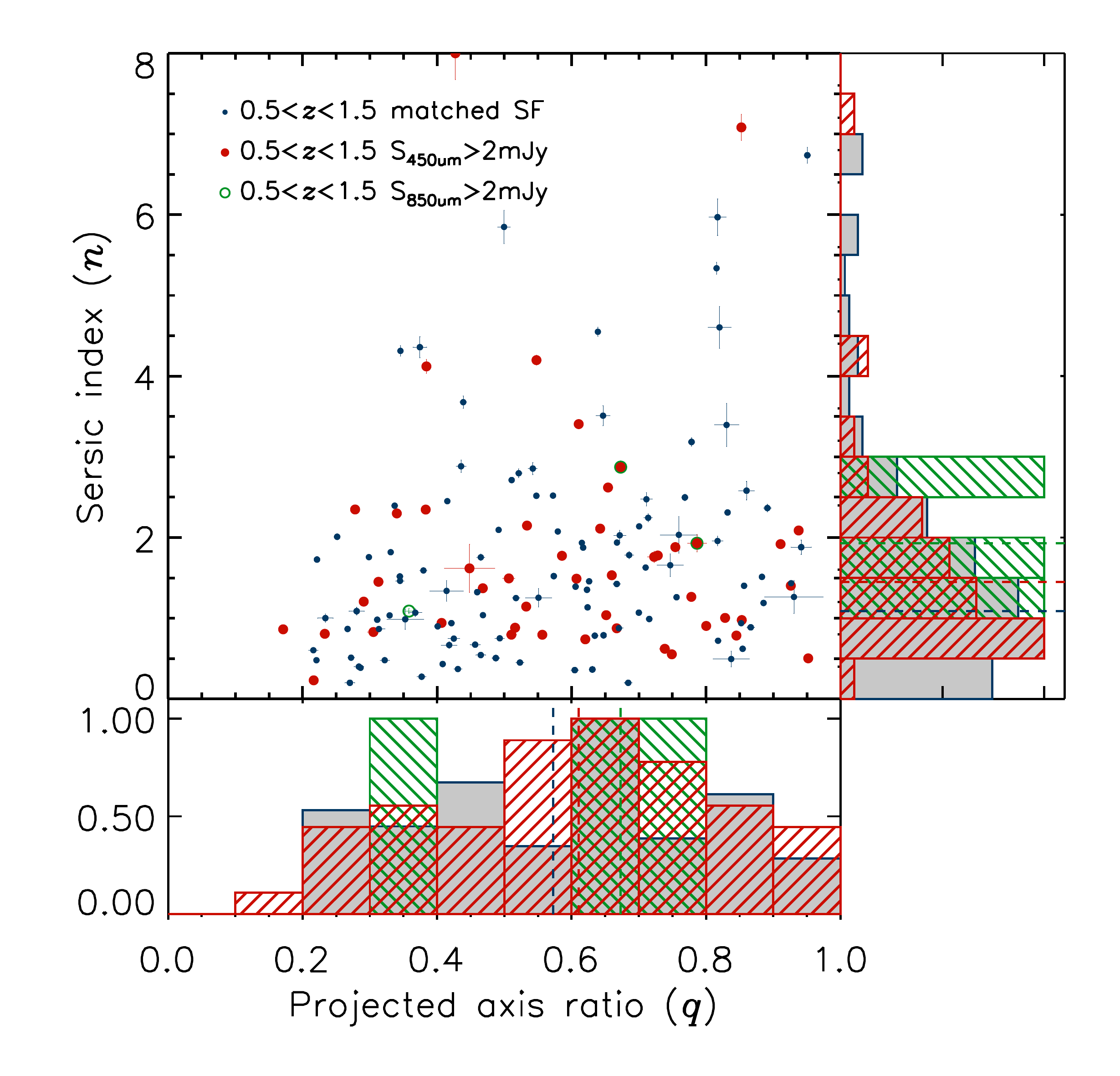} 
\includegraphics[width=0.49\textwidth]{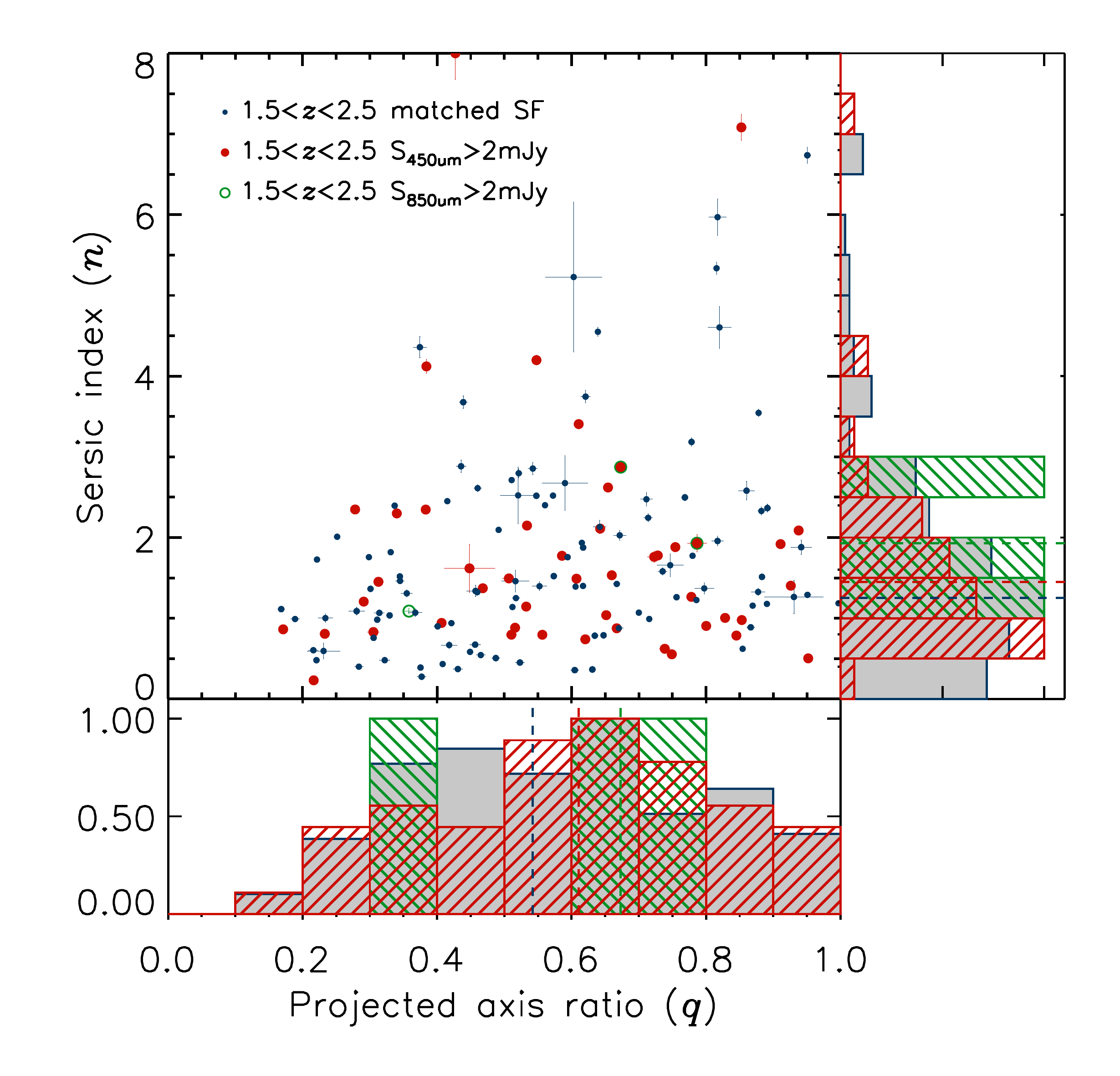} 
\caption{S\'ersic index and axis ratio measured in the $F160W$ band ($H<23.5$) for the 450-$\mu$m-selected sources, 850-$\mu$m-selected sources, and a comparison sample of stellar mass and SFR matched star-forming galaxies at $0.5<z<2.5$ and $1.5<z<2.5$. In the histograms, the dash lines show the median values. There is no significant differences ($P_{\rm K-S}>0.05$) between the SMGs and the matched sample.}
\label{studies_qn}
\end{figure*}

\begin{figure*}
\centering
\includegraphics[width=1.00\columnwidth]{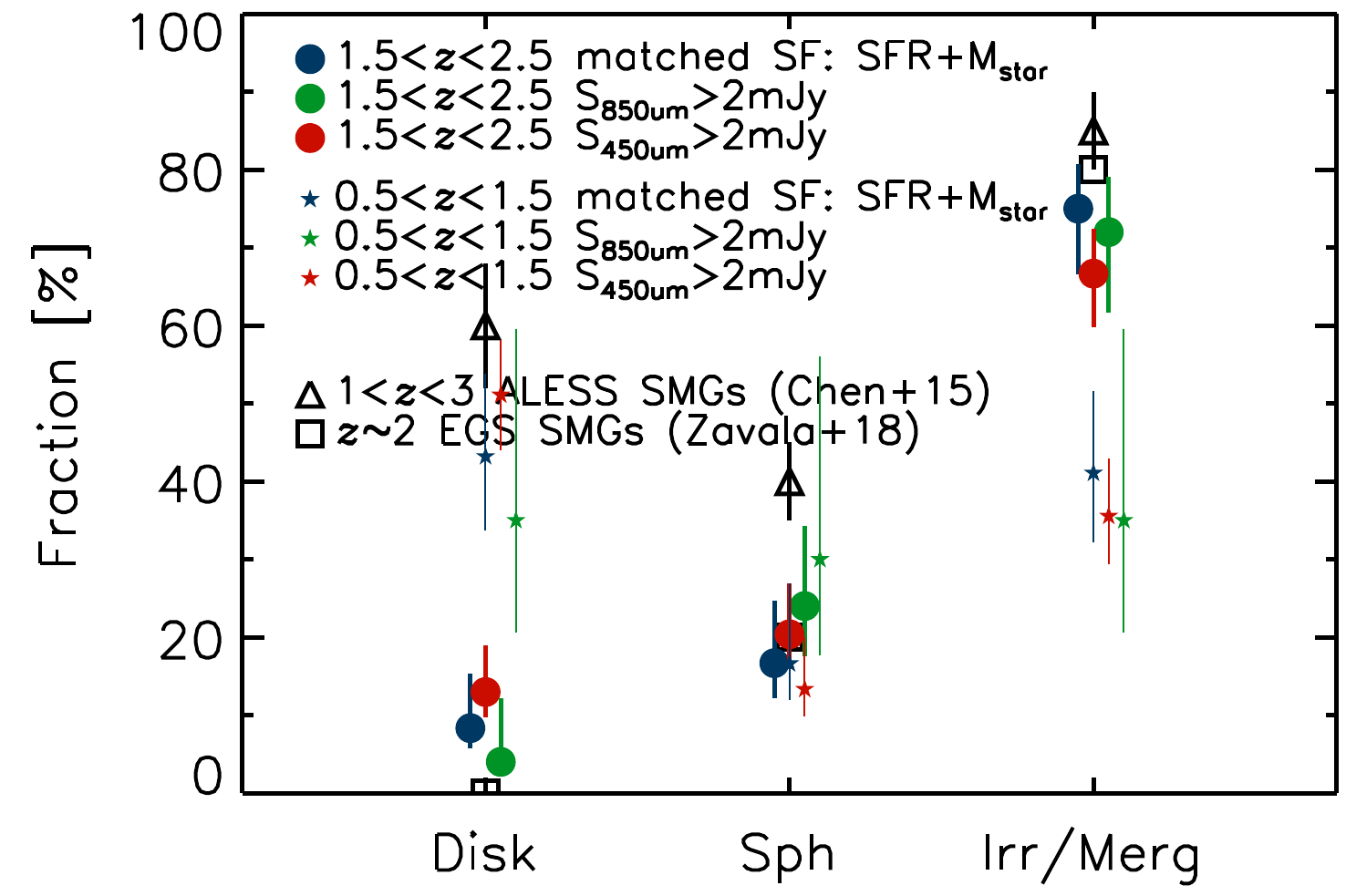} 
\includegraphics[width=1.00\columnwidth]{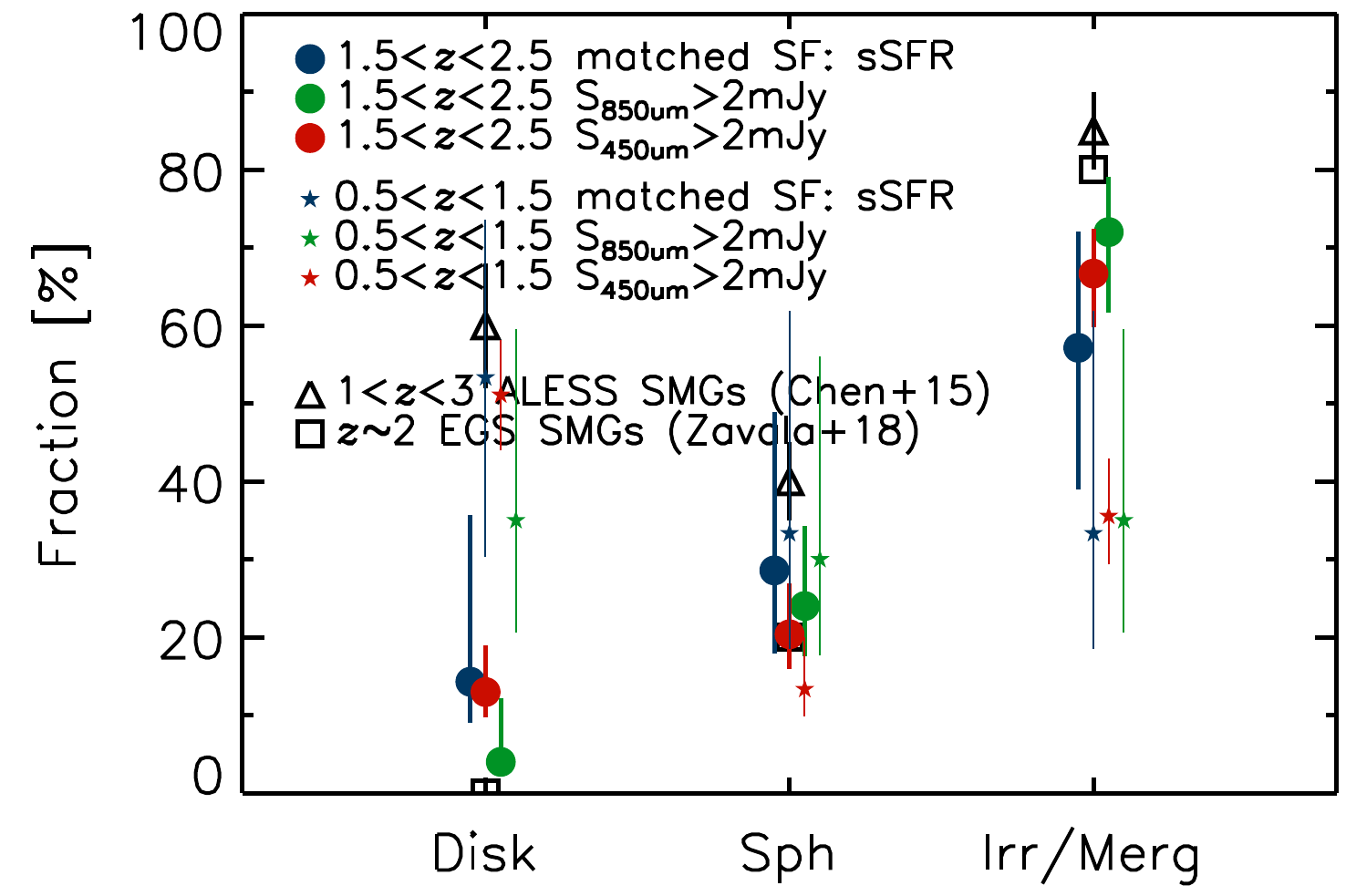} 
\caption{Visual classification of the 450-$\mu$m detected sources and 850-$\mu$m sources. The comparison star-forming galaxies are matched to stellar mass and SFR (left panel), as well as sSFR (right panel). Most of the SMGs contain merger features, similarly or slightly more frequently than the comparison sample. The error bars represent the  68.3\% binomial confidence limits, as described in \citet{2011PASA...28..128C}. For comparison, we show the visual classifications of SMGs providedby \citet{2015ApJ...799..194C} and \citet{2018MNRAS.475.5585Z}.The higher fractions of disks and spheroids in the literature are because of their non-mutually exclusive classifications.}
\label{studies_morph}
\end{figure*}

\begin{figure*}[t]
\centering
\includegraphics[width=1.00\columnwidth]{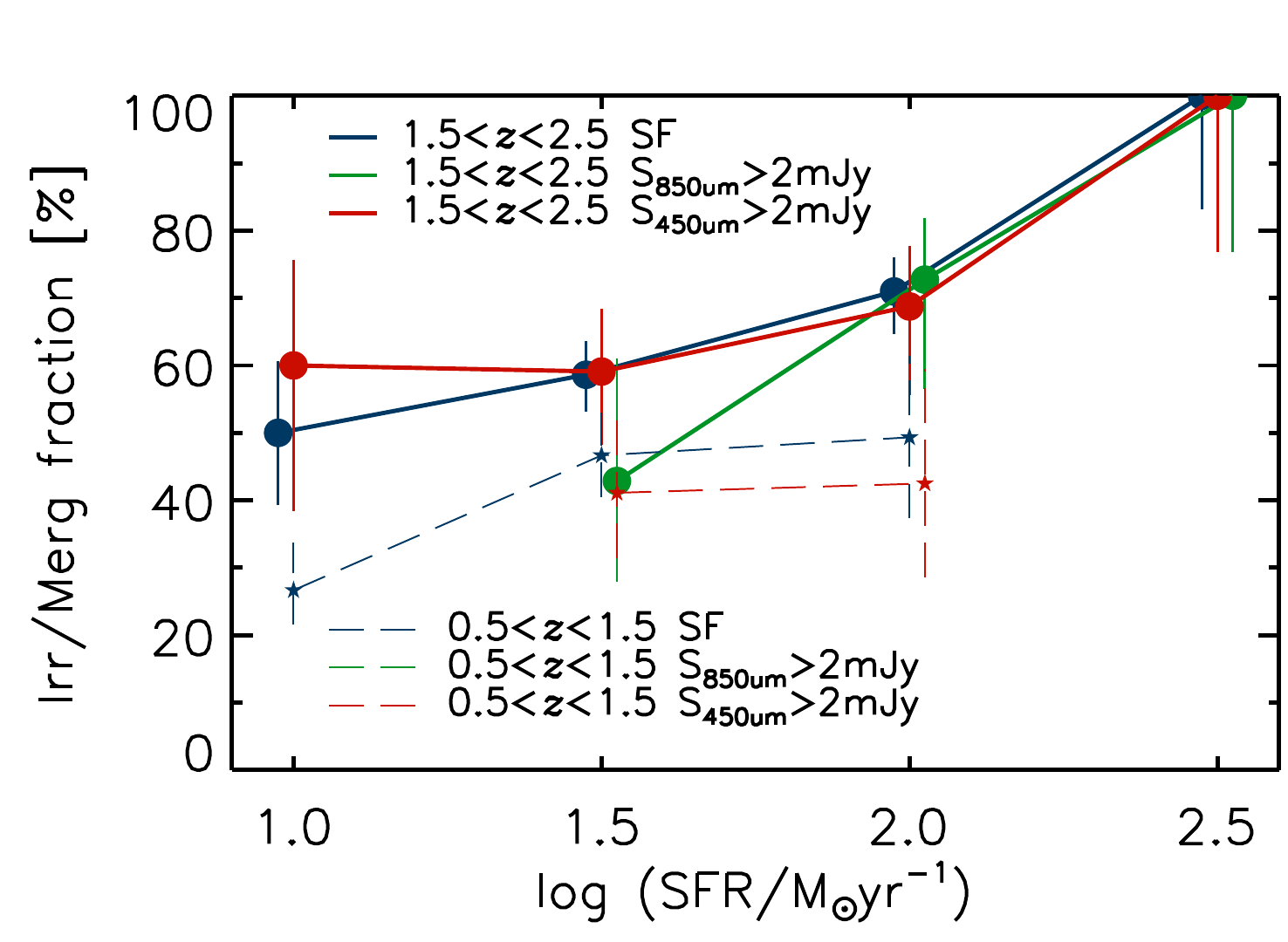} 
\includegraphics[width=1.00\columnwidth]{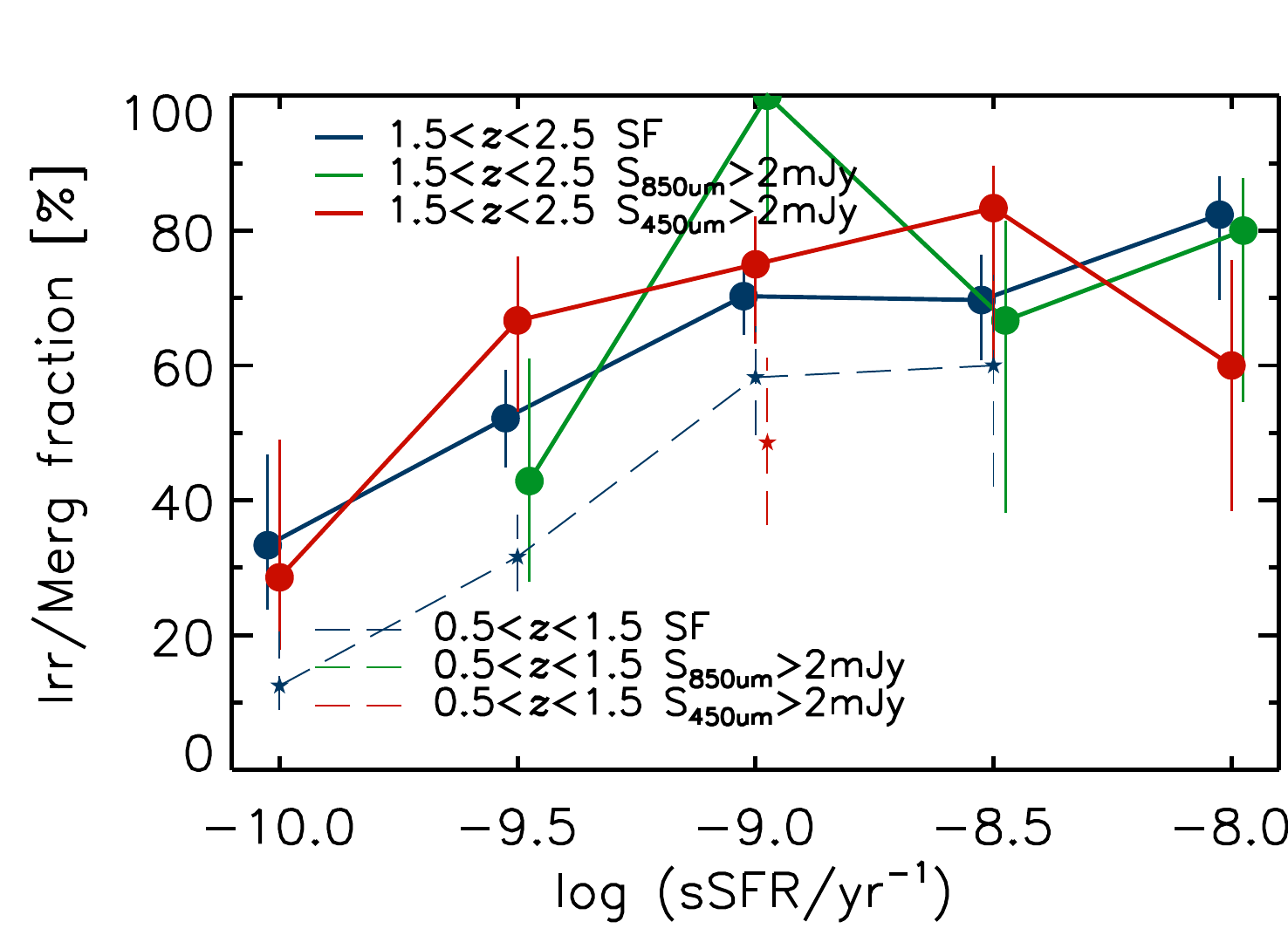} 
\caption{Disturbed feature fraction versus SFR and sSFR for 450-$\mu$m sources, 850-$\mu$m sources, and a comparison sample of $M_*>10^{10} {\rm M_\odot}$ star-forming galaxies at $1.5<z<2.5$. This shows that disturbed features depend on the SFRs for all types of galaxy. The error bars here represents the  68.3\% binomial confidence limits, as described in \citet{2011PASA...28..128C}. We also show the classifications of previous SMGs by \citet{2015ApJ...799..194C} and \citet{2018MNRAS.475.5585Z}.}
\label{studies_morph_sfr}
\end{figure*}

\section{Analysis}
\label{sec3}

\subsection{Star-Forming Sequence}
\label{sec3_1}

In order to investigate the star formation process in dusty galaxies compared to the general galaxy population, we compare our sample with the star-forming sequence \citep[stellar mass vs.\ SFR, also called ``main sequence;'' e.g., ][]{2004MNRAS.351.1151B,2007A&A...468...33E,2007ApJ...660L..43N,2007ApJ...670..156D,2015A&A...579A...2I,2015A&A...575A..74S,2015ApJS..219....8C}.
Based on our SED fitting results, we define our own star-forming sequence at different redshifts in Figure~\ref{studies_ms} (black solid lines). Here we only include galaxies with reasonable SED fitting results by visually inspecting them and deciding an empirical choice of reduced chi-square value ($\chi^2<50$) in the MAGPHYS output files. 
There are 159 (out of 188) such SMGs and the stellar masses and SFRs are listed in Table~\ref{tab_1}.
 
With the same sample selection for star-forming galaxies, the median difference between COSMOS2015 and our SED fitting results are small (-0.01 dex for stellar masses and 0.07 for SFRs).
We also show the main sequence relations from \citet{2014ApJ...795..104W} and \citet{2014ApJS..214...15S}.  However, our SFR estimations are generally lower than those in the literature by $\sim$ 0.7 dex.
The main reason might be selection of the star-forming sample since our median values are close to the COSMOS2015 catalogs with the same sample selection. Moreover, our SFRs would be higher if we consider 24-$\mu$m or \emph{Herschel} selected sample. To avoid bias from different selection criteria, we use the ${\rm NUV}-r$ vs. $r-J$ selection, but show the literature in Figure\,~\ref{studies_ms} for comparison.
The SFR differences between the various works are more significant at the high stellar mass end, so it is important to compare SMGs with star-forming galaxies using SFRs derived with the same method.

In Figure\,~\ref{studies_ms} and Table~\ref{tab_2}, we show that the SMGs are above or on our star-forming sequence, that is, a higher or similar SFR than most of the population at that stellar mass, for both 450-$\mu$m and 850-$\mu$m sources at all redshifts.
If we define the star-forming sequence spanning $\pm$0.3 dex around our median values, about half (48$\pm$5\%) of the faint SMGs (450-$\mu$m sources) lie above the main sequence, and half (43$\pm$4\%) of them are on the main sequence. 

\subsection{GALFIT Measurements}
\label{sec3_2}

We investigate the stellar structure of the STUDIES SMGs that lie in the CANDELS/COSMOS region. 
We adopt the structural parameters of \emph{HST}/$F160W$-selected sources in the CANDELS field for available NIR filters ($F160W$ and $F125W$) in \citet{2012ApJS..203...24V}.
The best-fitting single-component S\'ersic models are produced by \texttt{GALFIT} \citep{2010AJ....139.2097P}.
According to \citet{2012ApJS..203...24V}, a precision and accuracy of $20\%$ or better can be reached for measurements of the effective radius ($R_{\rm e}$) and projected axis ratio ($q$) for $H<24.5$ (75\% of our sample) and S\'ersic index ($n$; the degree of curvature of the  S\'ersic profile) for $H<23.5$ (60\% of our sample) at typical CANDELS depth. In our structural and morphological analyses, we only consider STUDIES sources detected at $H<24.5$ (for $R_{\rm e}$ and $q$) or $H<23.5$ (for $n$), ensuring that the desired properties both highly precise and accurate.

In order to show the rest-frame optical size dependence on magnitude and infrared luminosity, we chose two typical redshift ranges  in Figure\,~\ref{studies_size}.
Both the $z\simeq2.2$ sample at $F160W$ and the $z\simeq1.5$ sample at $F125W$ correspond to a rest-frame wavelength of 5000{\AA}. 
For the 450-$\mu$m sources, the $z\simeq2.2$ sample tends to have smaller sizes than the $z\simeq1.5$ sample. 
This trend of smaller sizes lower redshift is consisent with the known size evolution of the overall galaxy population from $z\simeq3$ to our local universe \citep[e.g.,][]{2013ApJ...765..104B,2014ApJ...791...52B,2014ApJ...788...28V}. 
The median sizes of 450-$\mu$m sources are slightly larger than normal star-forming galaxies at the same magnitude and infrared luminosity. In general, Figure\,~\ref{studies_size} shows that the general population of 450-$\mu$m sources follow the most luminous end of normal star-forming galaxies. Therefore, we derive the effective radius ($R_{\rm e}$) at a rest-frame wavelength of 5000{\AA} according to Eq.~1 
\footnote{$R_{\rm e}$=$R_{\rm e, F}(\frac{1+z}{1+z_p})^{\frac{\Delta\log R_{\rm e}}{\Delta\log\lambda}}$, where F denotes either F125W ($z<1.5$) or F160W ($z>1.5$), and $z_p$ is the pivot redshift for these respective filters (1.5 for F125W and 2.2 for F160W). And $\frac{\Delta\log R_{\rm e}}{\Delta\log\lambda}=-0.35+0.12z-0.25\log(\frac{M_*}{10^10 M_\odot}$), where $M_*$ is the stellar mass.} 
in \citet{2014ApJ...788...28V}, which considers the wavelength dependence of $R_{\rm e}$ as a function of redshift and galaxy stellar mass.
In Figure\,~\ref{studies_mr}, we compare our stellar mass to size relation with that inferred by \citet{2014ApJ...788...28V} (see their Figure 5). Most of the SMGs are similar in size (for their stellar mass) to star-forming galaxies at all redshifts. 

We then consider high accuracy objects ($H<24.5$ for the effective radius) over the range $1.5<z<2.5$. The median effective radius is 4.26$\pm$0.05 kpc, and the mean is 4.68$\pm$0.09 kpc for 450-$\mu$m sources, while the median is 4.38$\pm$0.07 kpc, and the mean is 5.28$\pm$0.10 kpc for 850-$\mu$m sources, compared with a median of 3.13$\pm$0.12 kpc, and a mean of 3.86$\pm$0.12 kpc for a stellar-mass- and SFR-matched sample to the 450-$\mu$m galaxies. 
The uncertainties here and below are estimated from a bootstrapping analysis.
A Kolmogorov-Smirnov (K-S) test shows that the matched star-forming sample is marginally different from both the 450-$\mu$m and the 850-$\mu$m samples ($P_{\rm K-S}\simeq 0.05$).
The 850-$\mu$m sample seems to be more extended than the 450-$\mu$m sample, probably because they are more luminous, and massive sources. The sample sizes are not large enough to show significant differences with the K-S test ($P_{\rm K-S}\simeq 0.95$).  

In the lower redshift range $0.5<z<1.5$, the median effective radius is 4.90$\pm$0.26 kpc, and the mean is 5.55$\pm$0.10 kpc for the 450-$\mu$m sources, while the median is 4.85$\pm$0.61 kpc, and the mean is 4.07$\pm$0.07 kpc for the 850-$\mu$m sources, compared with a median of 4.28$\pm$0.09 kpc, and a mean of 4.75$\pm$0.09 kpc for the 450-$\mu$m matched sample. The sizes of both the 450-$\mu$m and the matched sample at $z\simeq1$ are larger than those at $z\simeq2$, which is consistent with the usual size evolution of galaxies \citep[e.g., ][]{2007ApJ...656...66Z,2007ApJ...671..285T,2008ApJ...677L...5V,2012ApJ...746..162N,2014ApJ...788...28V}. However, the sample size of the 850-$\mu$m sources is too small to constrain the $z\simeq1$ population. On the other hand, the effective radius of the 450-$\mu$m sample seems to be larger than the matched sample, although the difference is not significant, ($P_{\rm K-S}\simeq 0.49$).

In Figure\,~\ref{studies_qn}, we show the S\'ersic index versus the projected axis ratio measured in the $F160W$ band for sources brighter than $H=23.5$ (the S\'ersic index measurement limit).  For the comparison sample, we matched the stellar masses and SFRs of star-forming galaxies with the available 450-$\mu$m sources ($H<23.5$) in the same diagram. 
At z$\simeq$2, the median S\'ersic index is 1.08$\pm$0.07, and the mean is 1.79$\pm$0.06 for the 450-$\mu$m sources, while the median is 0.86$\pm$0.06, and the mean is 1.67$\pm$0.06 for the 850-$\mu$m sources, compared with a median of 1.30$\pm$0.05, and a mean of 1.96$\pm$0.06 for the comparison sample. 
At z$\simeq$1, the median S\'ersic index is 1.45$\pm$0.03, and the mean is 1.80$\pm$0.04 for the 450-$\mu$m sources, while the median is 1.93$\pm$0.88, and the mean is 1.96$\pm$0.03 for the 850-$\mu$m sources, compared with a median of 1.22$\pm$0.03, and a mean of 1.60$\pm$0.04 for the comparison sample.
However, according to the K-S test, there are little differences ($P_{\rm K-S}>0.05$) between the SMGs (both the 450-$\mu$m and 850-$\mu$m selected sources) and the matched star-forming galaxies for their S\'ersic index and projected axis ratio.

\subsection{Visual Classification}
\label{sec3_3}

We create a stellar-mass- matched sample of star-forming galaxies to the STUDIES sources ($H<24.5$) to investigate their morphology visually. 
We use the \emph{HST} WFC3 $F160W$ band, WFC3 $F125W$ band, and ACS $F814W$ band images to study as presented in the Appendix.
Due to the limited sample size at $z\simeq3$, we focus on $z\simeq1$ and $z\simeq2$ sources in this subsection.

First, we compared visual classification of the SMGs with that of stellar mass and SFR matched star-forming galaxies. We classify them as disks, spheroids, and irregular/mergers.
The fractions of these classes are presented in the left panel of Figure~\ref{studies_morph}. 
These classes are mutually exclusive, so the classification represents the dominant morphology. 
All the sources are examined by five classifiers (Y.Y.C., F.F., W.H.W., C.F.L., and Y.T.). 
The error bar in each classification class represents the 68.3\% (1-$\sigma$) confidence limits, derived with the method described in \citet{2011PASA...28..128C}, which estimates the confidence intervals for a population with a Bayesian approach.
In the left panel of Figure\,~\ref{studies_morph}, most of the SMGs have merger or disturbed features (irregular galaxies). However, the stellar mass and SFR matched star-forming sample also show a comparably high disturbed feature fraction. 
The classifiers used a strict definition for the class of irregular/mergers, which includes weak perturbed features. We also checked that the disturbed feature fractions of the SMGs and the comparison sample are still comparable and high if we conducted a less strict classification. 
Such high fractions could be explained by early-to-mid-stage major mergers.  
There is a hint of an elevated disturbed feature fraction in the 850-$\mu$m sample, compared to the 450-$\mu$m sample and the normal star-forming galaxies. However, the difference between the 850-$\mu$m sample and the star-forming galaxies is 2$\sigma$, and thus statistically insignificant.
In Figure\,~\ref{studies_morph}, we also show previous $F160W$-band visual classification results. \citet{2015ApJ...799..194C} classified ALESS SMGs to five non-mutually exclusive classes: disk, spheroid, irregular, unresolved, and unclassified by four classifiers. \citet{2018MNRAS.475.5585Z} used a visual-like classification from \citet{2015ApJS..221....8H}, which is based on neutral networks trained to reproduce the visual mophologies by \citet{2015ApJS..221...11K}. Though the definition of classes are not the same, the high disturbed feature fractions are consistent with our results. 

In order to investigate whether the disturbed feature fraction depends on SFR, we consider a comparison sample of $M_*>10^{10} {\rm M_\odot}$ star-forming galaxies that lie at the same redshift as the SMGs. In the left panel of Figure\,~\ref{studies_morph_sfr}, we show that the fraction of disturbed sources correlates with SFR for the 450-$\mu$m, 850-$\mu$m, and comparison sample. 
At $z\simeq2$, fraction goes up mildly with SFR, although there are no clear distinctions among the disturbed feature fractions of the three samples at any given SFR.
Using the star-forming sample, we find the disturbed feature fraction as a function of SFR to be: 
$f_{\rm irr/merg} {\rm(\%)}= (32\pm 8) + (11\pm 15) \times \log\left[{\rm SFR} / ({\rm M_\odot}~\rm{yr}^{-1})\right] $
where $f_{\rm irr/merg}$ is the irregular/merger fraction. 
Moreover, the correlations are very strong (Pearson correlation coefficient $>0.96$) for all the samples. 
The disturbed fraction does not seem to depend on how dusty the starbursts are; the dependence only seems to be on SFR.
At $z\simeq1$, disturbed feature fractions are lower than those at $z\simeq2$, as shown in Figure\,~\ref{studies_morph}. A possible reason is that our selection at 450 $\mu$m and 850 $\mu$m identifies more massive and luminous sources at $z\simeq2$ (Figure\,~\ref{studies_zlir}). Though the disturbed feature fractions of $z\simeq1$ sources are slightly lower at fixed SFR, it is still difficult to conclude any redshift dependence due to the limited sample sizes. Nevertheless, both $z\simeq1$ and $z\simeq2$ samples show correlations between disturbed feature fractions and their star formation in Figure\,~\ref{studies_morph_sfr}.

In parallel, we checked an sSFR (=SFR/$M_*$) matched sample that is randomly selected using five star-forming galaxies within 0.1 dex in sSFR of every 450-$\mu$m SMG at the same redshift bin. In the right panel of Figure\,~\ref{studies_morph}, we find that the disturbed feature fractions of SMGs are higher than that of the sSFR matched sample. In the right panel of Figure\,~\ref{studies_morph_sfr}, the correlation of the 450-$\mu$m sources is still strong (Pearson correlation coefficient $>0.80$), but not as strong as that of the sSFR-matched galaxies (Pearson correlation coefficient $>0.98$). 


\section{Discussion}
\label{sec4}

\subsection{How do SMGs compare with normal galaxies in the star-forming sequence?}
\label{sec4_1} 

According to our stellar mass and SFR estimation, most of the SMGs are on or slightly above the star-forming sequence as shown in Figure\,~\ref{studies_ms}. 
Despite a decade of observational study, the location of the most luminous, 850-$\mu$m selected SMGs relative to the star-forming main sequence remains hotly debated. Indeed, various studies into the properties of luminous SMGs have concluded that these systems either represent starburst galaxies, which lie significantly above the main sequence \citep[e.g.,][]{2015ApJ...806..110D,2017ApJ...840...78D}, or, conversely, that they are simply represent the massive `tip’ of the known main sequence \citep{2016MNRAS.458.4321K,2017MNRAS.469..492M}. The reason for these discrepant results can typically be traced to systematic uncertainties on the measurement of stellar mass, which is strongly affected by different assumptions on the star formation history \citep{2011ApJ...740...96H,2012A&A...541A..85M,2014A&A...571A..75M}.
STUDIES allows us to extend such studies to a sample of faint 450-$\mu$m sources. 
In Table~\ref{tab_1}, the stellar masses and SFRs of the STUDIES 450-$\mu$m  sources are lower than those of 850-$\mu$m sources at $z<2.5$. The main reason is that the SED peak of typical z$\sim$2 SMGs is around 200-400 $\mu$m, and 450-$\mu$m observations can detect less luminous SMGs compared to 850-$\mu$m observations. However, 450-$\mu$m detected galaxies still have higher stellar mass and SFRs than normal star-forming galaxies. 

Can our result that SMGs lie slightly above the star-forming sequence a consequence of overestimated SFRs?
\emph{Herschel} observations may overestimate FIR fluxes (and hence SFRs) of dusty galaxies due to source clustering \citep{2010MNRAS.409...75H,2017ApJ...850...37W} within coarse resolution (15-35$\arcsec$ FWHM) of SPIRE imaging at 250-500 $\mu$m. Attempt to correct for this flux bias requires either a complete set of prior positions for deblending \citep[e.g.,][]{2014MNRAS.438.1267S}, or assumptions for the properties of the underlying population \citep[e.g.,][]{2012A&A...542A..58B,2016MNRAS.457.4179H}.

To test this, we conducted SED fitting by using only SPIRE (optical+\emph{Spitzer}+PACS+SPIRE) and only SCUBA-2 (optical+\emph{Spitzer}+PACS+SCUBA-2) data in the FIR bands. 
The resulting mean SFR offset is $4\%$ with a scatter of $9\%$ for SPIRE-detected sources (S/N $>3$ at 250 $\mu$m, 350 $\mu$m, or 500 $\mu$m). The difference is relatively small because MAGPHYS estimates SFRs by considering photometry from UV to FIR wavelengths. 
The overestimation can be larger if SFRs are derived monochromatically from SPIRE and SCUBA-2 fluxes. 
In order to avoid such a bias in the SED fitting of our comparison sample, we also considered the upper limits at 450 $\mu$m for them. 
On the other hand, because \emph{Herschel} fluxes are included in their SED fitting and the SCBUA-2 450 $\mu$m photometry is not deep enough for most of them, it is still possible that their SFRs are overestimated in Figure\,~\ref{studies_ms}. However, this scenario would further strengthen our finding that the SMGs from our deep 450 $\mu$m survey can be on or slightly above the star-forming galaxies on the SFR-$M_*$ plane.

We find that 450-$\mu$m selected SMGs ($S_{\rm 450}=2.8$--29.6~mJy; S/N$>4$ at $z<3$) are on or slightly above the star-forming sequence. This result seems robust against potential biases in the estimations of the SFR of our SMG and comparison samples. It is commonly assumed that galaxies above the sequence are undergoing merger-induced starbursts. However, \citet{2017MNRAS.467.1231C} show that dynamically-triggered star formation (e.g. merger/disc-instability) does not necessarily segregate galaxies on the SFR-$M_*$ plane, which may also help to explain the half-on half-off results on the star-forming sequence. Hence even for the SMGs on the star-forming sequence, there may be additional dynamical processes occurring, such as merging. Therefore in the next subsection, we will turn our focus to the stellar structure of SMGs and look for evidence of merging and interaction.

We examined the source density and SFR density per comoving volume for $z=1$--3. 
Above 200 $M_{\odot}$~yr$^{-1}$, the SMG sample dominates over the normal galaxy sample in terms of both source density and SFR density, but the sample sizes are small for both samples. When we go down to $>100 M_{\odot}$~yr$^{-1}$, the normal galaxy sample becomes roughly twice larger than the SMG sample, but their integrated SFR densities are comparable.  Below $100 M_{\odot}$~yr$^{-1}$, the normal galaxy sample strongly dominates in both the source density and SFR density.

It is now clear that once we probe down to SFR of $\sim100$ $M_{\odot}$~yr$^{-1}$, we see both obscured galaxies (appearing as SMGs) and unobscured galaxies (appearing in the optical sample).  Above this limit, SMGs are dominant, and below this, normal galaxies are dominant.  Therefore, from the points of view of morphology (the topic of this paper), SED (obscured vs. unobscured star formation as tested with stacking analyses), and comoving SFR density, we see that as we go deeper in the submillimeter, we start to enter the regime where normal galaxies play more important roles, or dusty galaxies become less important. This is also in concordance with our 450 $\mu$m counts \citep{2017ApJ...850...37W}, which suggest that we can fully account for the 450 $\mu$m background once we can detect faint sources of roughly 0.5--0.8 mJy.  As we further deepen and widen our 450 $\mu$m map, we will publish better constrained faint-end counts at 450 $\mu$m.  We also defer a complete SED analyses of 450 $\mu$m sources versus normal galaxies to a future paper.   All these should help to better understand how ultra luminous dusty galaxies are connected to normal star-forming galaxies and their relative contribution to the cosmic star formation history.

\subsection{Structures of Dusty Galaxies}
\label{sec4_2}

The stellar mass to size relation in Figure\,~\ref{studies_mr} shows that the sizes of SMGs are similar to those of star-forming galaxies, rather than passive galaxies. 
In general, 850-$\mu$m sources are more extended (larger and flatter) than the 450-$\mu$m sources, and 450-$\mu$m sources are more extended than normal star-forming galaxies. 
The larger spatial extent of the 850-$\mu$m sources can be understood through their higher luminosities and stellar masses. Extended stellar structures were also found in previous SMG studies \citep[e.g.,][]{2000ApJ...528..612S,2010MNRAS.405..234S,2013MNRAS.432.2012T,2015ApJ...799..194C}. 

The slight difference in size might be explained if the ${\rm NUV}-r$ vs. $r-J$ selections of the star-forming sample are contaminated by passive galaxies. 
However, such contamination can be removed using the SFR estimated from our SED fitting.
After matching the stellar mass and SFR, we still find a small difference in size between our SMGs and the comparison sample at $z\sim2$ as discussed in \S~\ref{sec3_2}. 
A plausible explanation for the mild size difference is dust extinction.  Recent high-resolution ALMA imaging shows that dust continuum emission from SMGs and massive star forming galaxies is quite compact, compared to their NIR stellar continuum emission \citep[e.g.,][]{2015ApJ...810..133I,2015ApJ...807..128S,2016ApJ...829L..10I, 2016ApJ...833..103H,2017ApJ...841L..25T}. Even if SMGs and normal star-forming galaxies are comparable in the sizes of their stellar components, the highly extincted cores caused by the compact dust components could bias the measured effective radii outward.  More sophisticated analyses are clearly required to further investigate this possibility, including spatially resolved SED fitting for dust extinction and stellar mass, and high-resolution ALMA imaging for low SFR galaxies, as well as multi-wavelength image simulations. Such studies may explain the lack of obvious difference in S\'ersic index and projected axis ratio between SMGs and the matched sample (as shown in Figure\,~\ref{studies_qn}).

Figure\,~\ref{studies_morph} shows that most SMGs (around 70\%) contain irregular/merger features. We find that the irregular/merger fraction is positively correlated with the SFR (Figure\,~\ref{studies_morph_sfr}).
Moreover, the comparison sample, which is plausibly less obscured, behaves identically to the submillimeter selected sample.
Given the high SFRs of 850-$\mu$m sources (as shown in \S~\ref{sec3_1}), it is thus natural to see them having the highest disturbed feature fraction in Figure\,~\ref{studies_morph}. This is consistent with previous morphological studies of submillimeter samples \citep[e.g.,][]{2003ApJ...596L...5C,2003ApJ...599...92C,2010MNRAS.405..234S,2014ApJ...785..111W, 2015ApJ...799..194C}. 

The dependence on sSFR is consistent with that of  \citet{2013ApJ...778..129H}, who showed that the fraction of interacting merger systems increases with the deviation from star-forming sequence. Moreover, \citet{2011A&A...535A..60H} also demonstrated that galaxy-galaxy interactions and mergers have been strongly affecting SFRs by using \emph{Herschel} data.
Unlike the result for the SFR, we see slightly different behaviors of the irregular/merger fractions with sSFR between the SMGs and the matched sample.
The disturbed feature fraction of SMGs seems to be higher than a sSFR- matched sample, as shown in the right panels of Figures~\ref{studies_morph} and \ref{studies_morph_sfr}. 
What this implies is that for galaxies of the same sSFR, those in merging/disturbed systems tend to be more luminous at 450-$\mu$m or 850-$\mu$m, while the undisturbed ones tend to have lower dust obscuration.
A naive explanation is that merging systems tend to have more compact star-forming regions in their cores (as revealed in many recent ALMA observations), while undisturbed systems tend to have disk-wide star formation.  The small spatial extent of dusty star-forming regions in the merging/disturbed systems then lead to stronger extinction in the UV and thus stronger dust re-radiation in the FIR and submillimeter.  This scenario again remains to be tested with more observations and simulations.  We also caution that the differences in irregular/merger fractions are far from huge (72$_{-10}^{+7}$\%, 67$_{-7}^{+6}$\%, 57$_{-18}^{+15}$\% for 850-$\mu$m sources, 450-$\mu$m sources, and sSFR matched star-forming galaxies, respectively), and are statistically insignificant, indicating that even if merging events play a role in triggering SMGs among galaxies with the same sSFR, they are probably not the only factor \citep{2011ApJ...743..159H}.

As well as having the high SFRs and sSFRs, SMGs also have globally low dust temperature and high attenuation (according to our SED fitting, see also \citealt{2012A&A...539A.155M}). Therefore, we checked dependence of the frequency of merger related features on dust temperature and attenuation. 
We found that the Pearson correlation coefficients are not high (0.05 for dust temperature and 0.11 for attenuation), as opposed to the value for SFR versus disturbed feature fraction ($>0.96$). 
Most SMGs do have disturbed features, but the disturbed feature fraction mainly depends on the SFR . 
This suggests that galaxy merging takes place in bright galaxies with high SFRs and can be related to star formation activity. According to our structural and morphological analyses, dusty galaxies are very similar to star-forming galaxies in the rest-frame optical bands.

Recently, several SMGs were imaged at high resolution by ALMA and the results appear to be mixed.  Some of them show clumpy and extended structures \citep[i.e., disk-like, e.g.,][]{2016ApJ...829L..10I}, while others show starbursts in compact regions \citep[e.g.,][]{2017ApJ...837..182O,2017ApJ...850...83F} or irregular morphologies \citep[e.g.,][]{2017A&A...606A..17M}. 
These results show a great variation in the structure of dusty emitting regions in SMGs, and future observations are required to quantify the prevalence of different morphologies in a thorough manner.
Moreover, recent findings show that the stellar morphologies of luminous SMGs appear significantly more extended and disturbed than their ALMA dust images at $z\sim2.5$ \citep{2016ApJ...833..103H,2017ApJ...846..108C}.  
Given these diverse results, it is clear that further investigations of the dust and stellar morphologies of SMGs are necessary.

To summarize, we have found that faint SMGs selected with deep 450-$\mu$m observations have stellar structures similar to those of less luminous star-forming galaxies in the optical sample in terms of S\'ersic index, projected axis ratio, and fraction of galaxies with perturbed features. The 450-$\mu$m sources are slightly more extended than normal star-forming galaxies and also lie on or slightly above the star-forming sequence, but these small differences might be a consequence of various selection effects or dust extinction.  
There is less similarity between the normal star-forming galaxies and the more luminous 850-$\mu$m selected SMGs, in terms of sizes of the stellar distribution.  These results show that as our submillimeter surveys approach the lower luminosity end ($<10^{12}{\rm L_\odot}$), we start to detect normal galaxies on the main sequence statistically. 


\section{Summary}
\label{sec5}

In this paper, we have investigated physical and structural properties of SMGs in the NIR, especially for a faint 450-$\mu$m sample selected from our extremely deep STUDIES image. Our main findings are as follows.

\begin{enumerate}

\item 450-$\mu$m selected SMGs are located above or on the star-forming sequence at $z<3$. If we define the star-forming sequence as being within $\pm$0.3 dex around the median values, about half (48$\pm$5\%) of the faint SMGs (450-$\mu$m-selected sources) lie above the main sequence, and half (43$\pm$4\%) are on the main sequence. 

\item SMGs are similar to star-forming galaxies in the size-mass relation at $z<3$. 

\item As a result of the intrinsic luminosity of each sample, the 850-$\mu$m sources are typically extended than 450-$\mu$m sources, and 450-$\mu$m sources are more extended than normal star-forming galaxies, in terms of the apparent sizes of their stellar components. For the stellar-mass- and SFR-matched sample, the size differences are only marginal between faint SMGs and the comparison galaxies. Such a minor difference may be explained  by the sizes of their dusty regions.

\item SMGs have similar S\'ersic index and projected axis ratio to star-forming galaxies with the same stellar mass and SFR at $z\simeq2$.

\item Both SMGs and the matched star-forming sample show high fraction ($\sim$80\%) of disturbed features, and the irregular/merger fractions of both SMGs and normal star-forming galaxies show similar SFR dependence. 

\item Our results suggest that galaxy merging can be related to star formation activity, and stellar structures of SMGs are similar to normal star-forming galaxies of comparable stellar mass.

\item Among SMGs and normal star-forming galaxies of similar sSFR, merging/disturbed systems tend to appear in the submillimeter sample as dusty sources, while undisturbed systems tend to show up in the optical sample. However, the tendency is not strong, indicating that galaxy merging is not the only factor in the triggering of SMGs.

\item Our results based on the STUDIES data show that as submillimeter surveys approach lower luminosities ($<10^{12}{\rm L_\odot}$). We start to detect large samples of normal galaxies that lie on the main sequence at $z<3$. 

\end{enumerate}


\begingroup 
\setlength{\tabcolsep}{2pt}
\LTcapwidth=\textwidth
\begin{longtable*}{ccccccccccccc}
\caption{List of 188 SMGs sources (169 450-$\mu$m and 80 850-$\mu$m optical-matched sources). Here we show their IDs (COSMOS2015), photometric redshifts (COSMOS2015), coordinates (COSMOS2015), 450-$\mu$m fluxes, 450-$\mu$m signal-to-noise ratios, 850-$\mu$m flux, 850-$\mu$m signal-to-noise ratios, stellar masses, SFRs, effective radii, projected axis ratios, and S\'ersic indices.}\\
\hline
\hline
ID & $z$ & RA & DEC & $S_{\rm 450}$ & ${\rm S/N}_{\rm 450}$ & $S_{\rm 850}$ & ${\rm S/N}_{\rm 850}$ & $M_*$ & SFR & ${\rm R}_e$ & $q$ & $n$\\
 &  &  [deg] & [deg] & [mJy] &  &  [mJy]  &  & [${\rm M}_\odot$] & [${\rm M}_\odot {\rm yr}^{-1}$] & [kpc] &  &  \\
\hline 
611035 & 1.03 & 150.07354 & 2.22639 & 22.19 & 4.48 & - & - & 10.83$\pm$0.10 & 1.47$\pm$0.12 & 7.56$\pm$0.07 & 0.65$\pm$0.00 & 1.04$\pm$0.02\\
616608 & 2.15 & 150.12496 & 2.23698 & 8.29 & 3.52 & 2.23 & 6.18 & 10.33$\pm$0.06 & 2.03$\pm$0.10 & 6.84$\pm$0.17 & 0.22$\pm$0.01 & 0.20$\pm$0.02\\
619287 & 2.03 & 150.17187 & 2.24070 & - & - & 2.84 & 6.46 & 11.13$\pm$0.03 & 1.53$\pm$0.14 & 3.22$\pm$0.36 & 0.82$\pm$0.02 & 1.64$\pm$0.15\\
623330 & 1.85 & 150.09406 & 2.24590 & 8.60 & 3.81 & 2.87 & 8.86 & 10.17$\pm$0.00 & 2.26$\pm$0.00 & 5.81$\pm$0.08 & 0.29$\pm$0.01 & 0.51$\pm$0.02\\
623536 & 2.31 & 150.05248 & 2.24560 & 22.74 & 6.19 & 6.38 & 12.21 & 11.37$\pm$0.09 & 2.51$\pm$0.08 & - & - & -\\
... & ... & ... & ... & ... & ... & ... & ... & ... & ... & ... & ... & ... \\
\hline
\label{tab_1}
\end{longtable*}
\endgroup

\begingroup 
\LTcapwidth=\textwidth
\begin{longtable*}{ccccccc}
\caption{The median values of stellar masses and SFRs of  $M_*>10^{10} {\rm M_\odot}$  galaxies in Figure\,~\ref{studies_ms}. The uncertainties are estimated by bootstrapping.}\\
\hline
\hline
Samples & $0<z<0.5$ & $0.5<z<1$ & $1<z<1.5$ & $1.5<z<2$ & $2<z<2.5$ & $2.5<z<3$ \\ 
\hline 
 & \multicolumn{6}{c}{$\log(M_*/{\rm M_\odot})$}\\
\cline{2-7}
Star-Forming Galaxies & 10.32$\pm$0.01 & 10.39$\pm$0.02 & 10.42$\pm$0.02 & 10.38$\pm$0.02 & 10.35$\pm$0.02 & 10.27$\pm$0.01\\
850-$\mu$m sources & 10.68$\pm$0.00 & 10.81$\pm$0.30 & 11.39$\pm$0.10 & 10.95$\pm$0.00 & 10.97$\pm$0.01 & 10.93$\pm$0.01\\
450-$\mu$m sources & 10.68$\pm$0.09 & 10.82$\pm$0.01 & 10.97$\pm$0.02 & 10.94$\pm$0.01 & 10.94$\pm$0.04 & 10.76$\pm$0.00\\
\cline{2-7}
 & \multicolumn{6}{c}{$\log({\rm SFR}/{\rm M_\odot} yr^{-1})$}\\
\cline{2-7}
Star-Forming Galaxies & 0.17$\pm$0.02 & 0.78$\pm$0.02 & 1.01$\pm$0.03 & 1.24$\pm$0.03 & 1.36$\pm$0.03 & 1.33$\pm$0.04\\
850-$\mu$m sources & 0.55$\pm$0.00 & 1.49$\pm$0.16 & 1.81$\pm$0.20 & 1.88$\pm$0.06 & 1.90$\pm$0.01 & 2.00$\pm$0.01\\
450-$\mu$m sources & 0.59$\pm$0.02 & 1.32$\pm$0.01 & 1.73$\pm$0.01 & 1.69$\pm$0.03 & 1.83$\pm$0.01 & 2.00$\pm$0.00\\
\hline
\label{tab_2}
\end{longtable*}
\endgroup

\acknowledgments

We thank the JCMT/EAO staff for the observational support and the data/survey management, and the contributions of the entire COSMOS collaboration.
Y.Y.C., W.H.W., and C.F.L. acknowledge financial support from  the Ministry of Science and Technology of Taiwan (105-2112-M-001-029-MY3). WIC acknowledges financial support from the ERC consolidator grant 681627 BUILDUP. LCH was supported by the National Key R\&D Program of China (2016YFA0400702) and the National Science Foundation of China (11473002, 11721303). M.J.M.~acknowledges the support of the National Science Centre, Poland through the POLONEZ grant 2015/19/P/ST9/04010; this project has received funding from the European Union's Horizon 2020 research and innovation programme under the Marie Sk{\l}odowska-Curie grant agreement No. 665778. IRS acknowledges support from STFC (ST/P000541/1), the ERC Advanced Investigator programme DUSTYGAL 321334 and a Royal Society/Wolfson Merit Award. X.S. acknowledges the support from Chinese NSF through grant 11573001, and National Basic Research Program 2015CB857005. YT acknowledges support from JSPS KAKENHI (Grant No.18J01050).

This work is based on observations taken by the CANDELS Multi-Cycle Treasury Program with the NASA/ESA HST, which is operated by the Association of Universities for Research in Astronomy, Inc., under NASA contract NAS5-26555. The submillimeter data used in this work include archival data from the S2CLS program (program code MJLSC01) and
the PI program of Casey et al. (2013, program code M11BH11A, M12AH11A and M12BH21A).

The James Clerk Maxwell Telescope is operated by the East Asian Observatory on behalf of The National Astronomical Observatory of Japan, Academia Sinica Institute of Astronomy and Astrophysics, the Korea Astronomy and Space Science Institute, the National Astronomical Observatories of China, and the Chinese Academy of Sciences (Grant No. XDB09000000), with additional funding support from the Science and Technology Facilities Council of the United Kingdom and participating universities in the United Kingdom and Canada.





\appendix

\section{NIR imaging of SMGs}
Here we show cutouts of the SMGs in Table~\ref{tab_1} if the NIR images are available.
The five panels are true color images \citep[modified from the code by][]{2004PASP..116..133L}, \emph{HST}/ACS $I$-band, along with the \emph{HST}/WFC3 $F125W$-band image, \emph{HST}/WFC3 $F160W$-band, and IRAC (3.6 $\mu$m) images. The box size is 6 arcsec $\times$ 6 arcsec, and the center is the optical counterpart in COSMOS2015. The positions of VLA 3GHz (x symbol) and 24-$\mu$m counterparts (plus symbol) for 450-$\mu$m sources are also labeled in the true color images. 
The COSMOS2015 ID, photometric redshift ($z$), effective radius ($R_{\rm e}$), projected axis ratio ($q$) , S\'ersic index ($n$) , 450-$\mu$m flux density, and 850-$\mu$m flux density are also given.

\begin{figure}
\centering
\includegraphics[width=0.95\columnwidth]{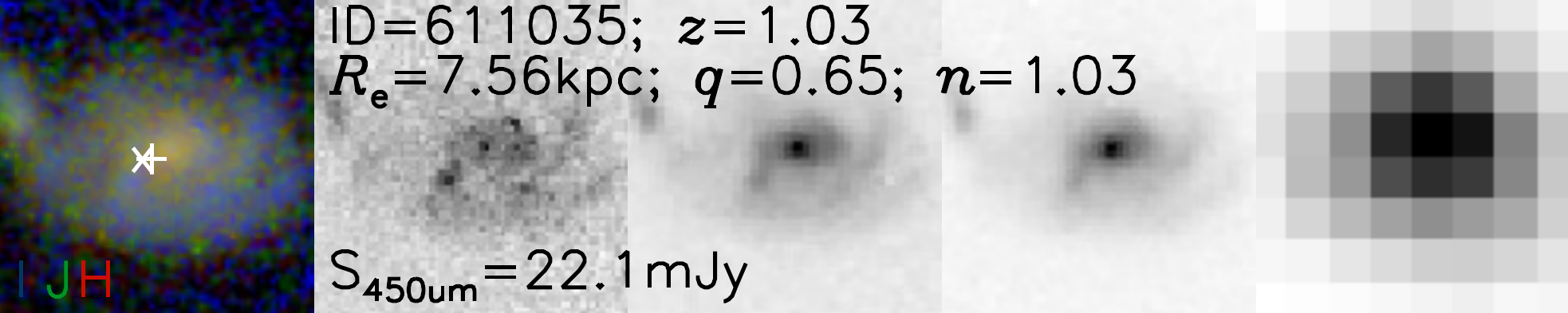}
\includegraphics[width=0.95\columnwidth]{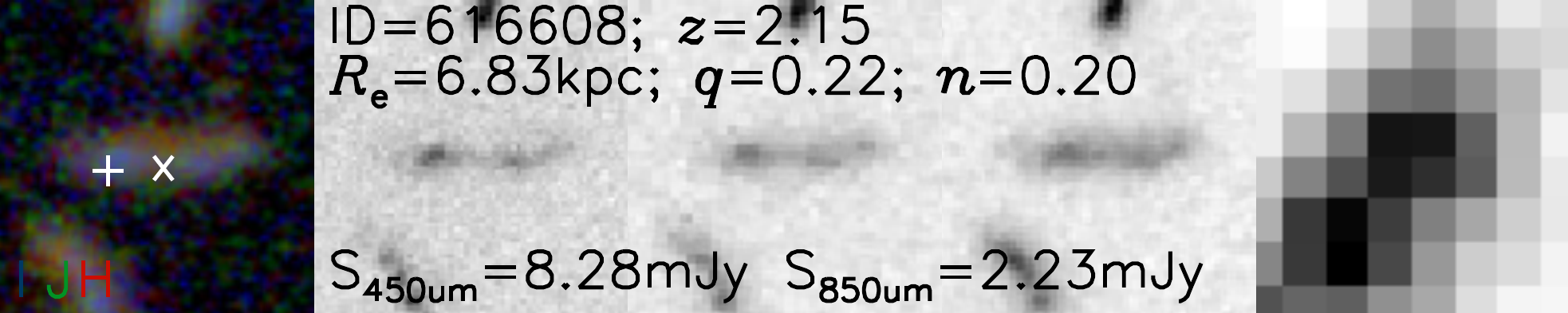}
\includegraphics[width=0.95\columnwidth]{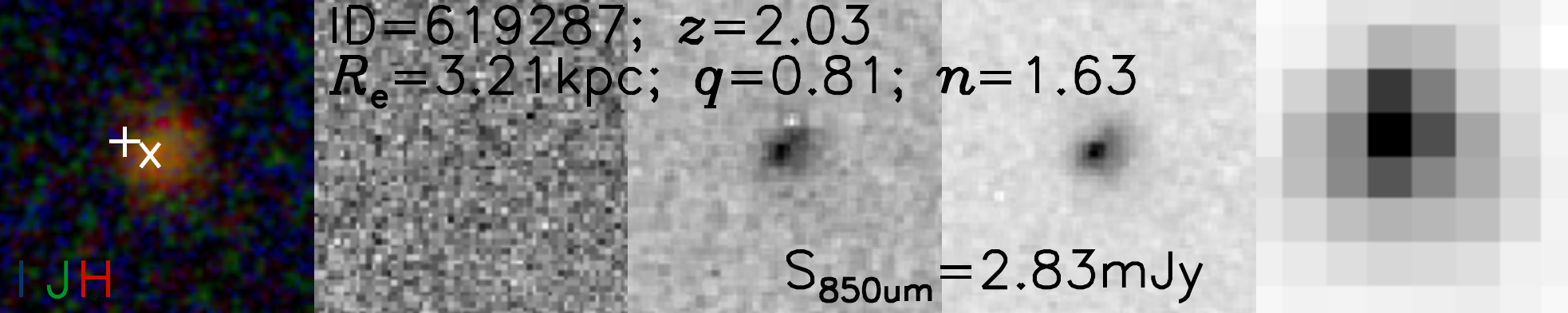}
\includegraphics[width=0.95\columnwidth]{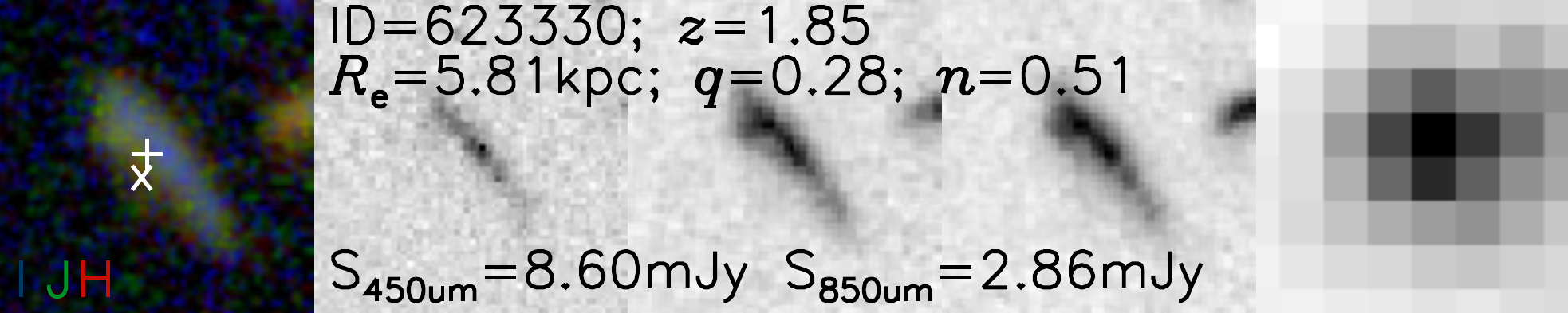}
\includegraphics[width=0.95\columnwidth]{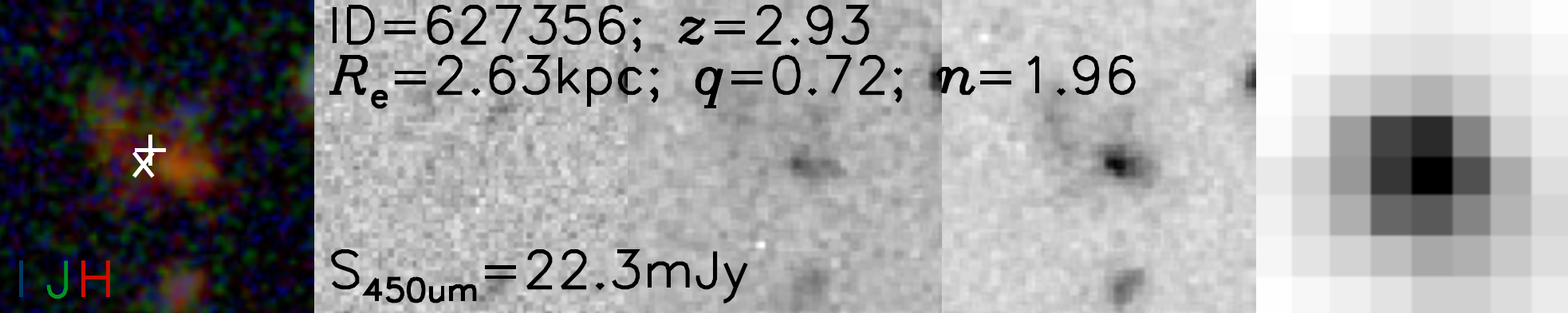}
\includegraphics[width=0.95\columnwidth]{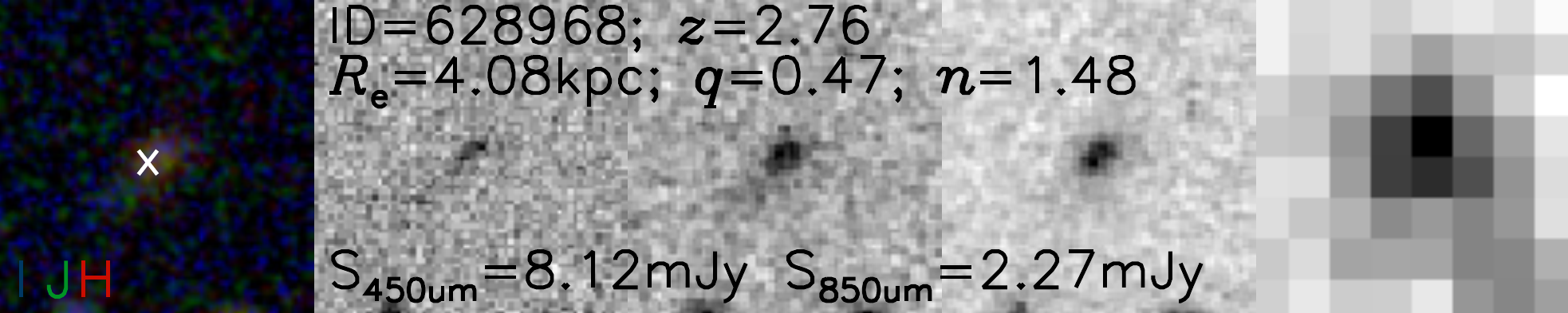}
\includegraphics[width=0.95\columnwidth]{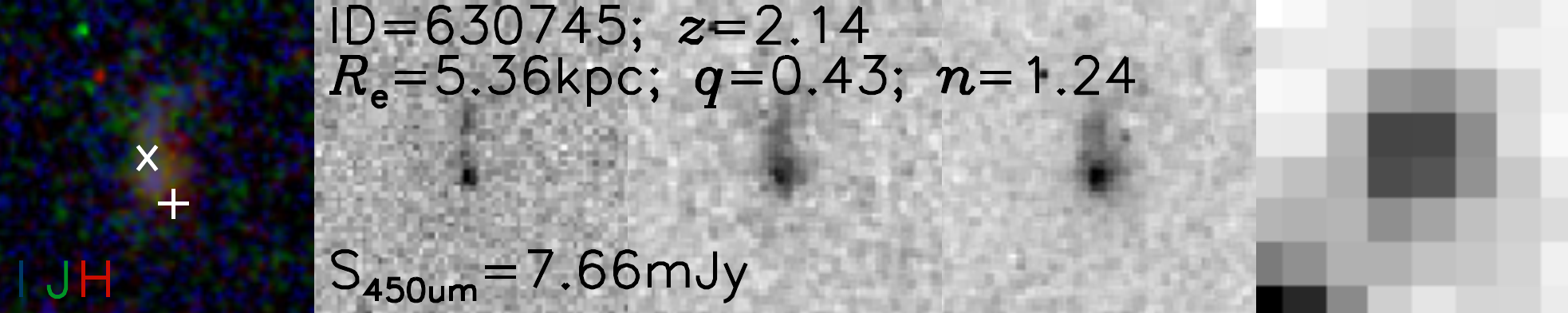}
\includegraphics[width=0.95\columnwidth]{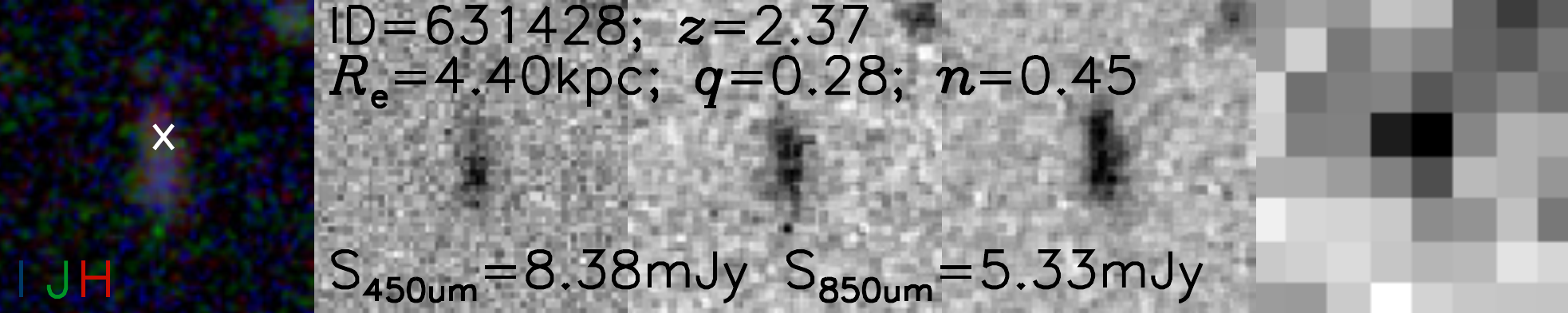}
\includegraphics[width=0.95\columnwidth]{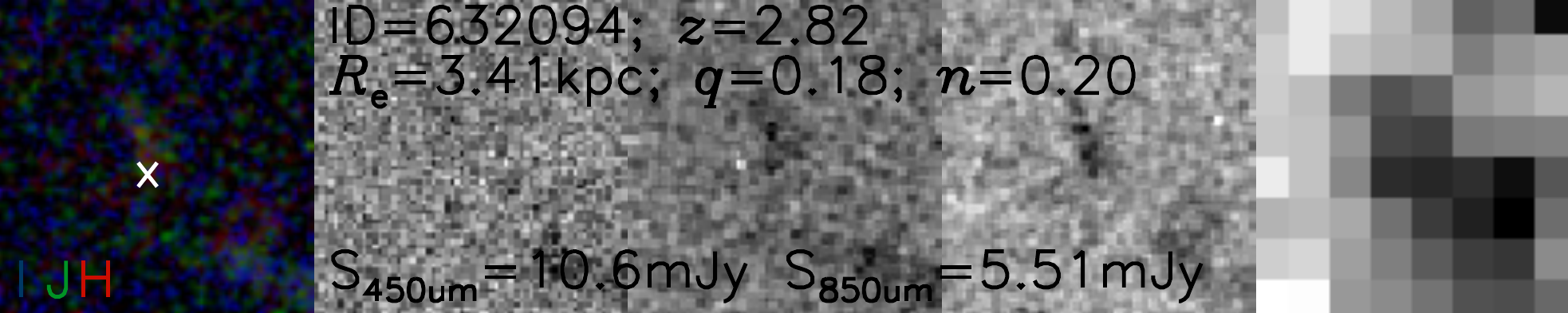}
\includegraphics[width=0.95\columnwidth]{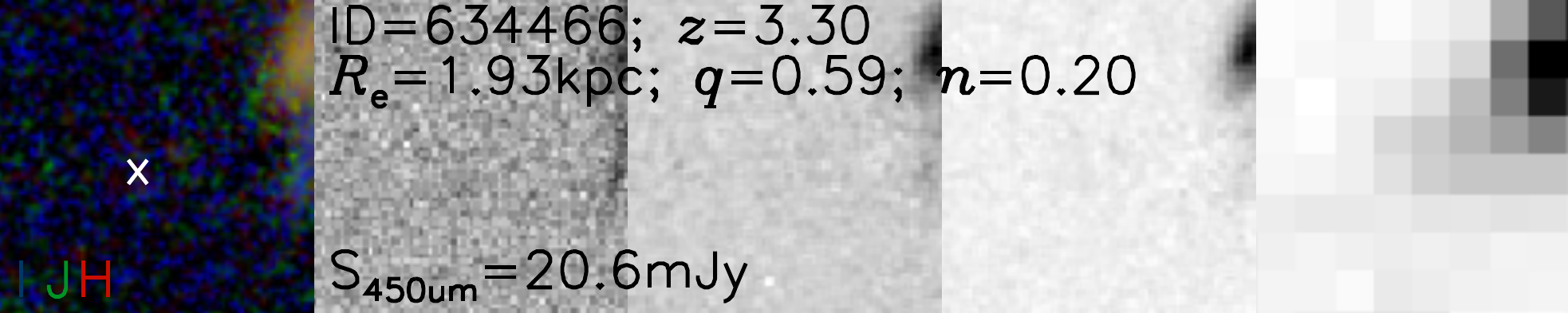}
\includegraphics[width=0.95\columnwidth]{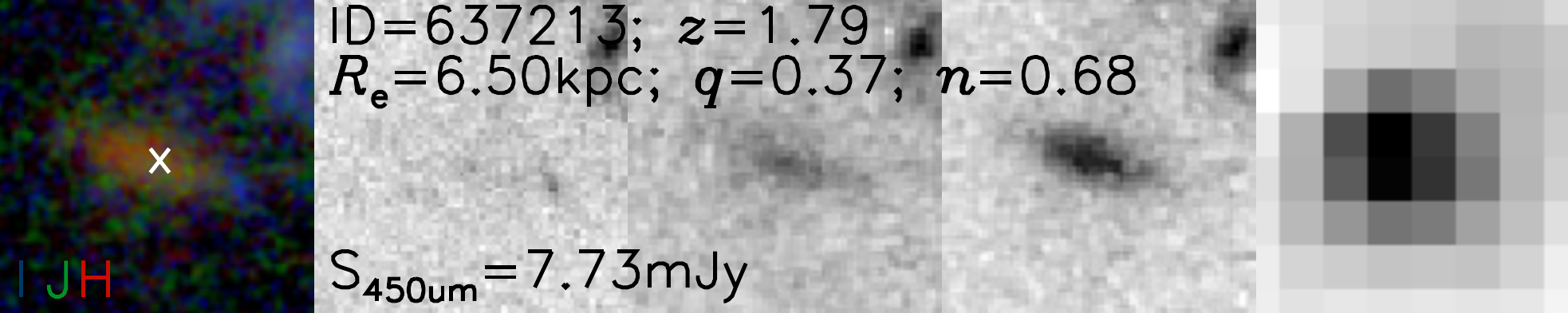}
\includegraphics[width=0.95\columnwidth]{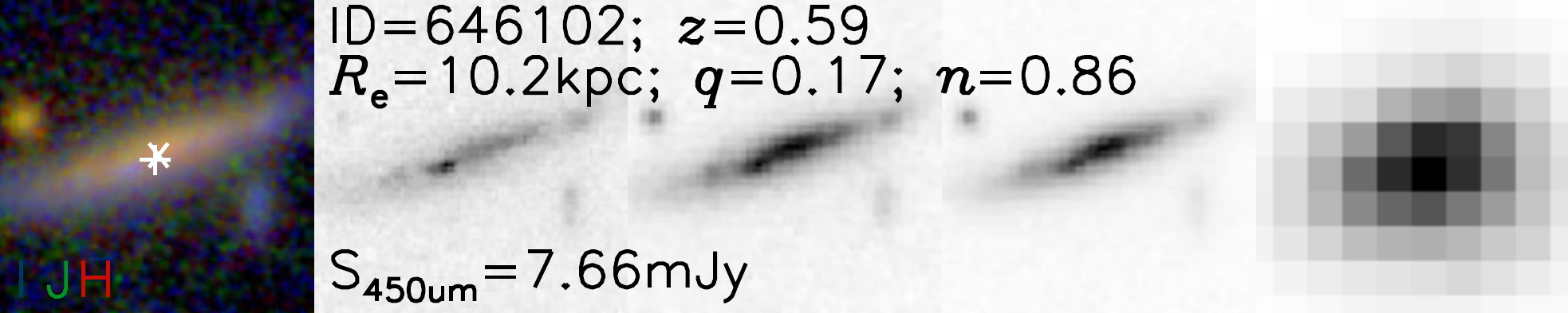}
\includegraphics[width=0.95\columnwidth]{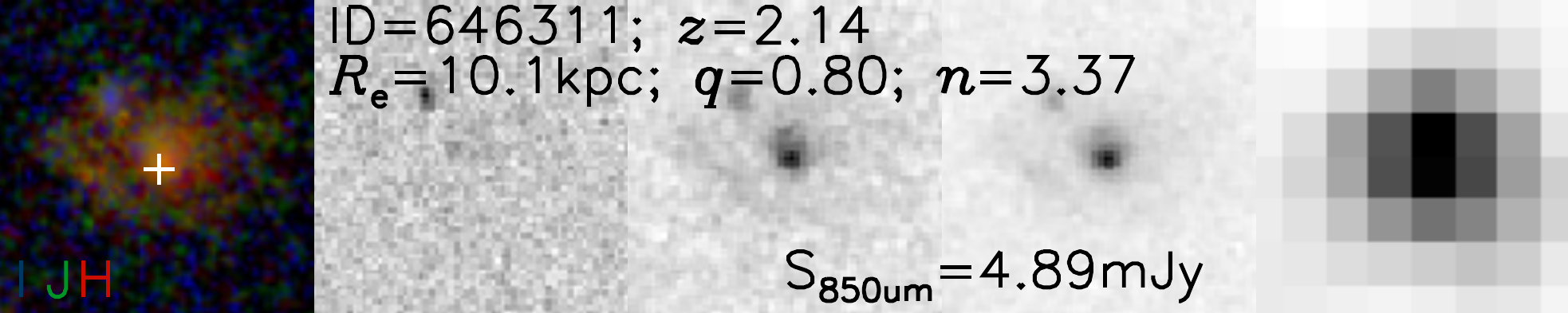}
\includegraphics[width=0.95\columnwidth]{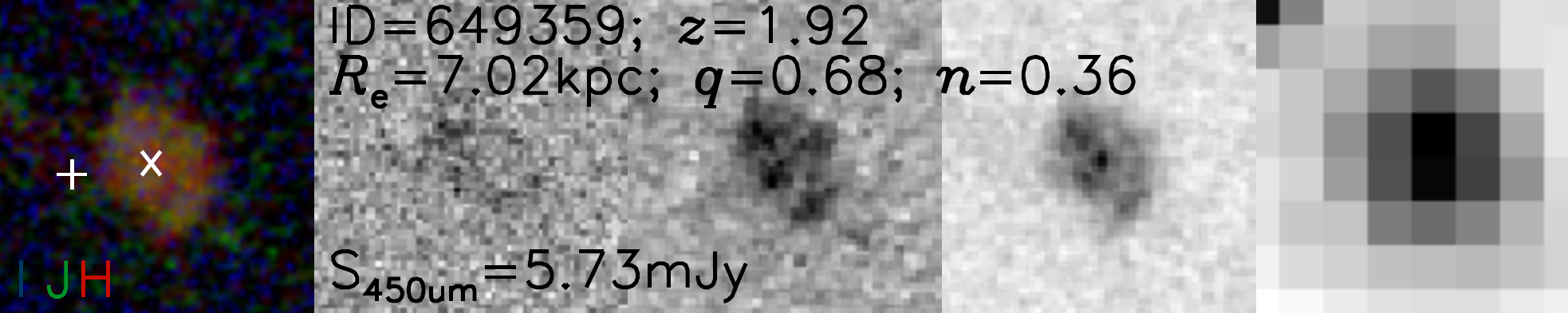}
\includegraphics[width=0.95\columnwidth]{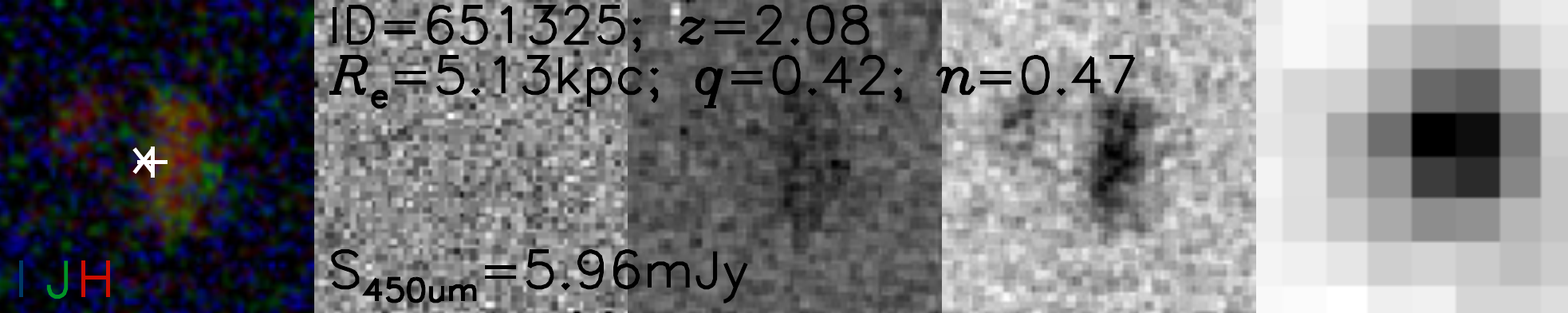}
\end{figure}
\begin{figure}
\centering
\includegraphics[width=0.95\columnwidth]{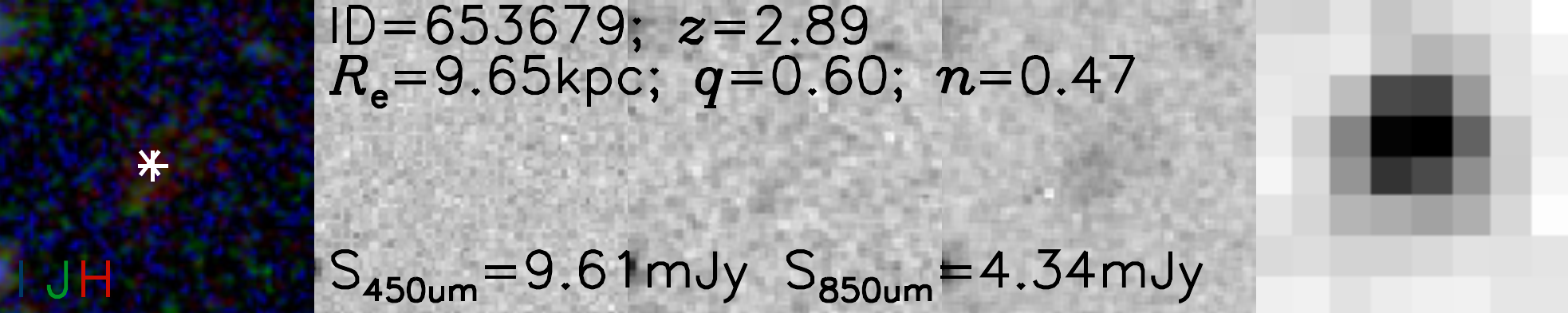}
\includegraphics[width=0.95\columnwidth]{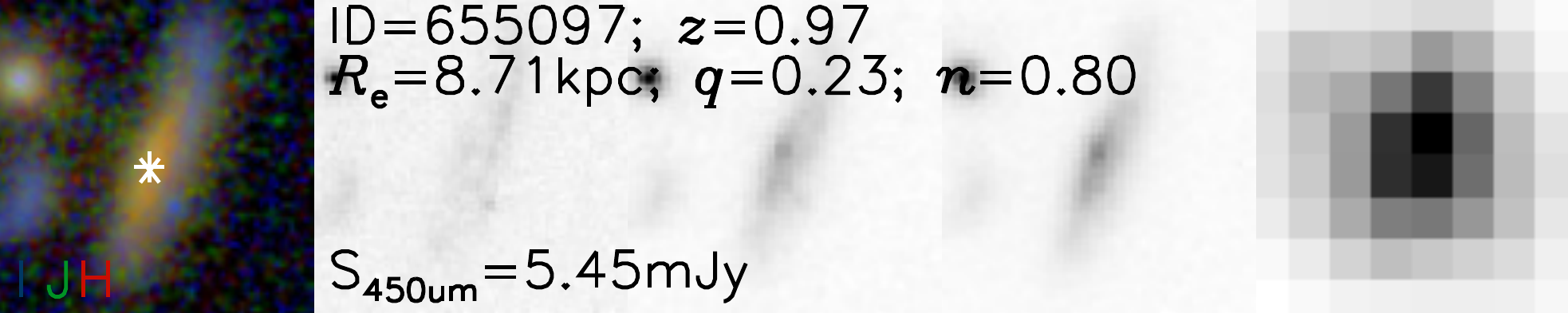}
\includegraphics[width=0.95\columnwidth]{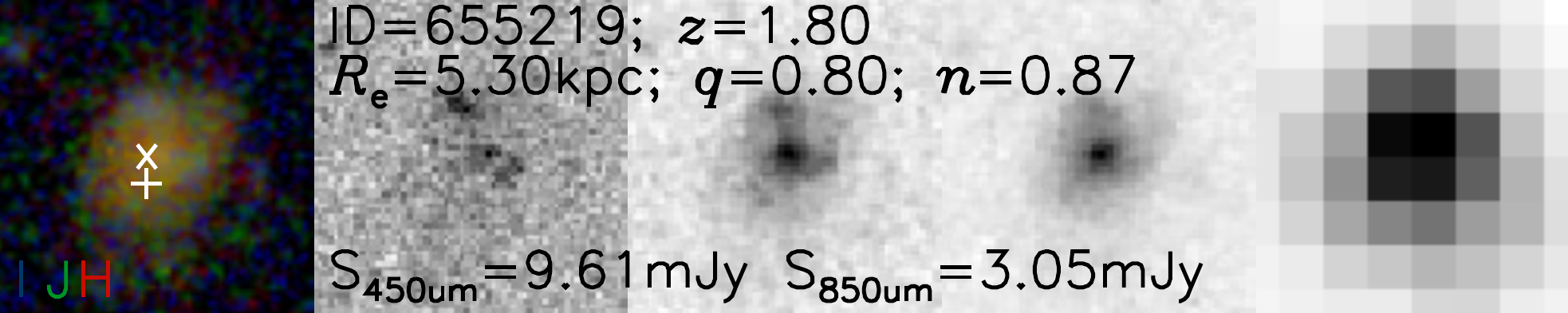}
\includegraphics[width=0.95\columnwidth]{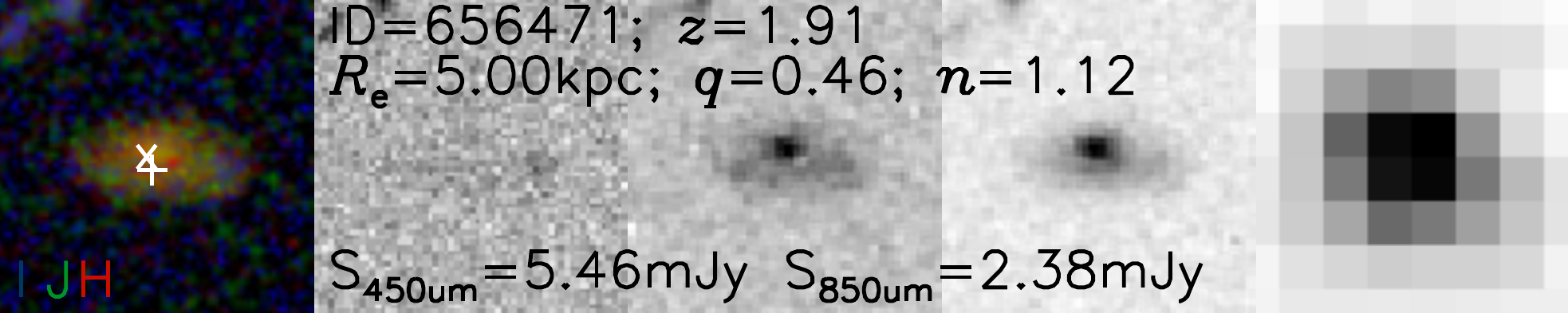}
\includegraphics[width=0.95\columnwidth]{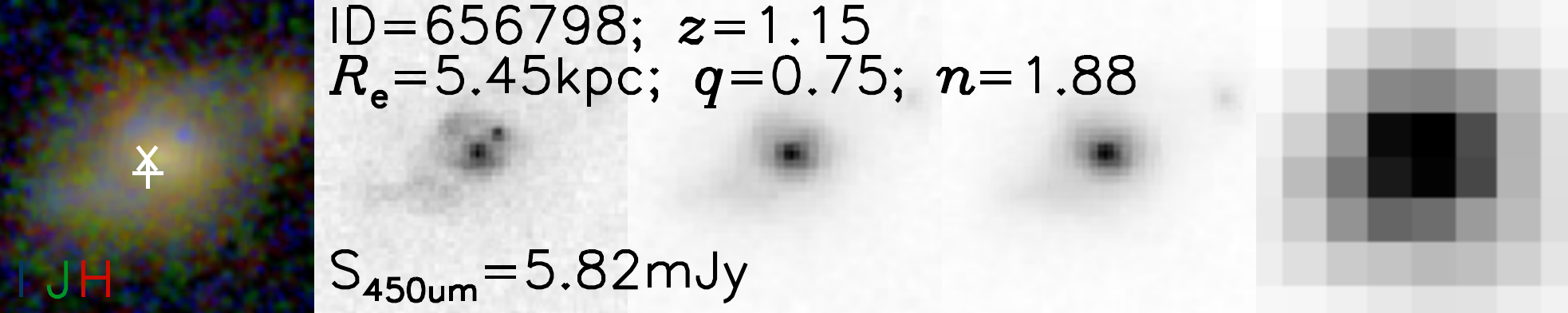}
\includegraphics[width=0.95\columnwidth]{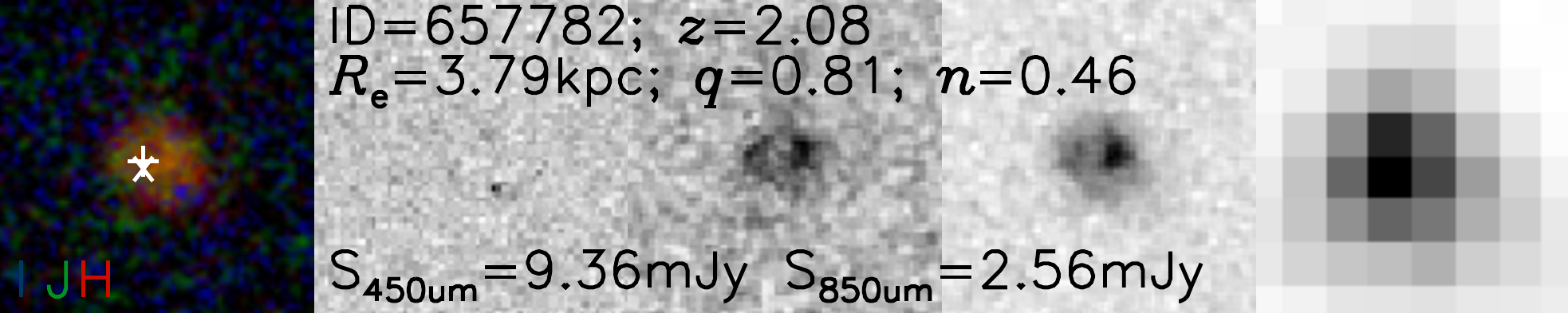}
\includegraphics[width=0.95\columnwidth]{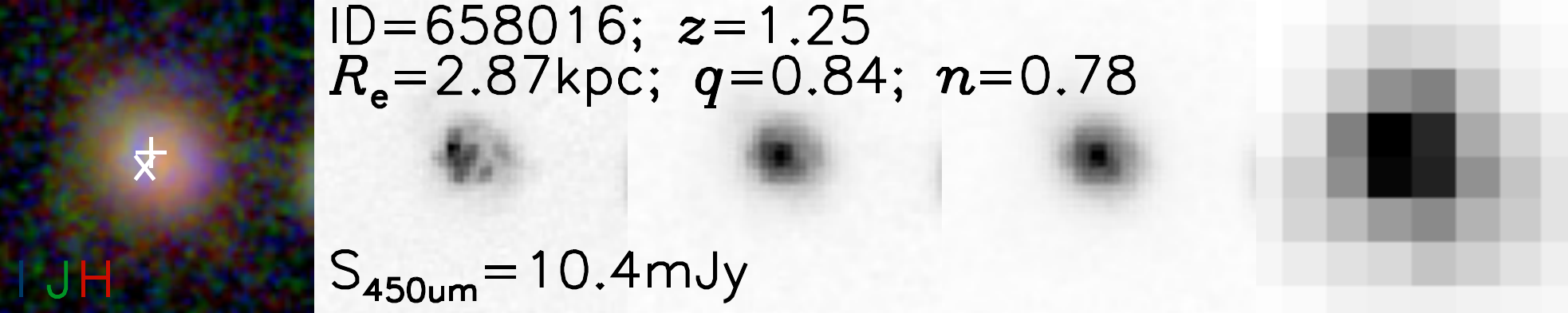}
\includegraphics[width=0.95\columnwidth]{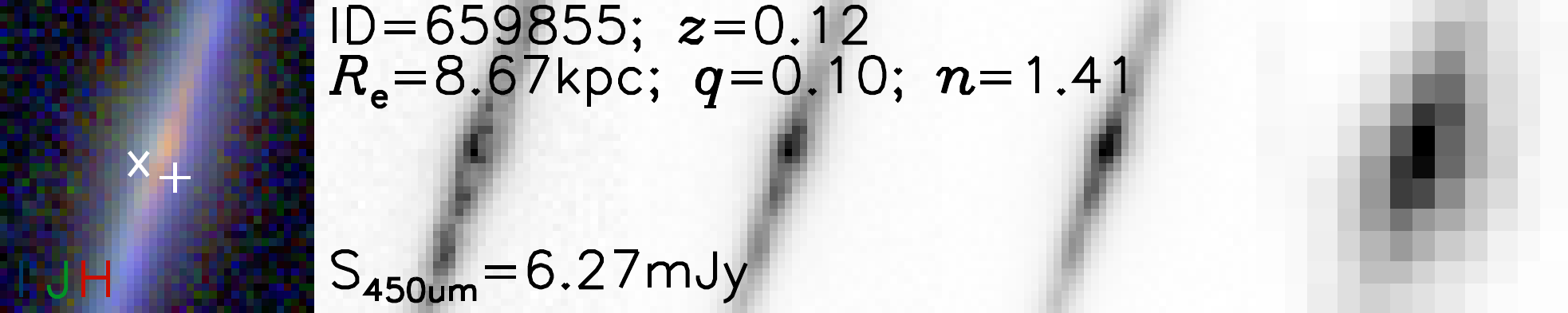}
\includegraphics[width=0.95\columnwidth]{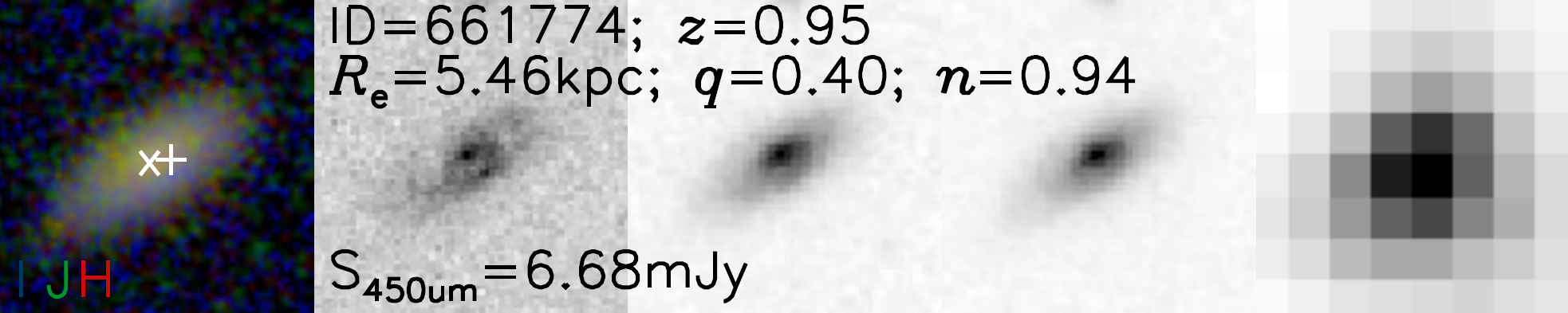}
\includegraphics[width=0.95\columnwidth]{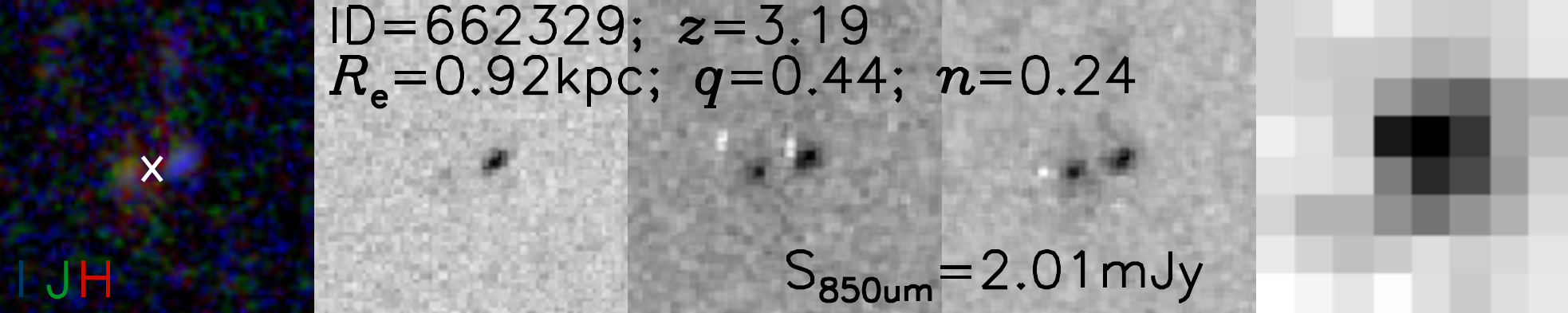}
\includegraphics[width=0.95\columnwidth]{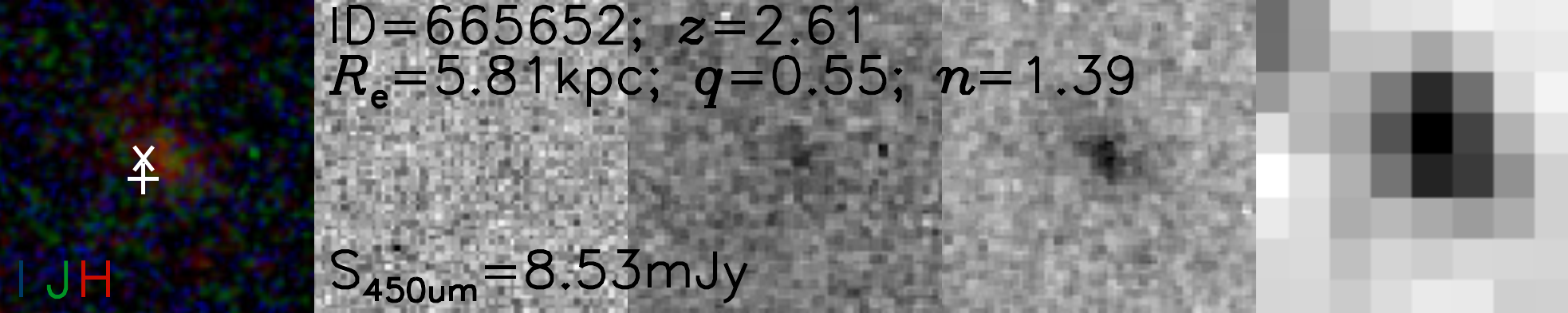}
\includegraphics[width=0.95\columnwidth]{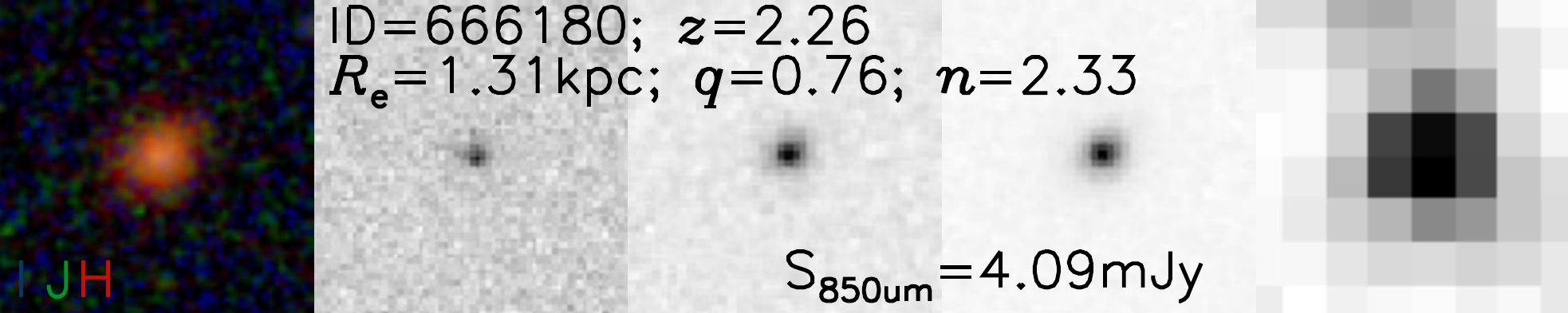}
\includegraphics[width=0.95\columnwidth]{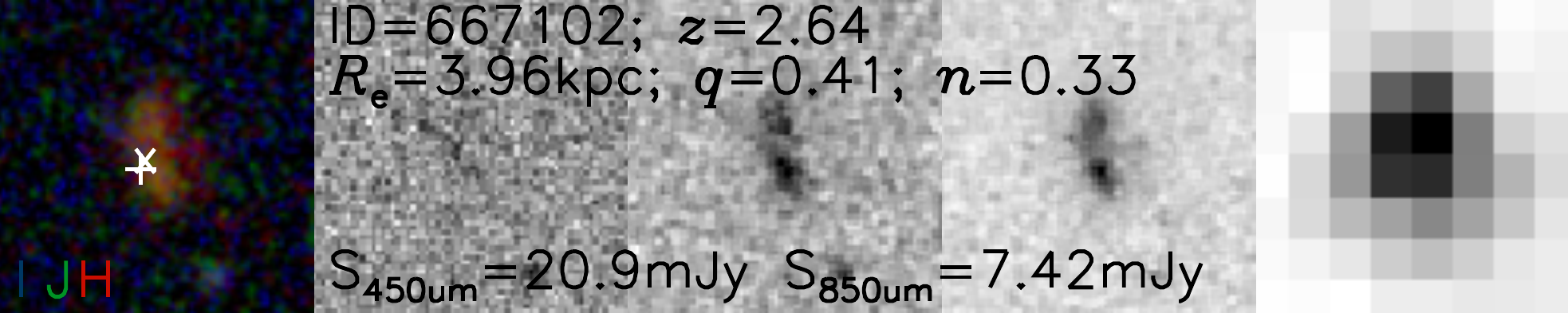}
\includegraphics[width=0.95\columnwidth]{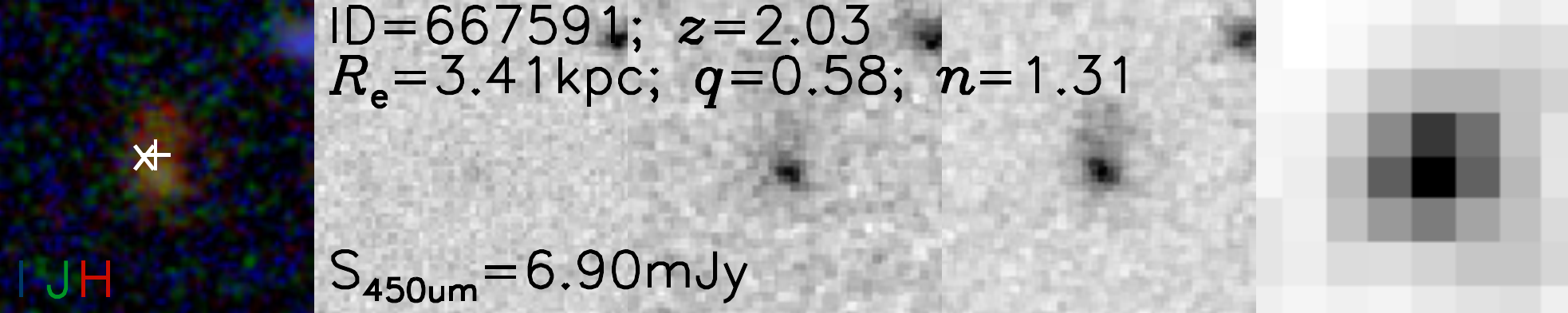}
\includegraphics[width=0.95\columnwidth]{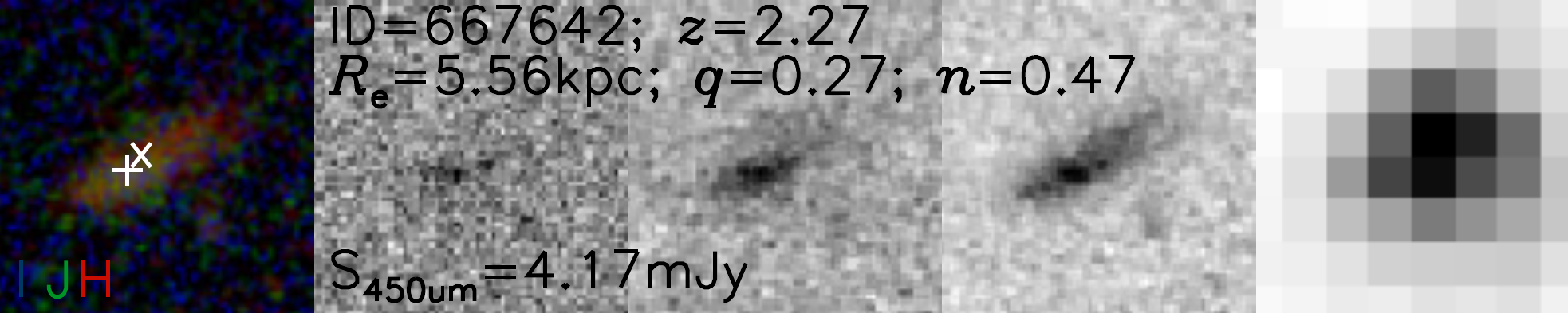}
\end{figure}
\begin{figure}
\centering
\includegraphics[width=0.95\columnwidth]{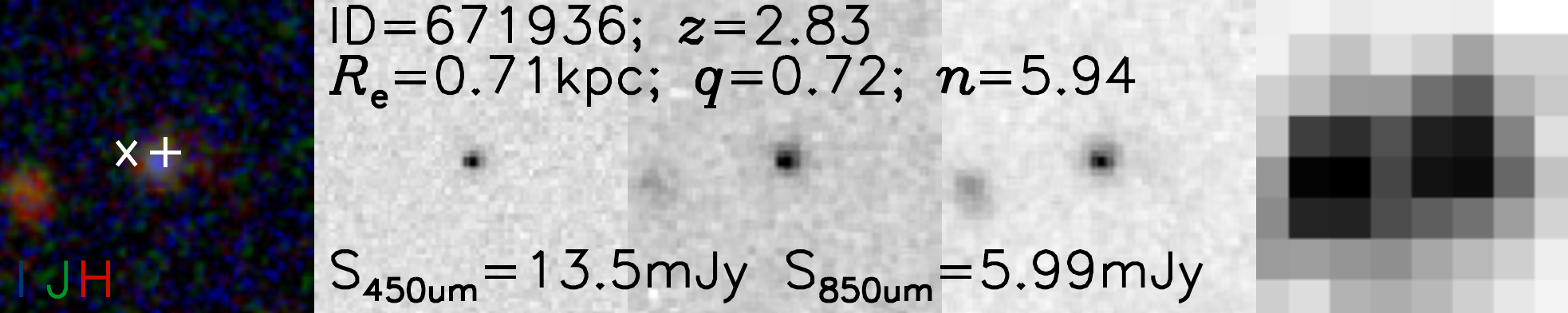}
\includegraphics[width=0.95\columnwidth]{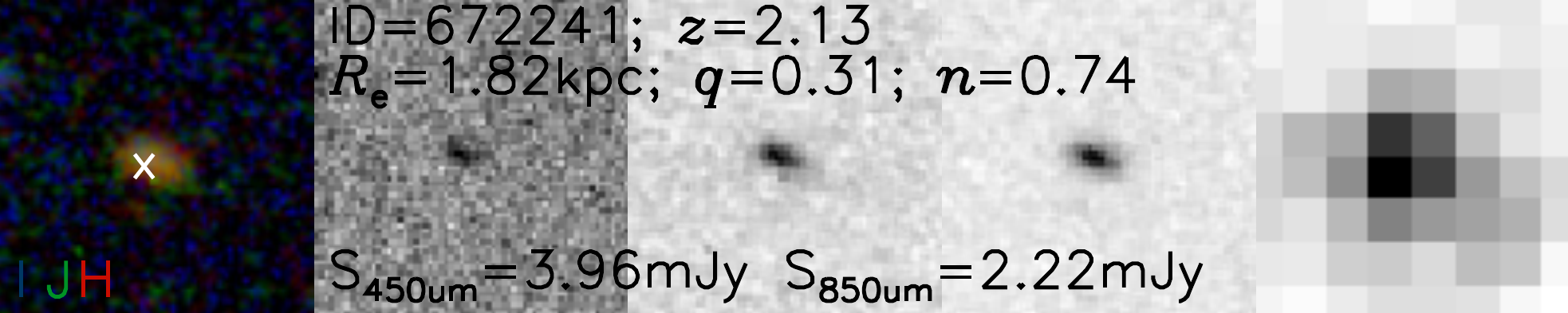}
\includegraphics[width=0.95\columnwidth]{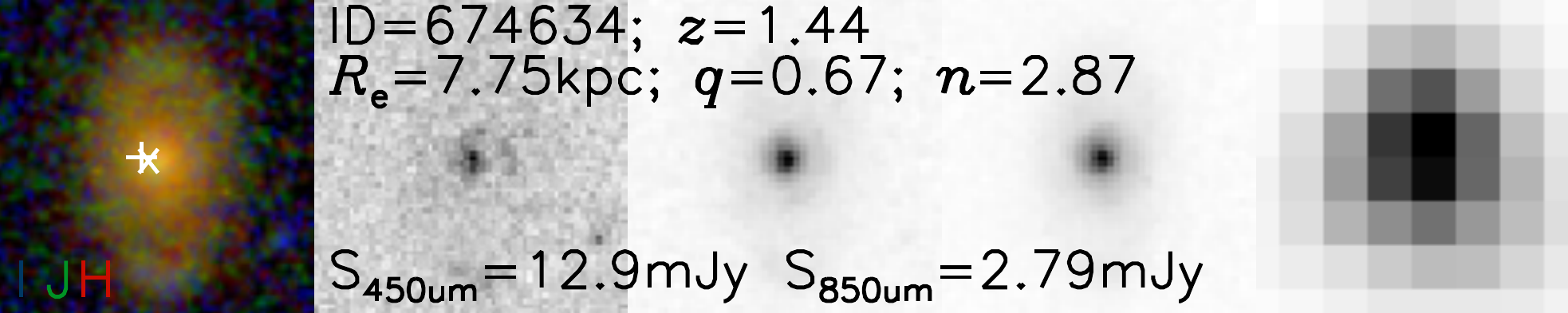}
\includegraphics[width=0.95\columnwidth]{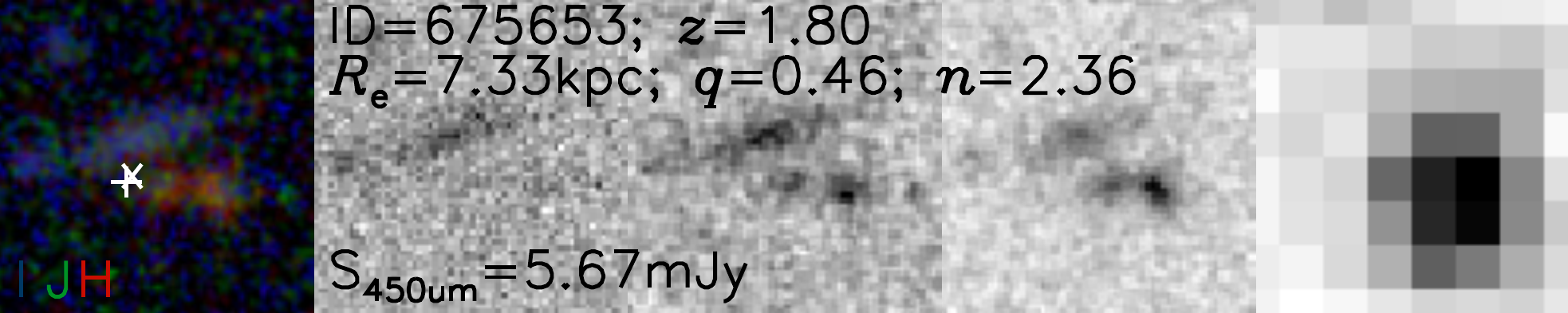}
\includegraphics[width=0.95\columnwidth]{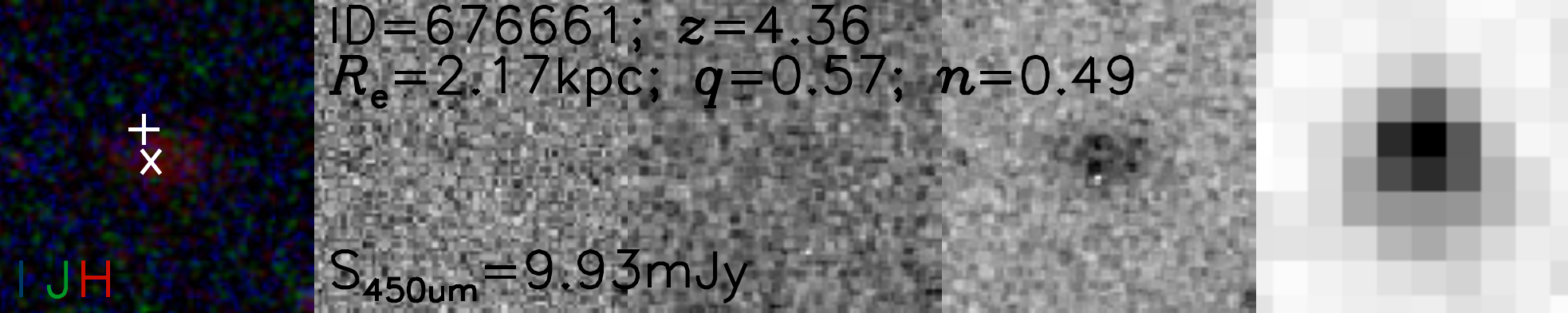}
\includegraphics[width=0.95\columnwidth]{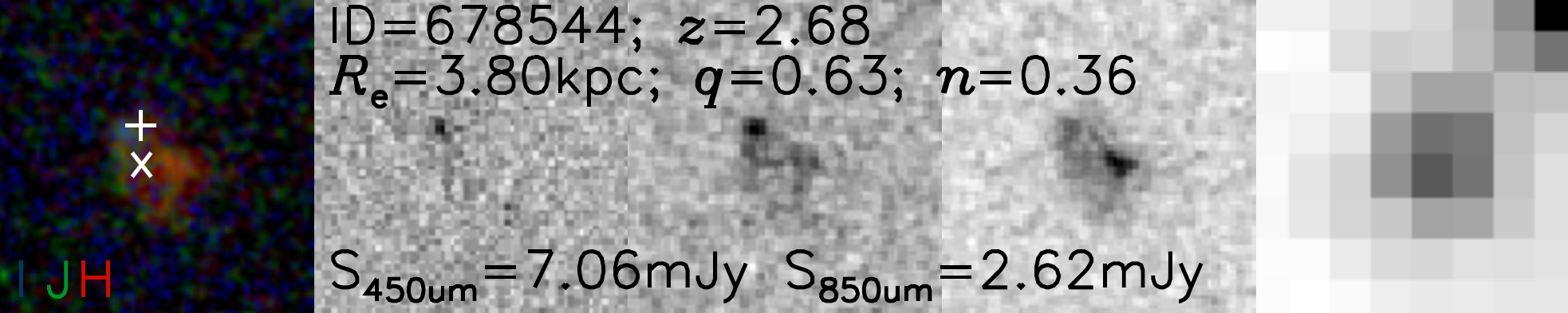}
\includegraphics[width=0.95\columnwidth]{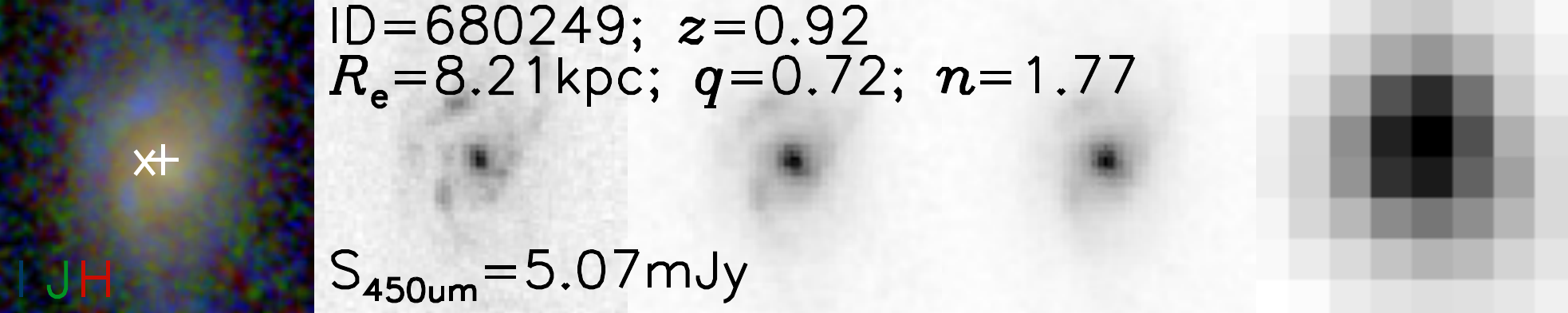}
\includegraphics[width=0.95\columnwidth]{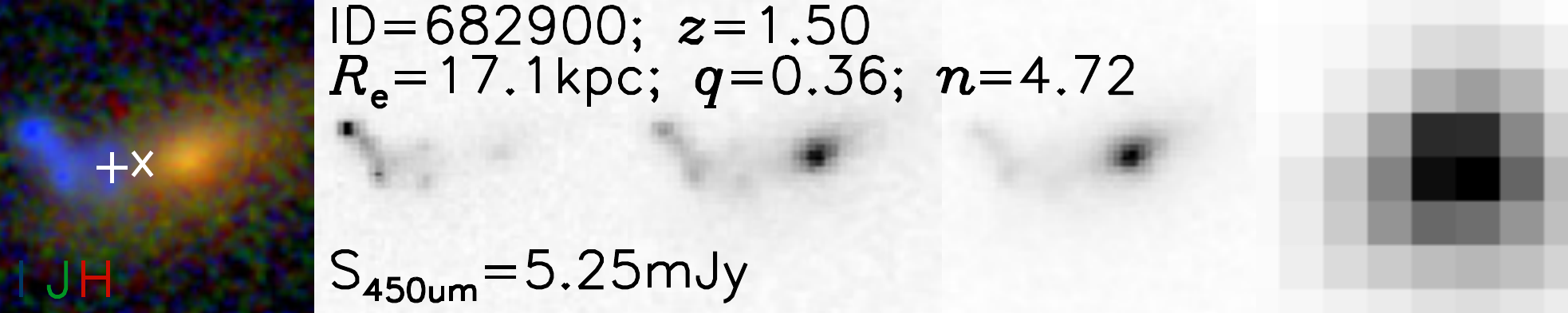}
\includegraphics[width=0.95\columnwidth]{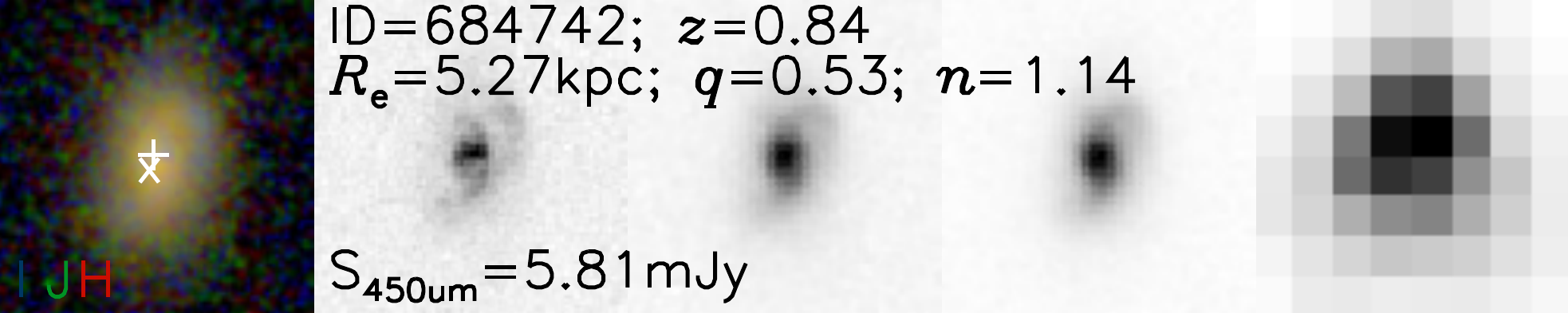}
\includegraphics[width=0.95\columnwidth]{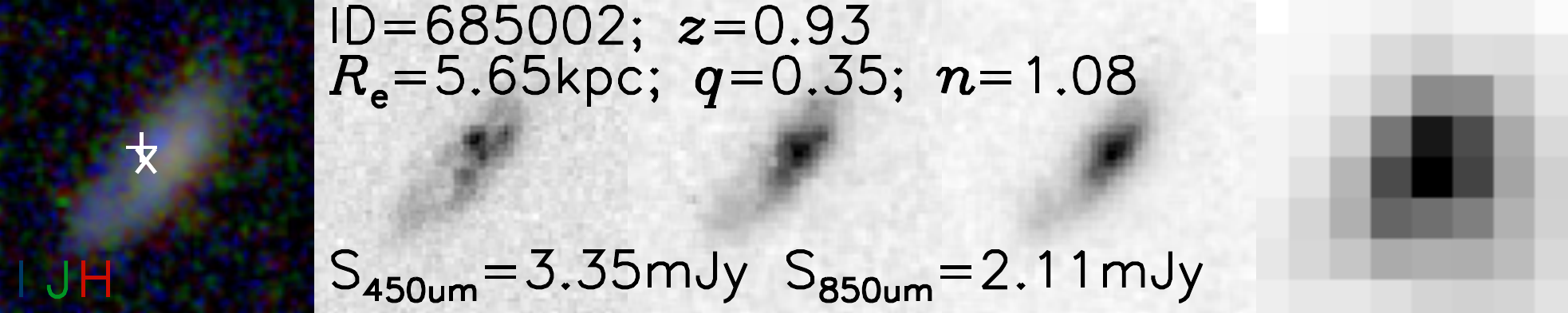}
\includegraphics[width=0.95\columnwidth]{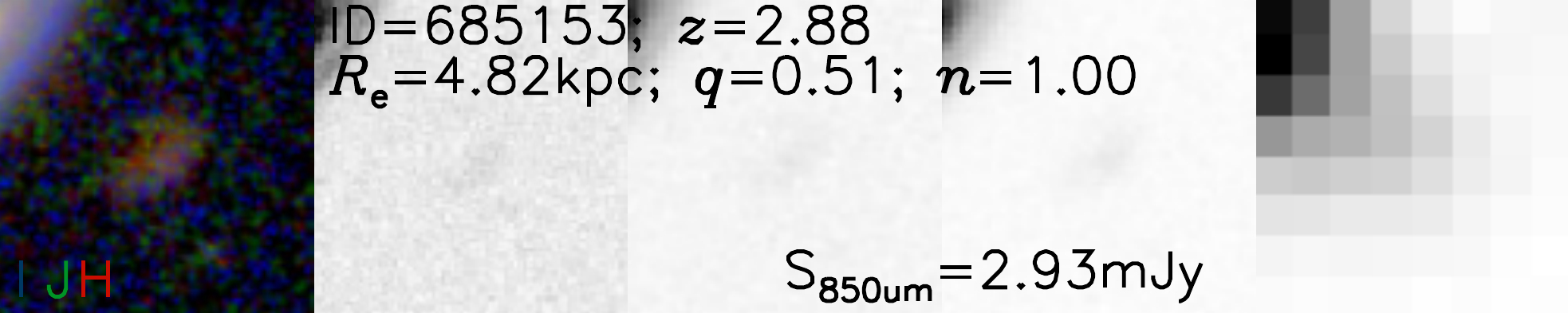}
\includegraphics[width=0.95\columnwidth]{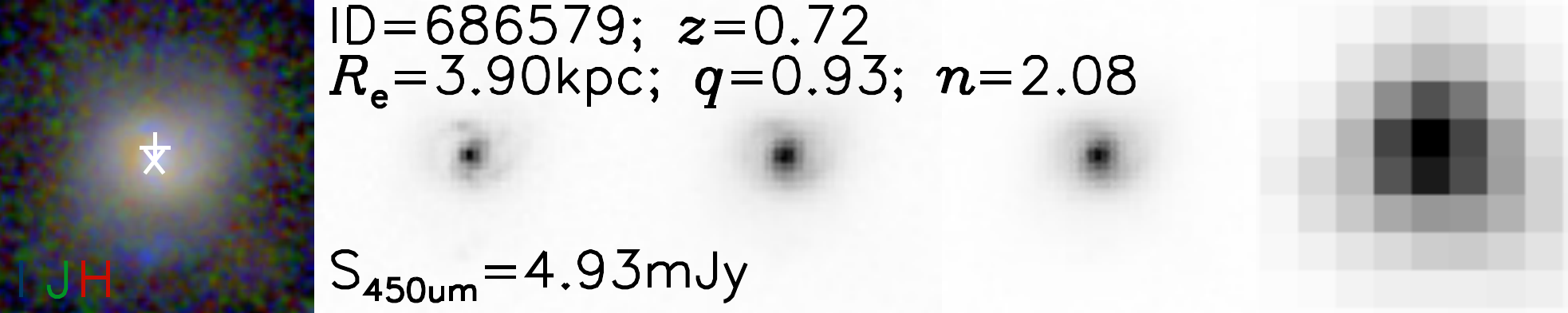}
\includegraphics[width=0.95\columnwidth]{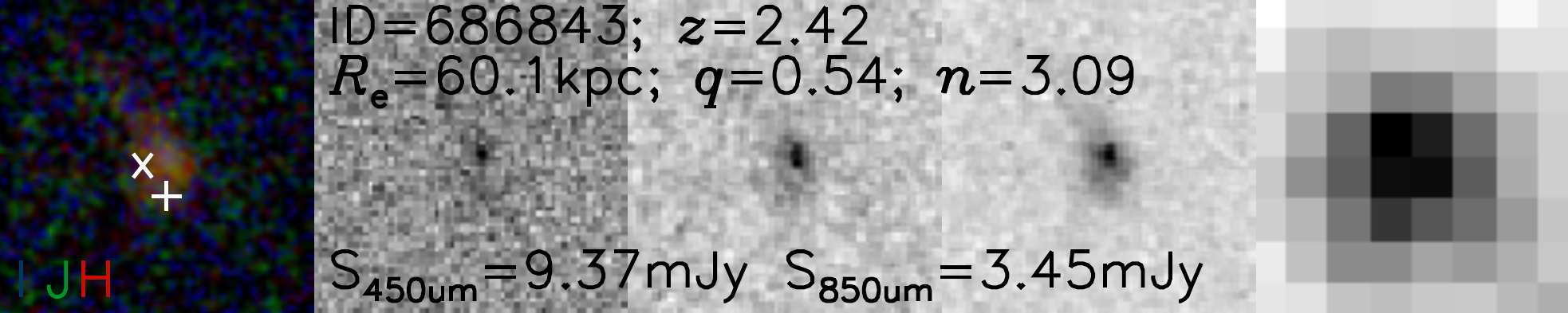}
\includegraphics[width=0.95\columnwidth]{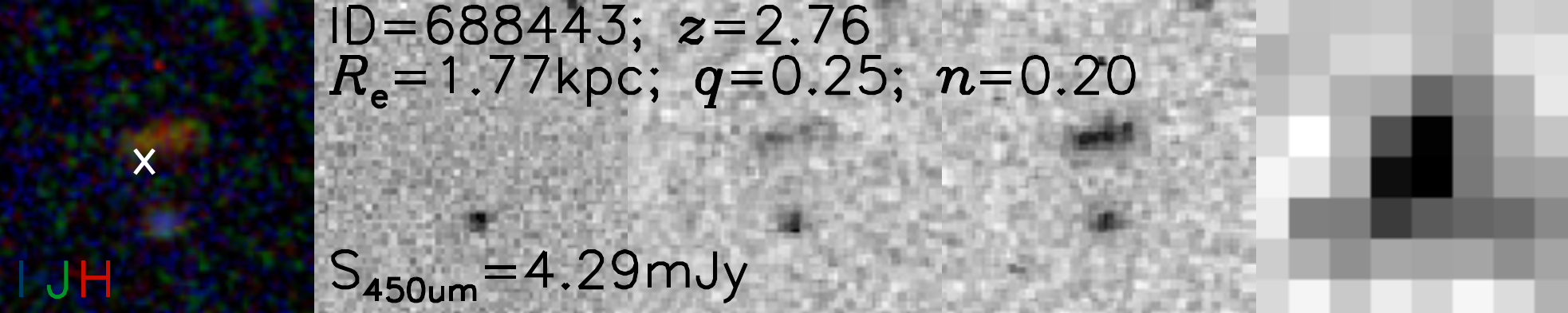}
\includegraphics[width=0.95\columnwidth]{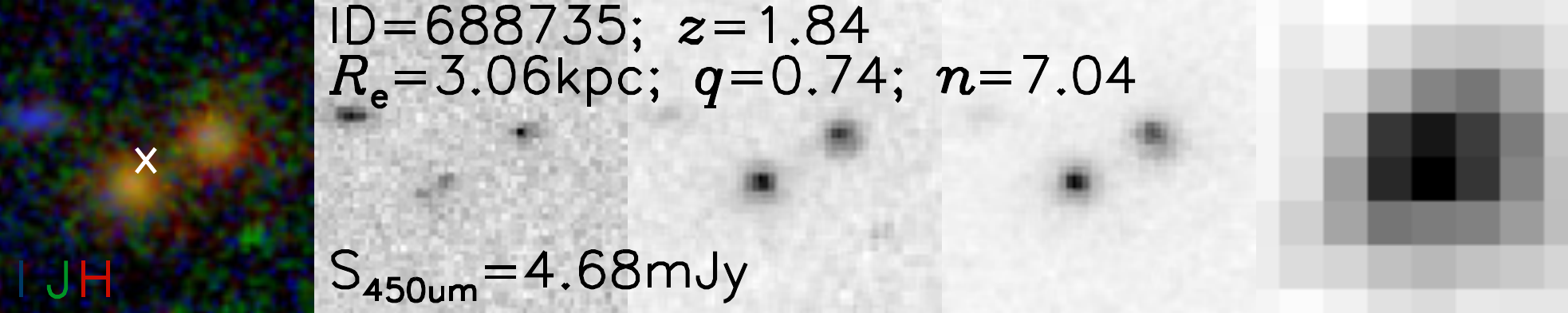}
\end{figure}
\begin{figure}
\centering
\includegraphics[width=0.95\columnwidth]{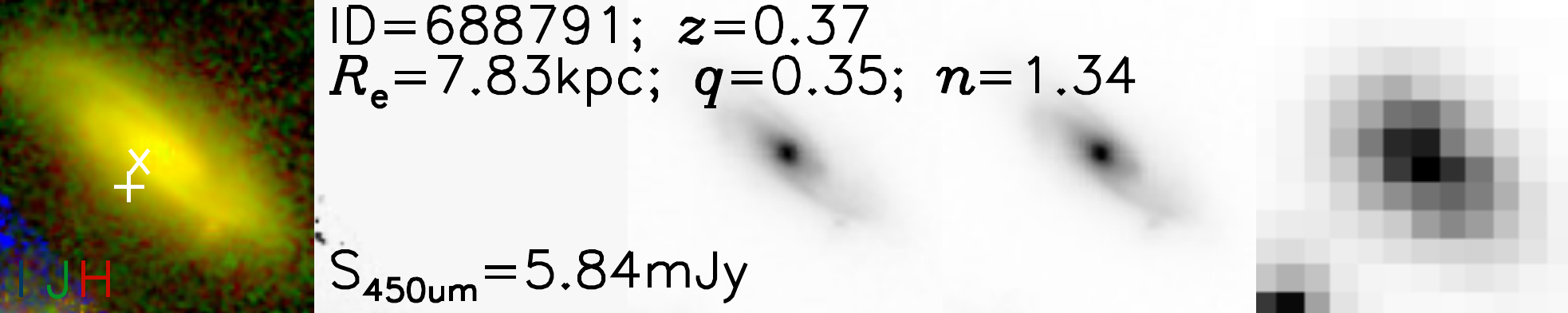}
\includegraphics[width=0.95\columnwidth]{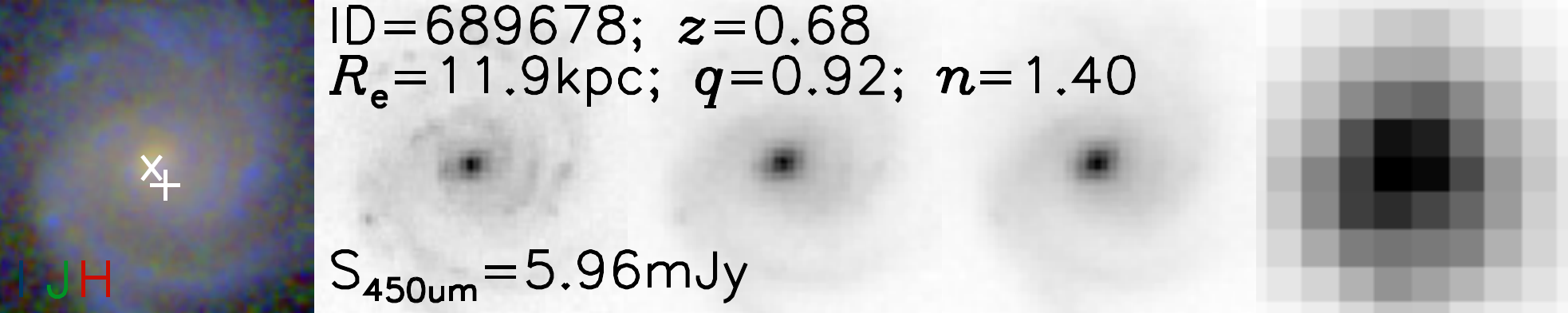}
\includegraphics[width=0.95\columnwidth]{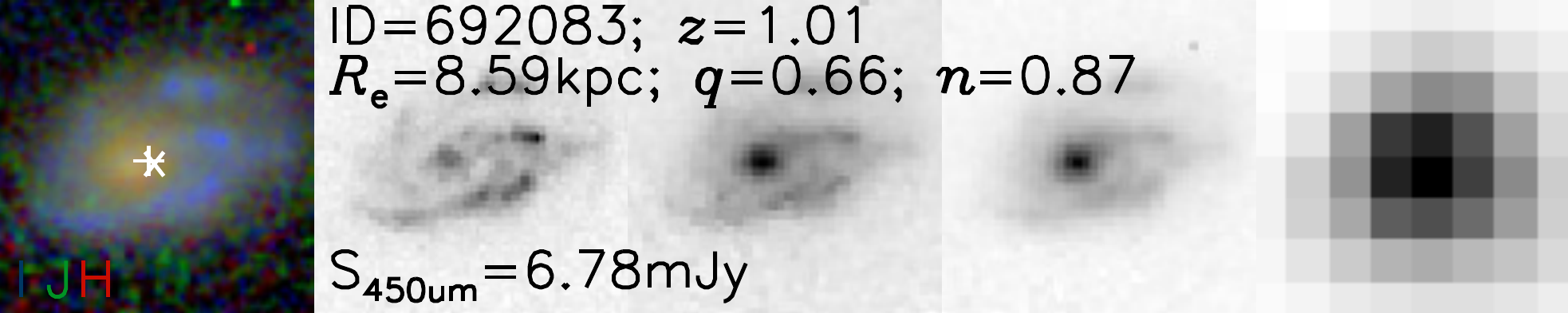}
\includegraphics[width=0.95\columnwidth]{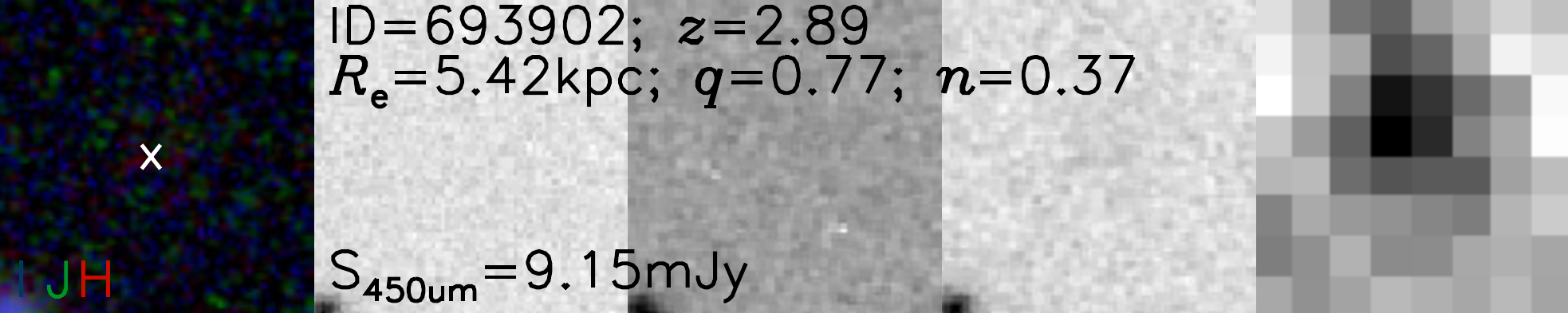}
\includegraphics[width=0.95\columnwidth]{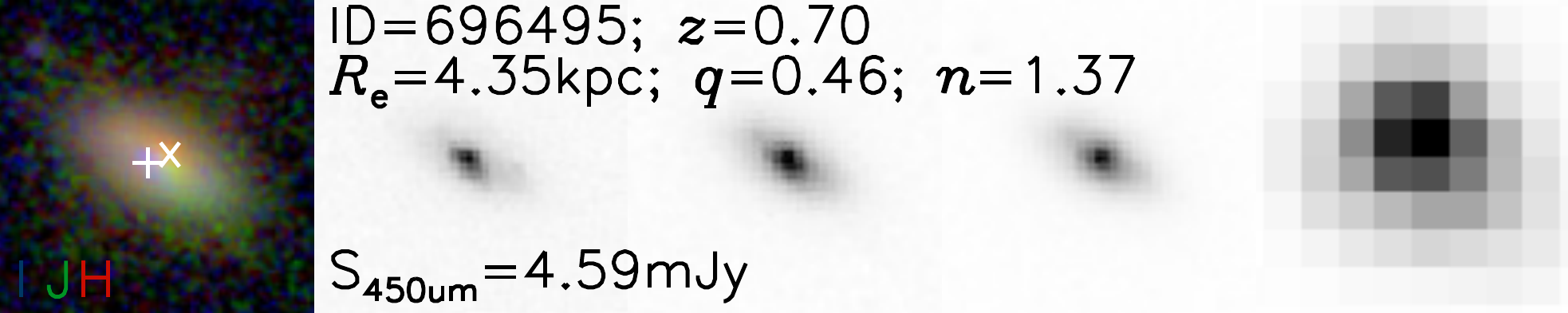}
\includegraphics[width=0.95\columnwidth]{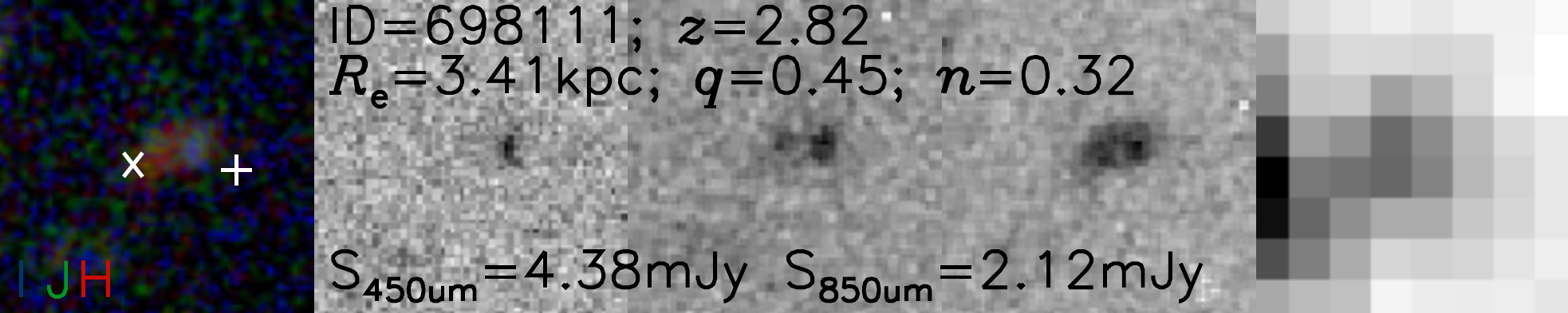}
\includegraphics[width=0.95\columnwidth]{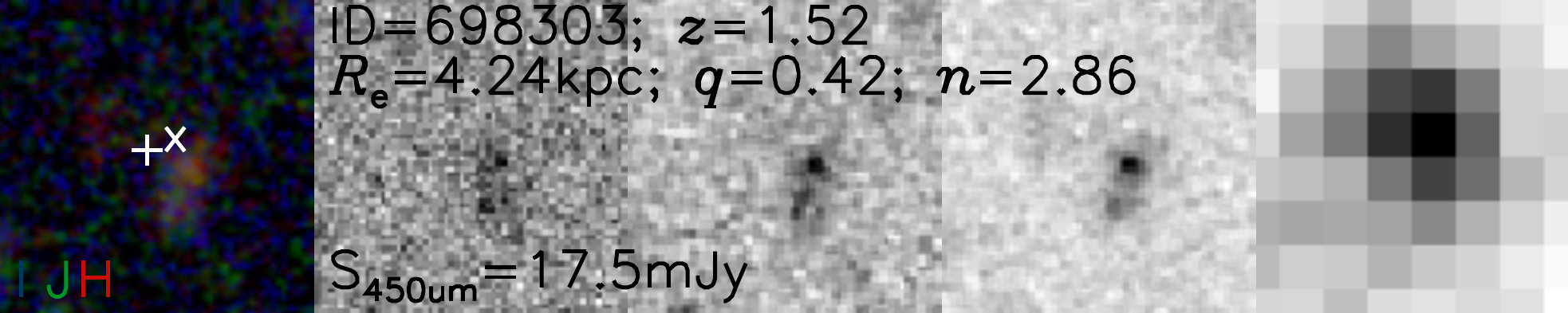}
\includegraphics[width=0.95\columnwidth]{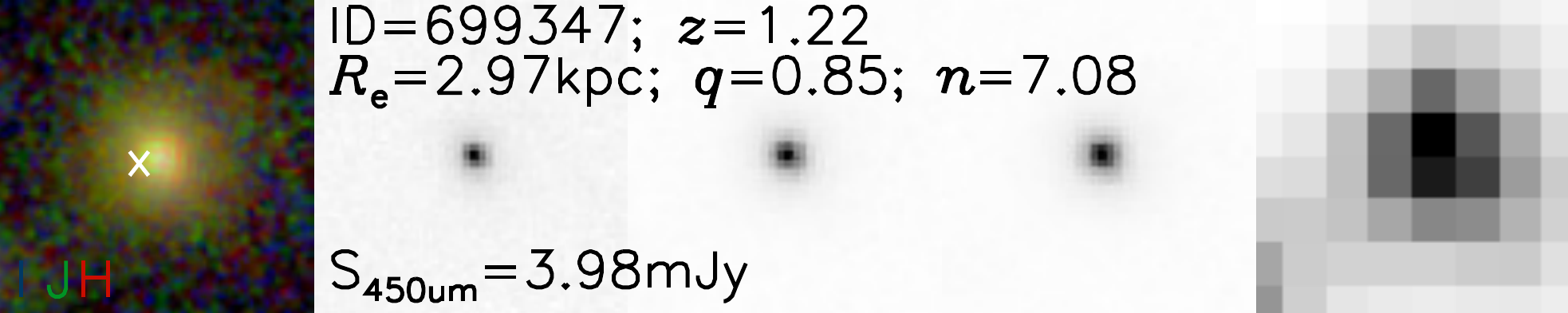}
\includegraphics[width=0.95\columnwidth]{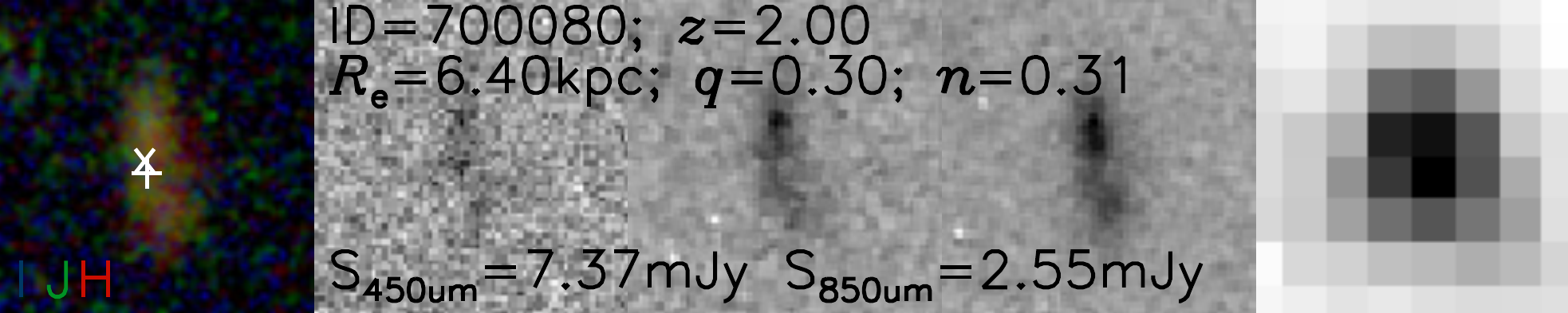}
\includegraphics[width=0.95\columnwidth]{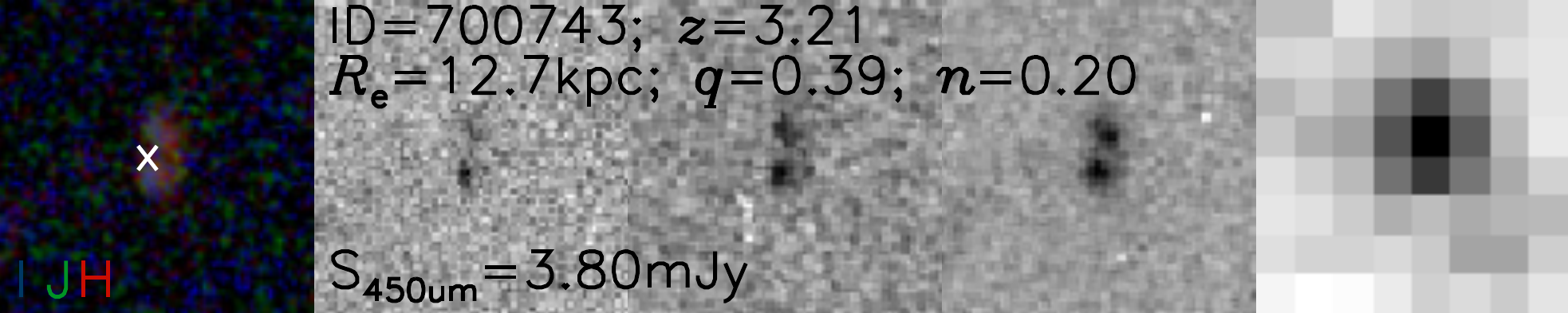}
\includegraphics[width=0.95\columnwidth]{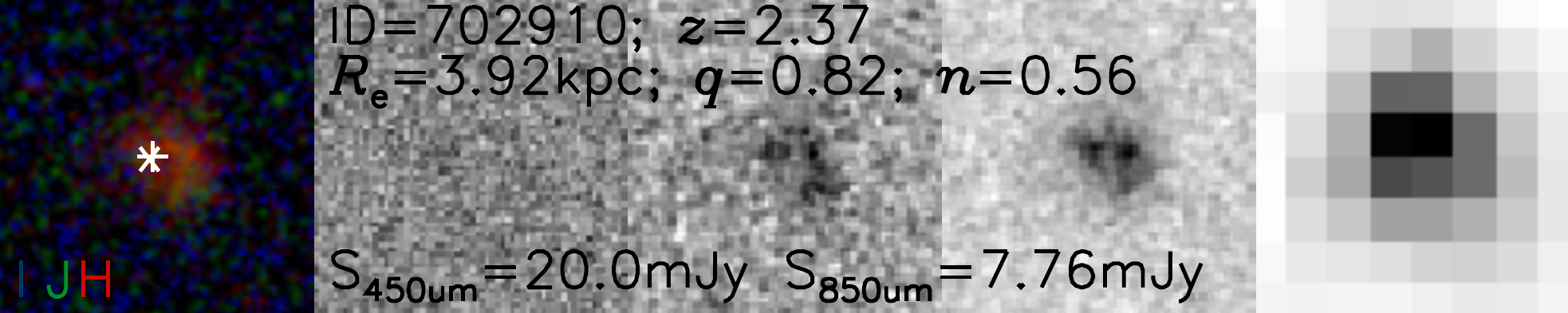}
\includegraphics[width=0.95\columnwidth]{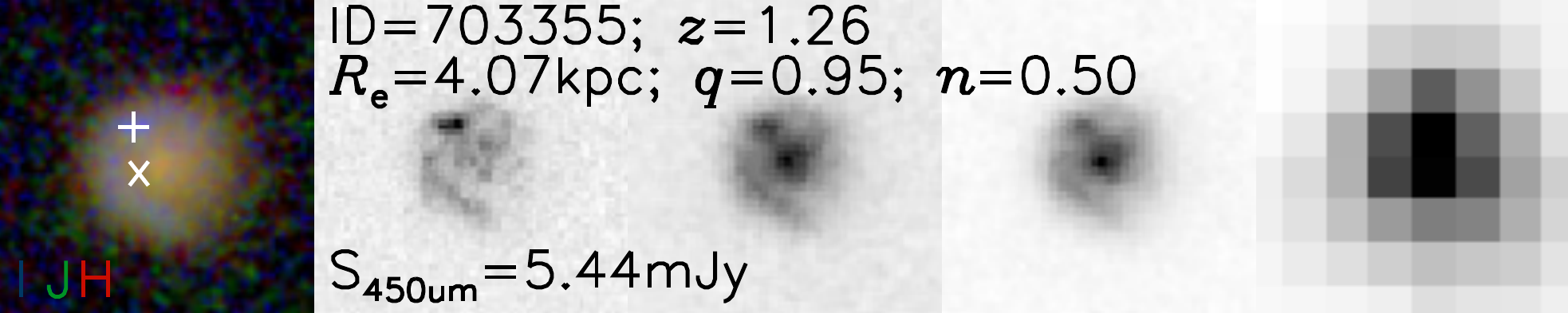}
\includegraphics[width=0.95\columnwidth]{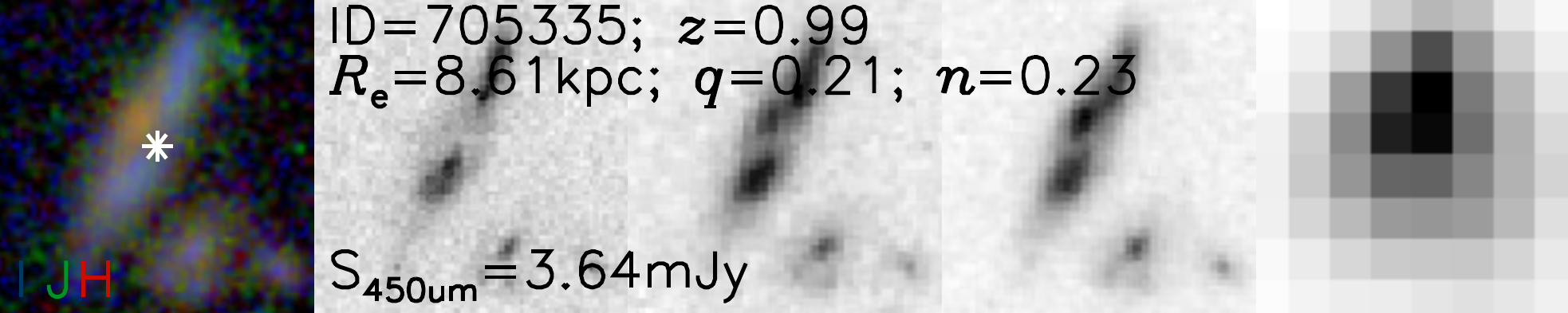}
\includegraphics[width=0.95\columnwidth]{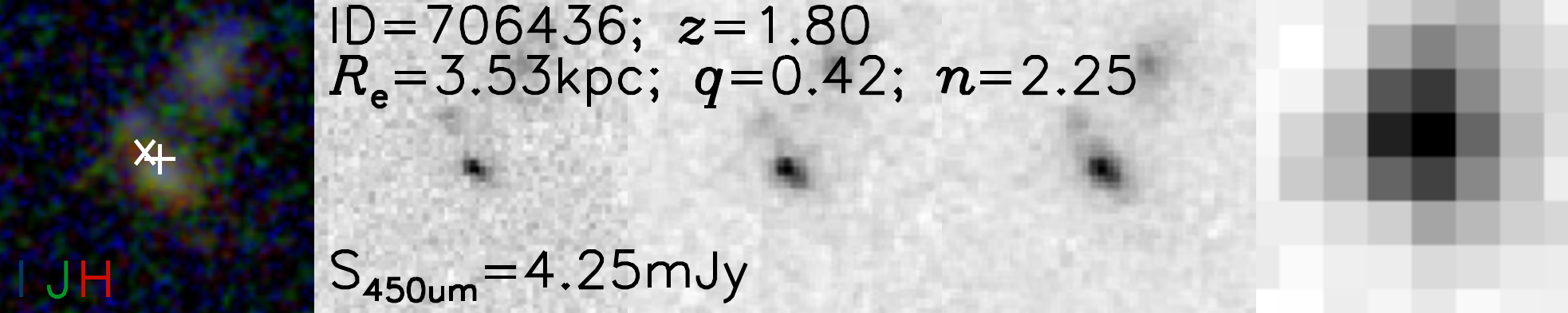}
\includegraphics[width=0.95\columnwidth]{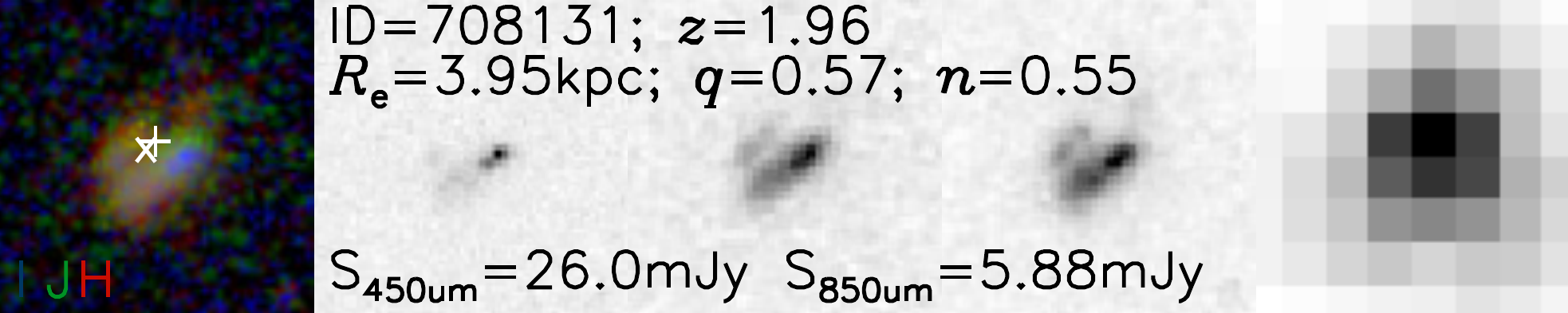}
\end{figure}
\begin{figure}
\centering
\includegraphics[width=0.95\columnwidth]{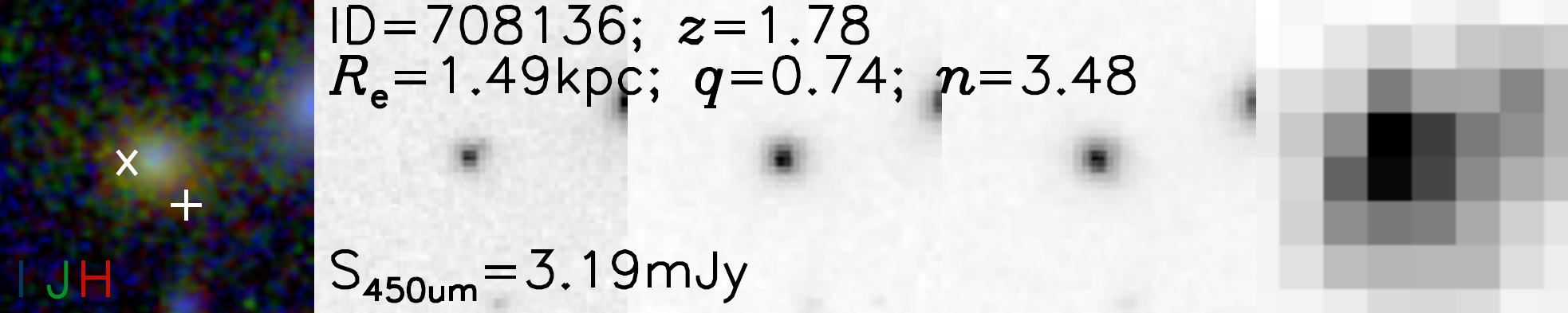}
\includegraphics[width=0.95\columnwidth]{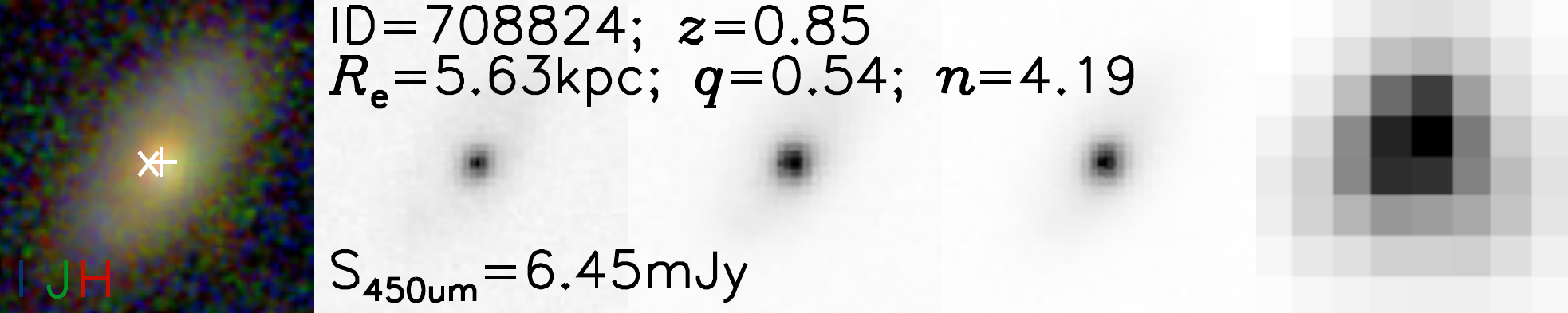}
\includegraphics[width=0.95\columnwidth]{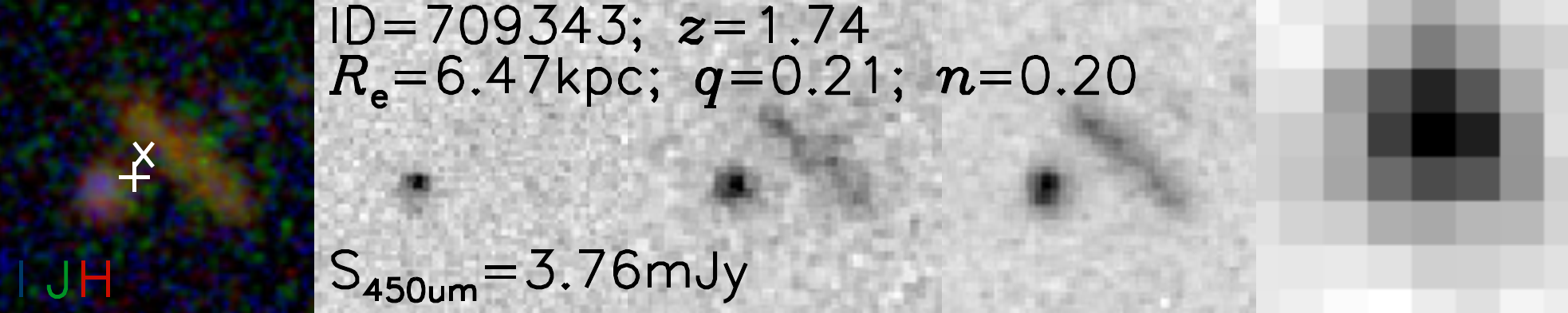}
\includegraphics[width=0.95\columnwidth]{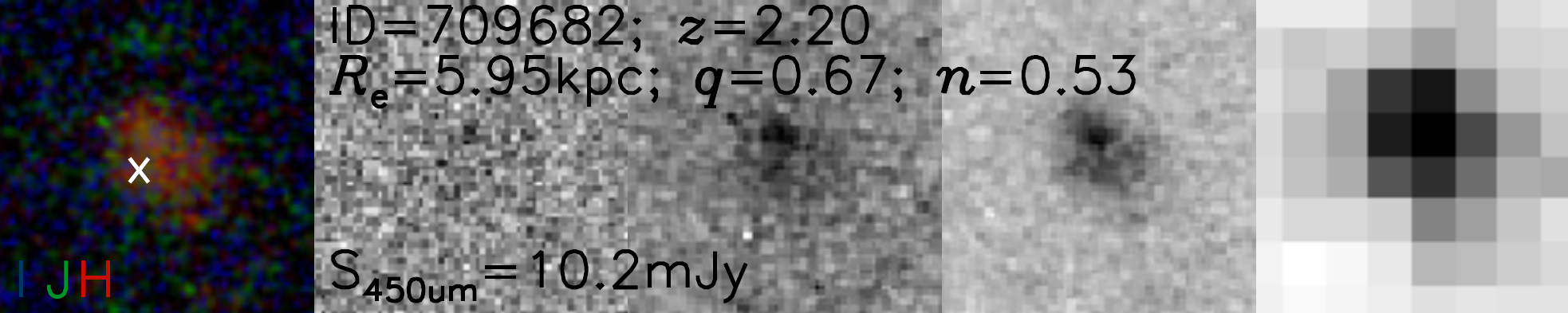}
\includegraphics[width=0.95\columnwidth]{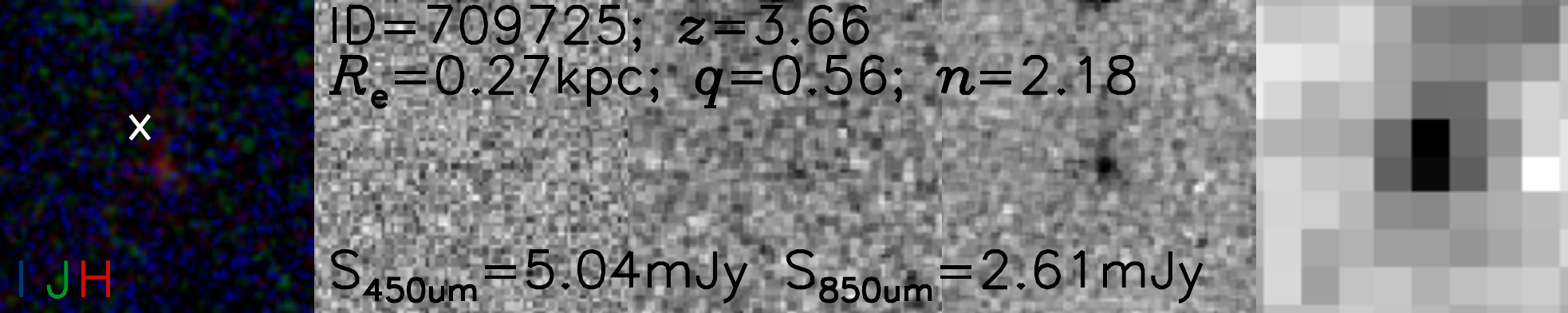}
\includegraphics[width=0.95\columnwidth]{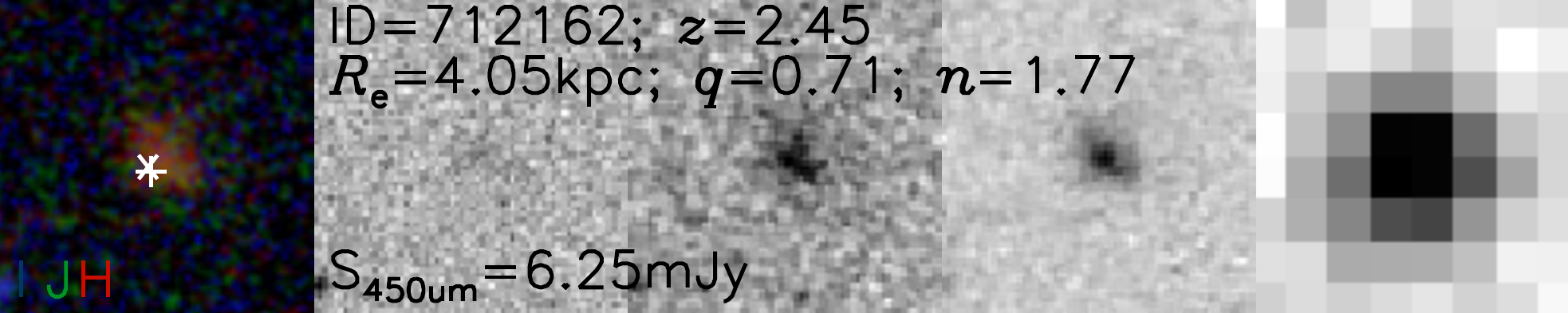}
\includegraphics[width=0.95\columnwidth]{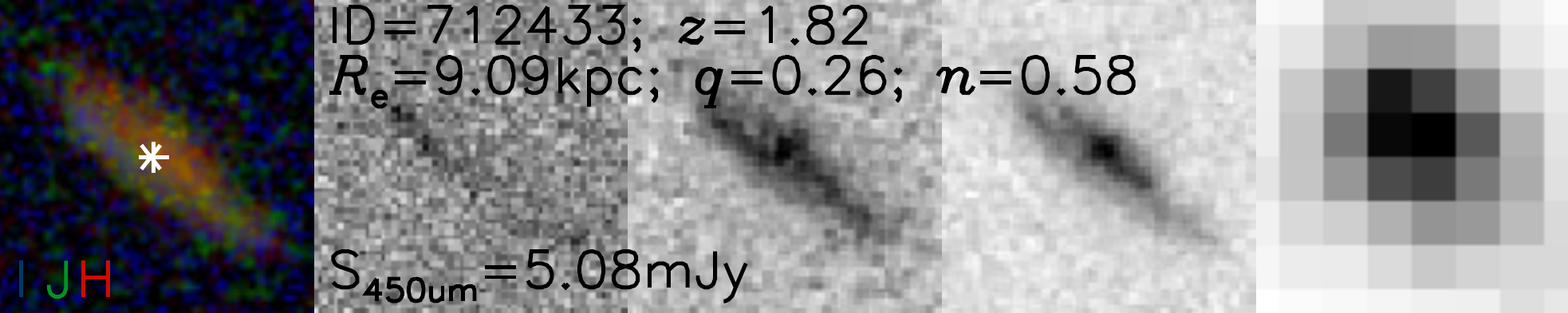}
\includegraphics[width=0.95\columnwidth]{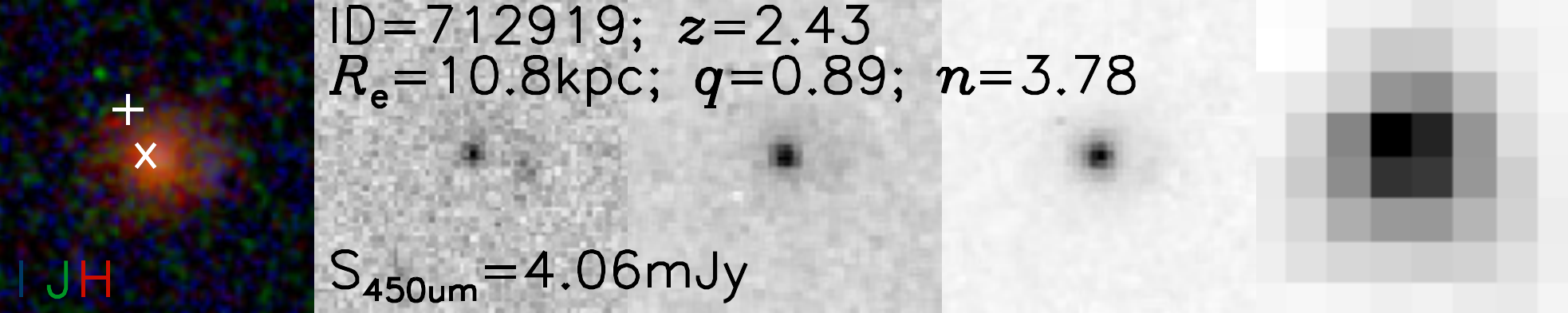}
\includegraphics[width=0.95\columnwidth]{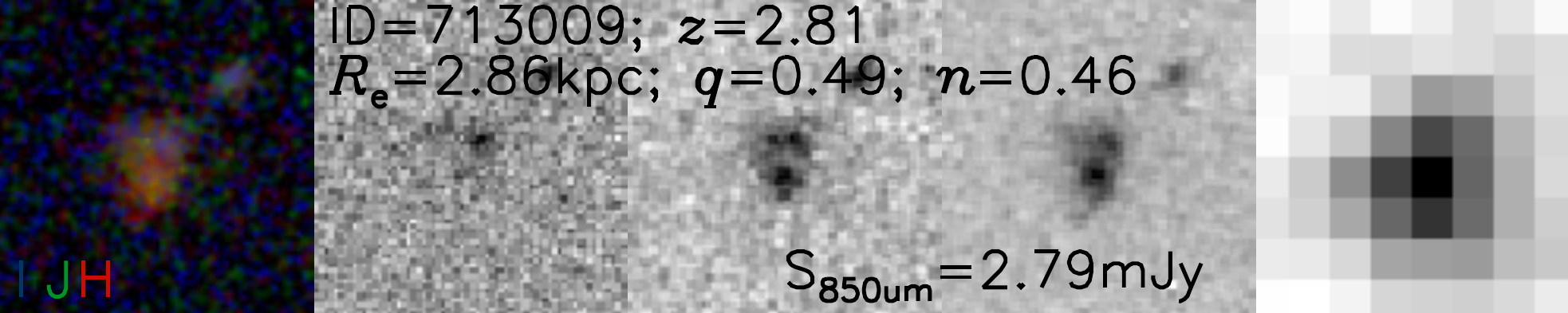}
\includegraphics[width=0.95\columnwidth]{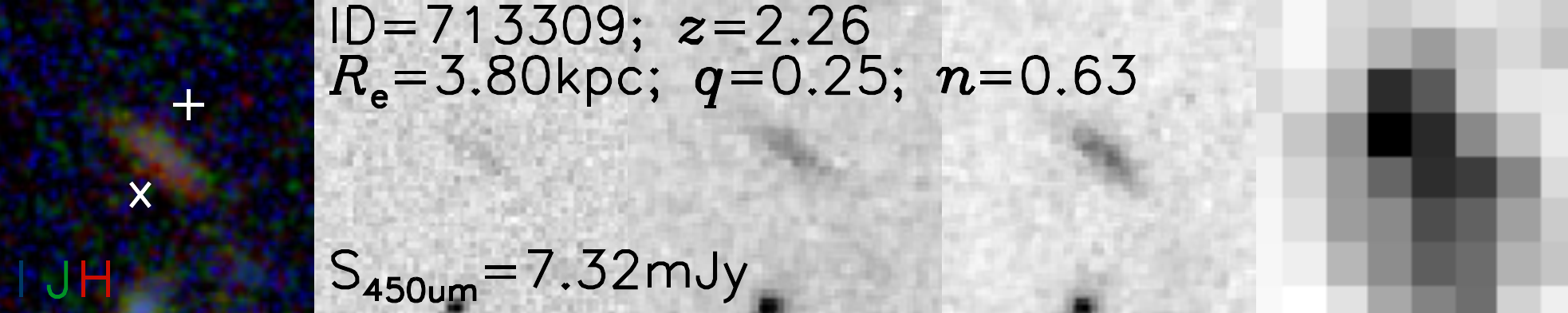}
\includegraphics[width=0.95\columnwidth]{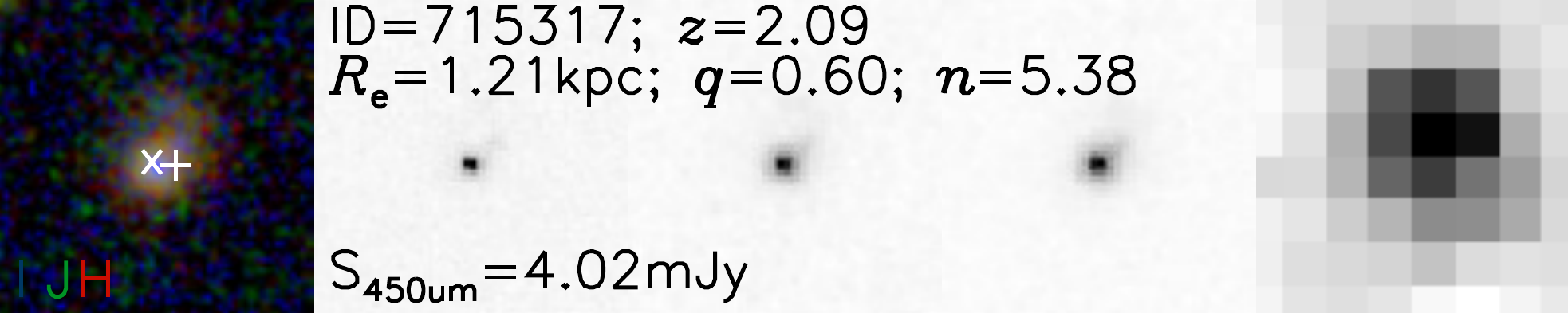}
\includegraphics[width=0.95\columnwidth]{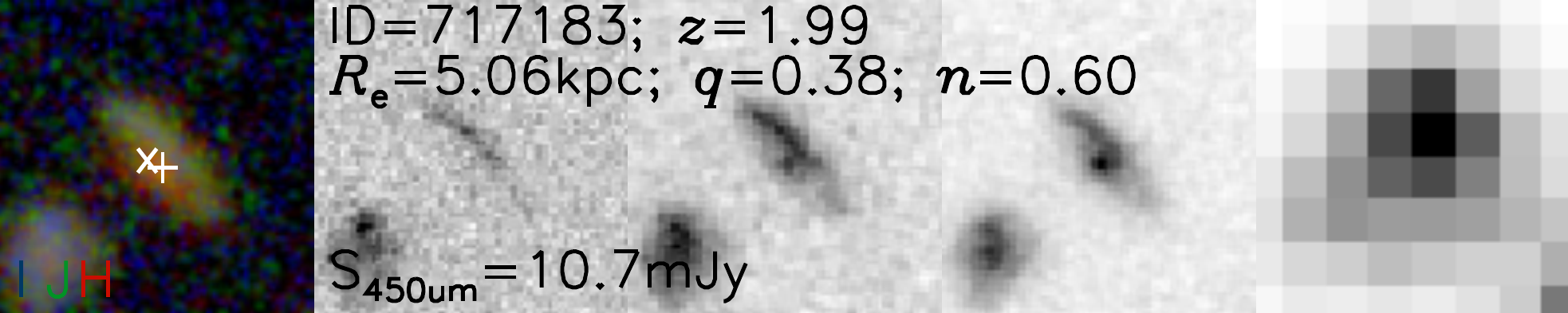}
\includegraphics[width=0.95\columnwidth]{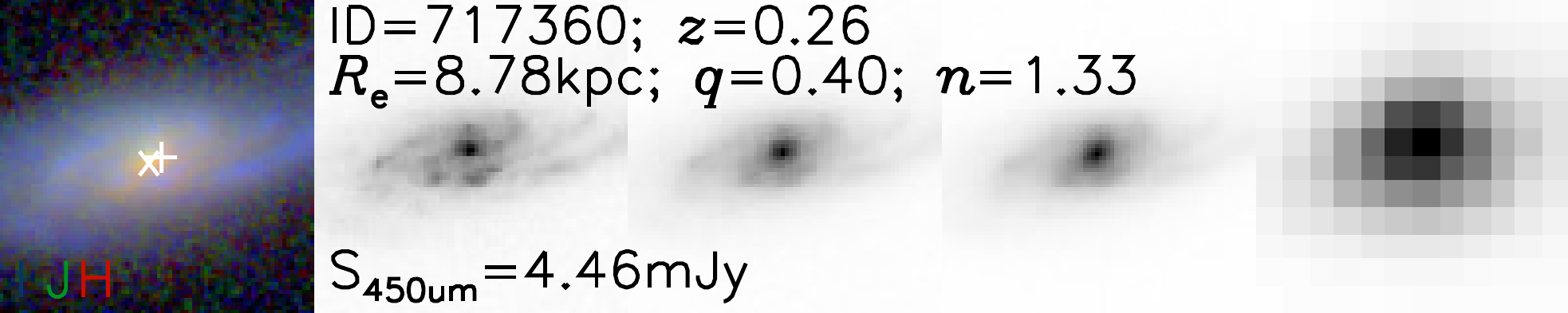}
\includegraphics[width=0.95\columnwidth]{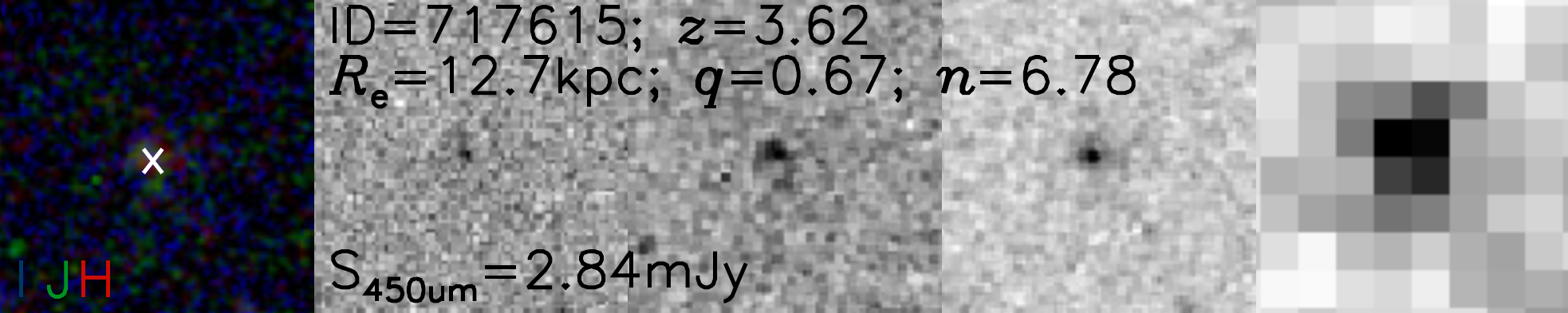}
\includegraphics[width=0.95\columnwidth]{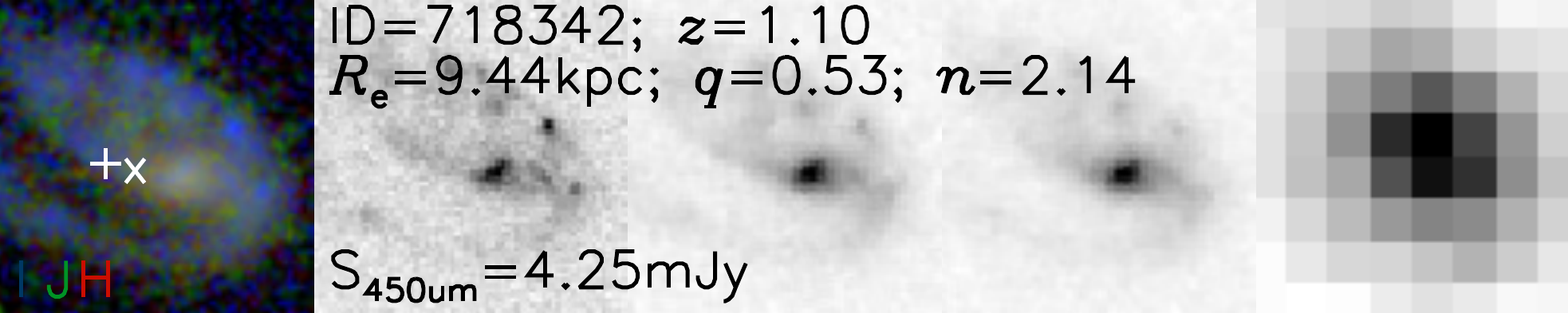}
\end{figure}
\begin{figure}
\centering
\includegraphics[width=0.95\columnwidth]{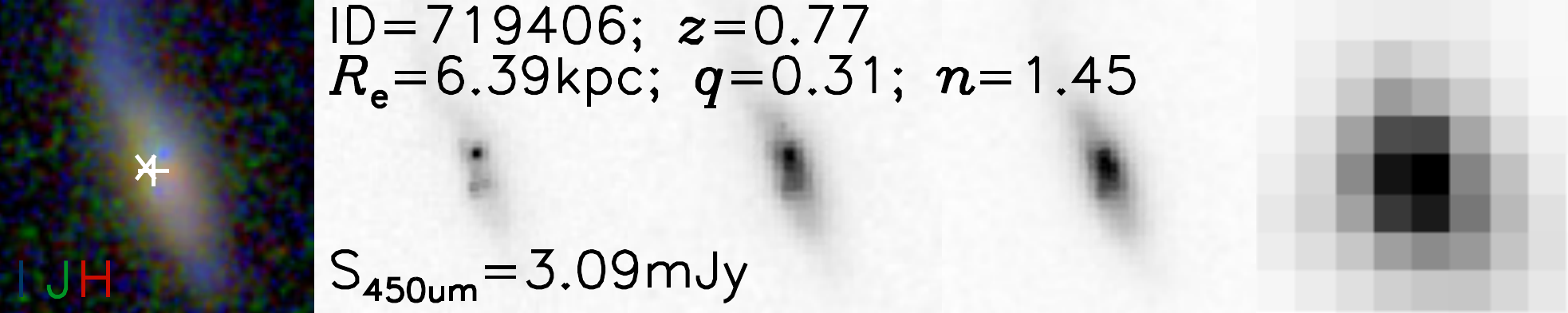}
\includegraphics[width=0.95\columnwidth]{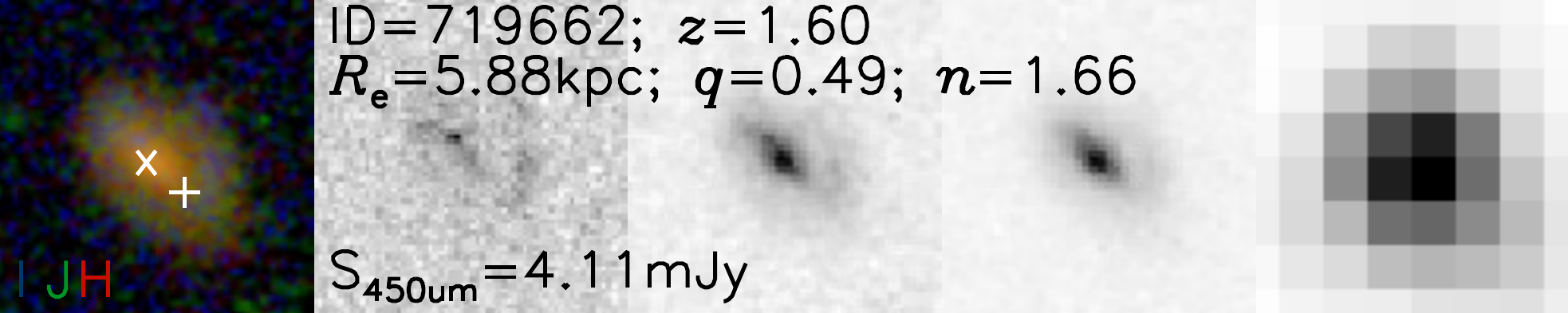}
\includegraphics[width=0.95\columnwidth]{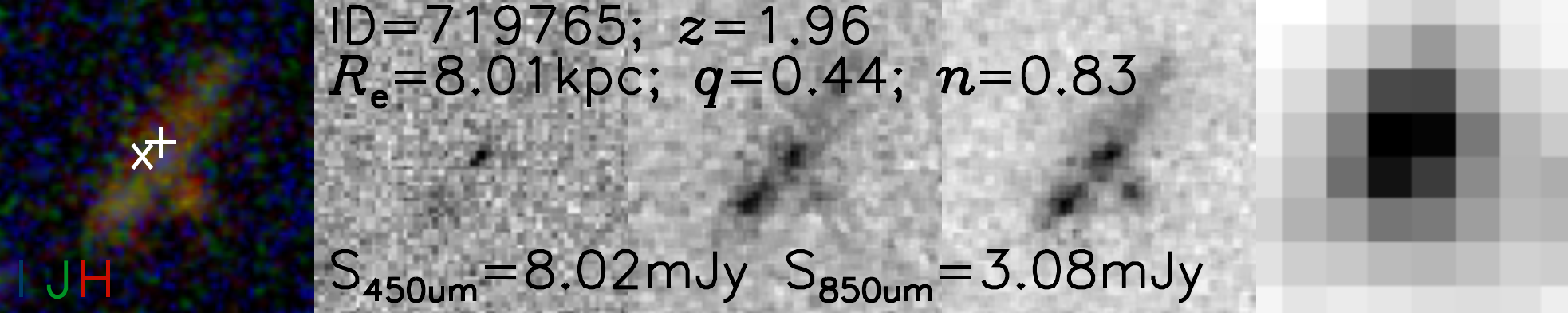}
\includegraphics[width=0.95\columnwidth]{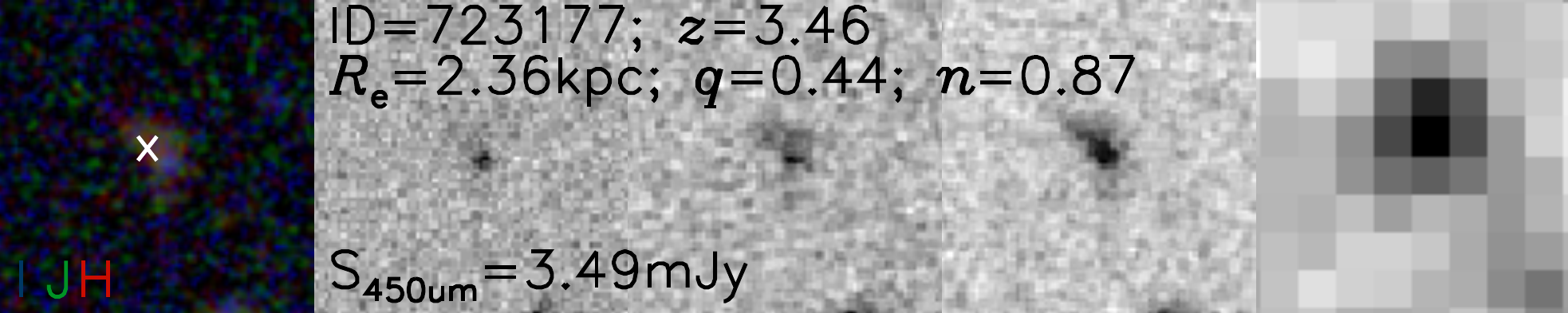}
\includegraphics[width=0.95\columnwidth]{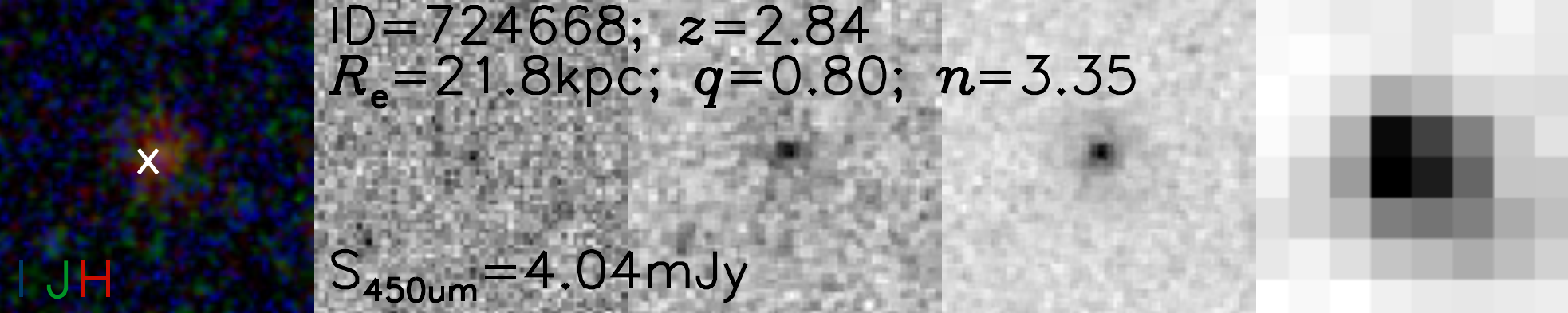}
\includegraphics[width=0.95\columnwidth]{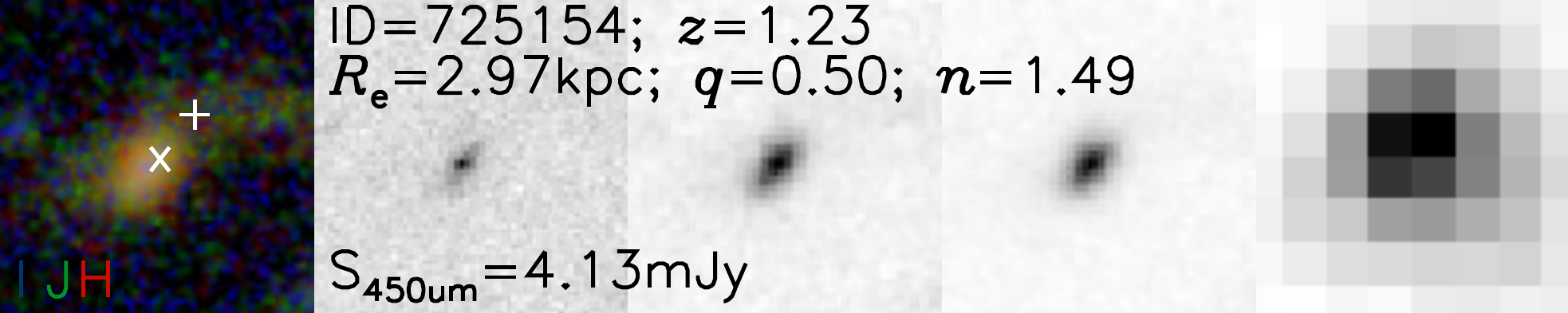}
\includegraphics[width=0.95\columnwidth]{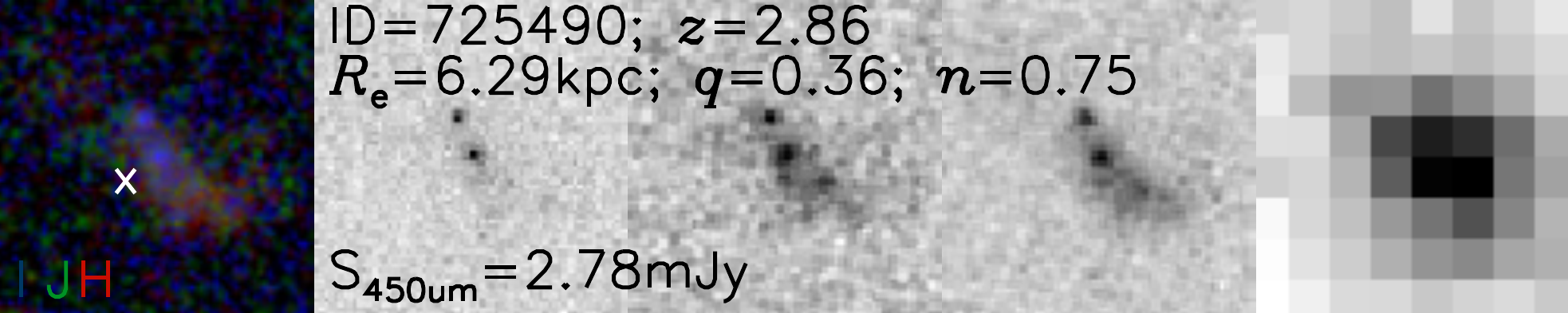}
\includegraphics[width=0.95\columnwidth]{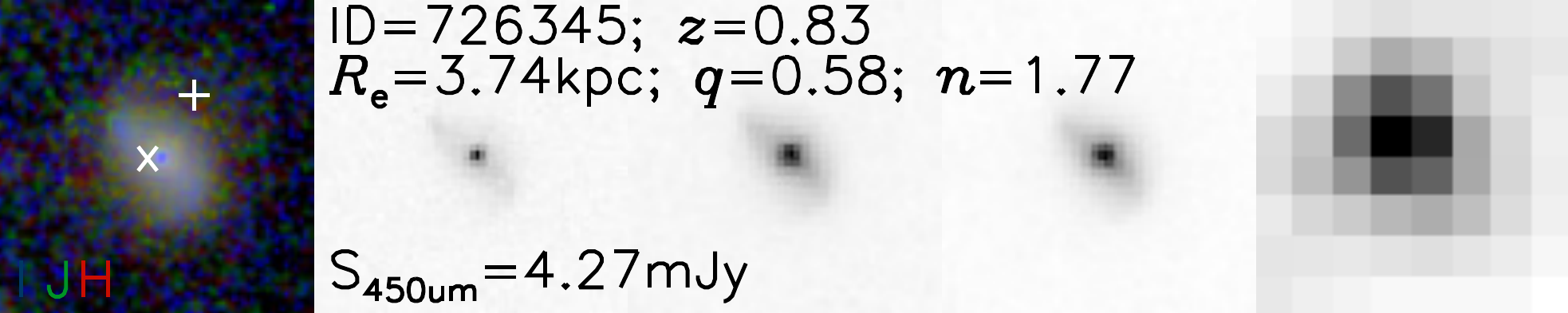}
\includegraphics[width=0.95\columnwidth]{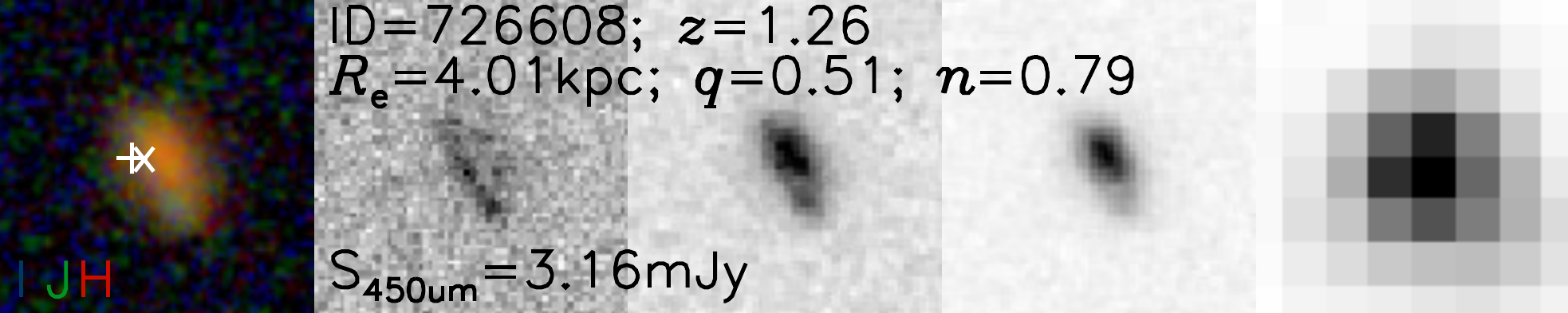}
\includegraphics[width=0.95\columnwidth]{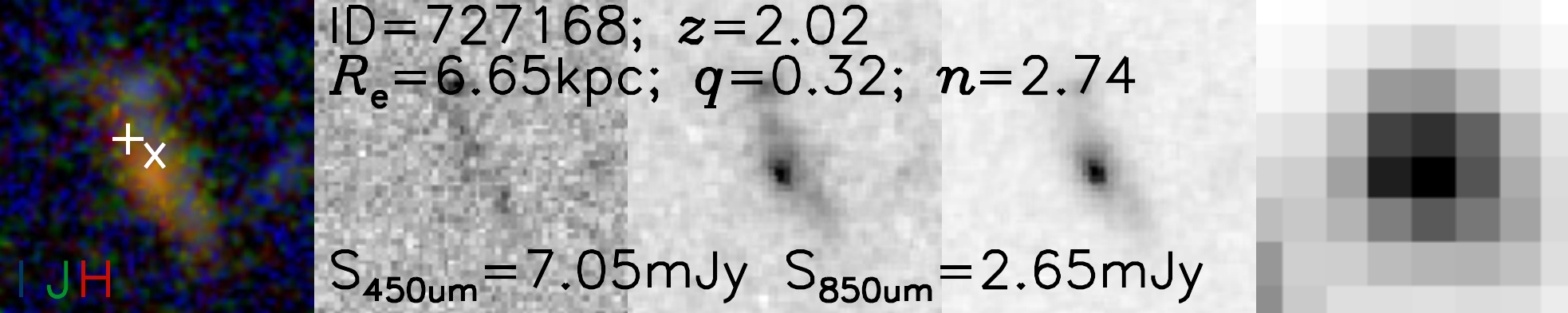}
\includegraphics[width=0.95\columnwidth]{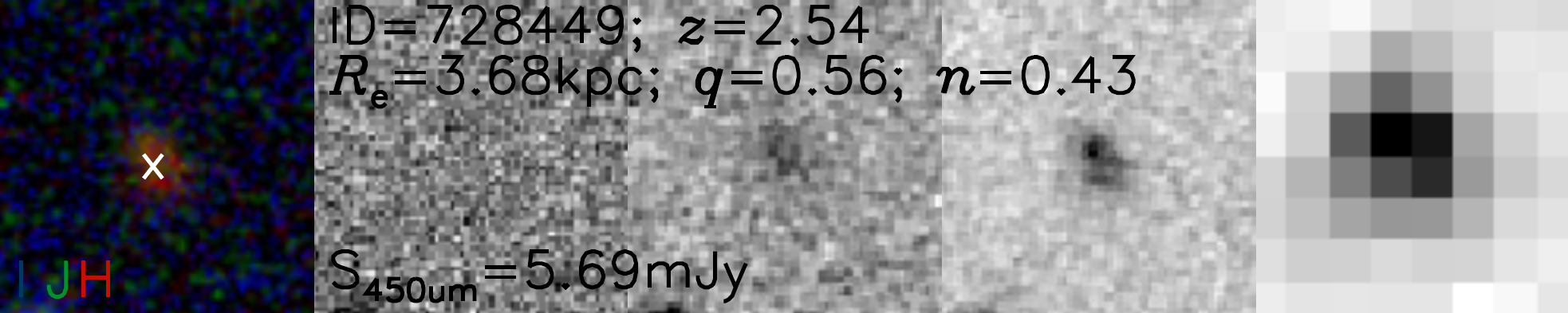}
\includegraphics[width=0.95\columnwidth]{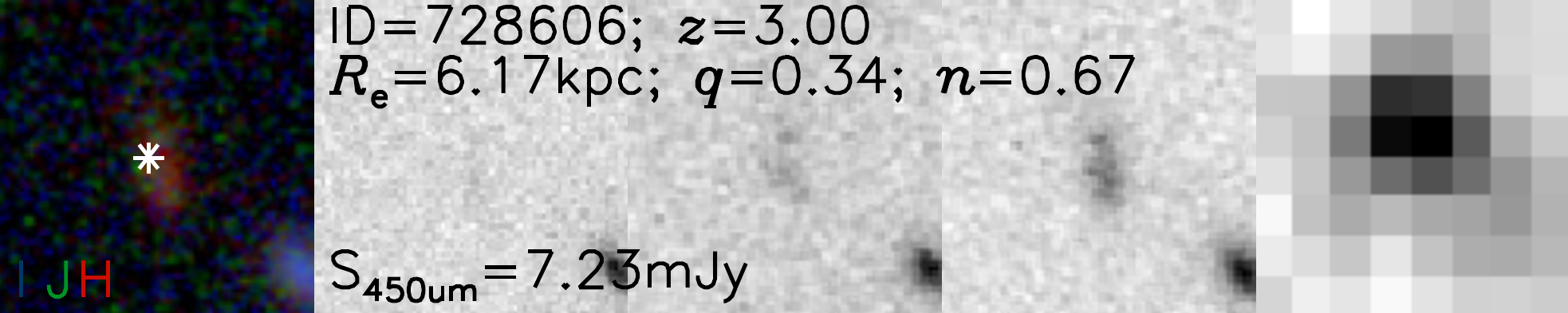}
\includegraphics[width=0.95\columnwidth]{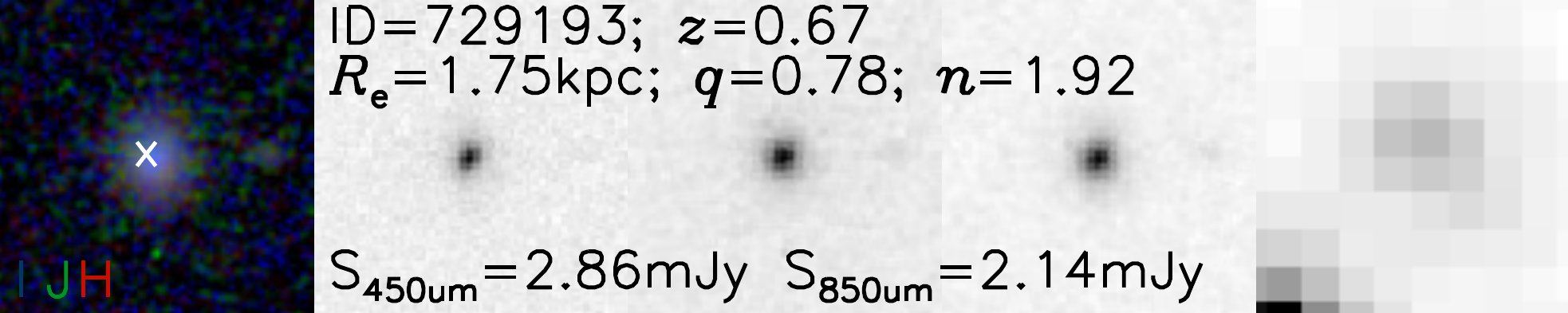}
\includegraphics[width=0.95\columnwidth]{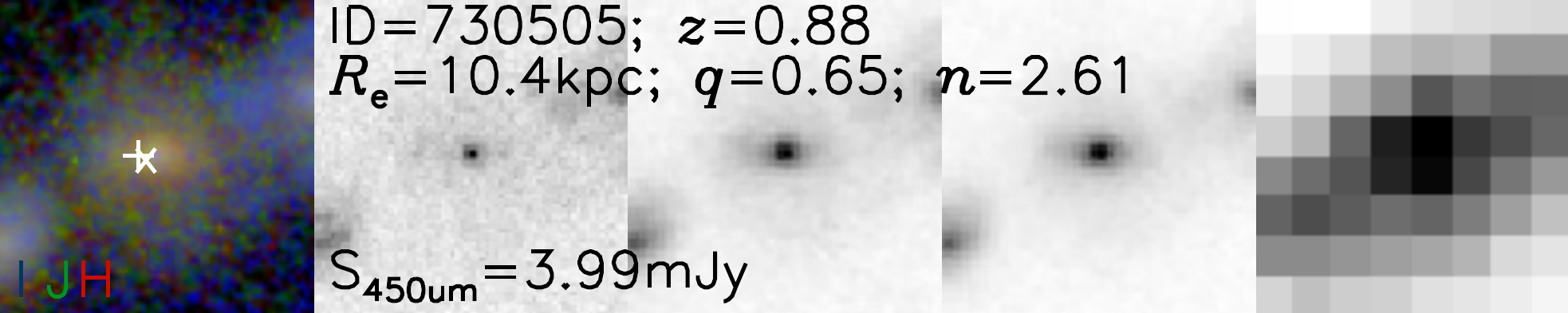}
\includegraphics[width=0.95\columnwidth]{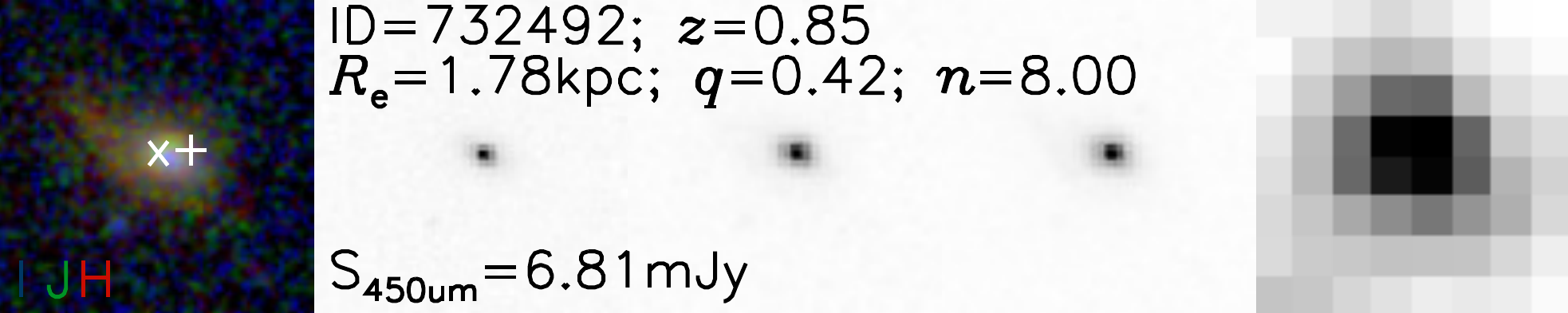}
\end{figure}
\begin{figure}
\centering
\includegraphics[width=0.95\columnwidth]{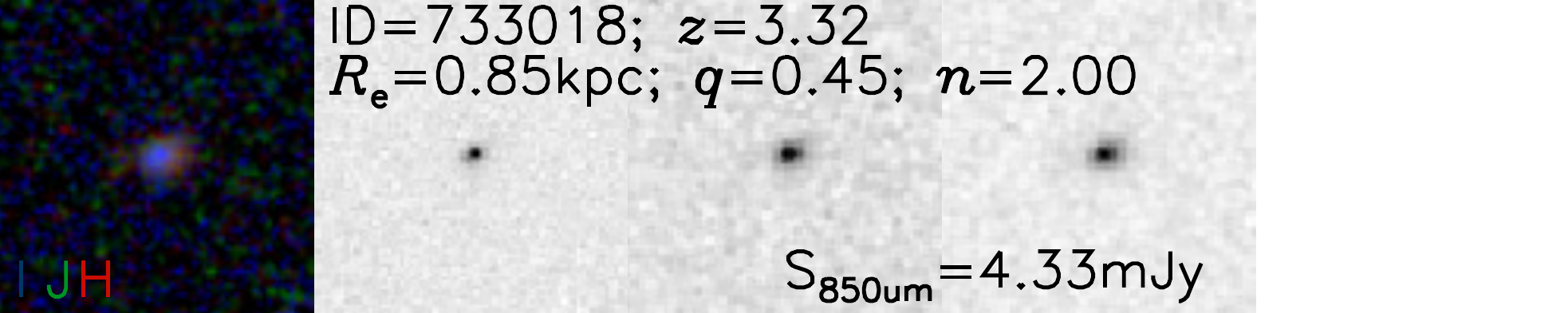}
\includegraphics[width=0.95\columnwidth]{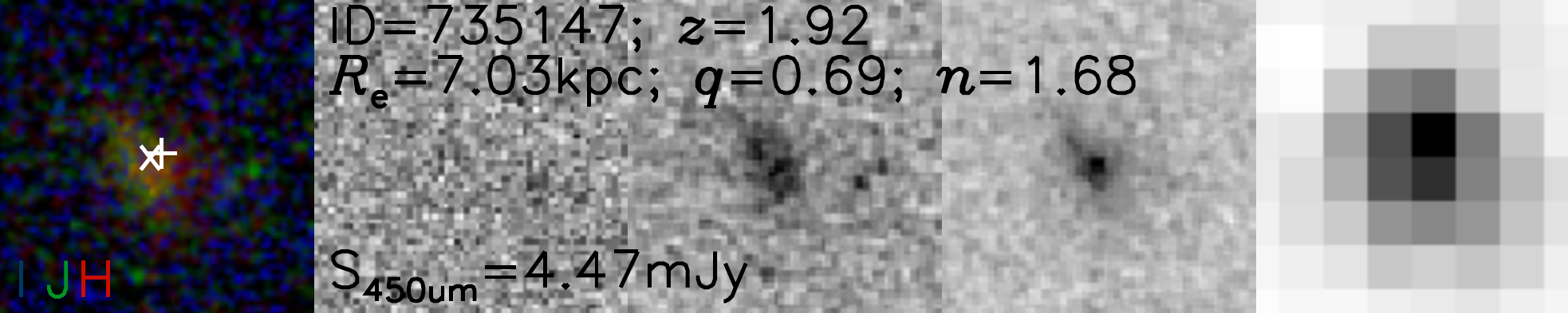}
\includegraphics[width=0.95\columnwidth]{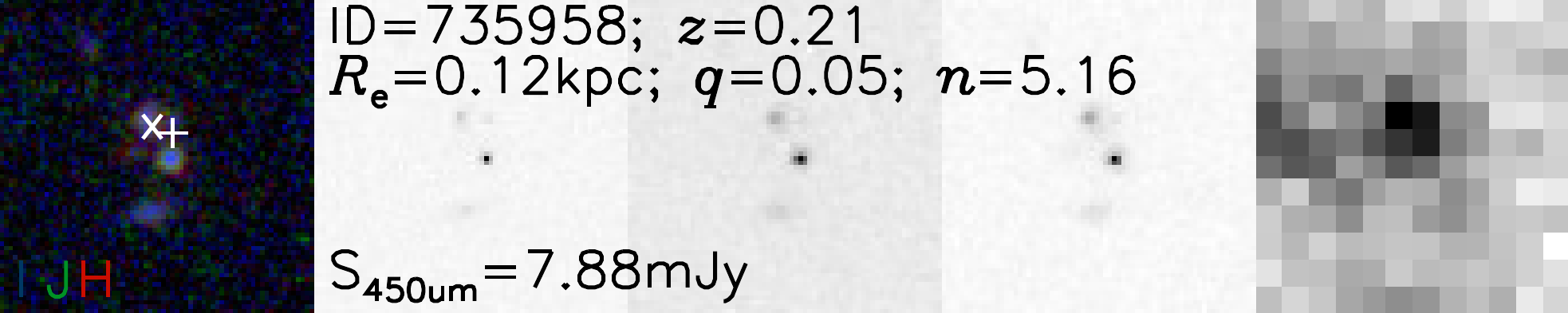}
\includegraphics[width=0.95\columnwidth]{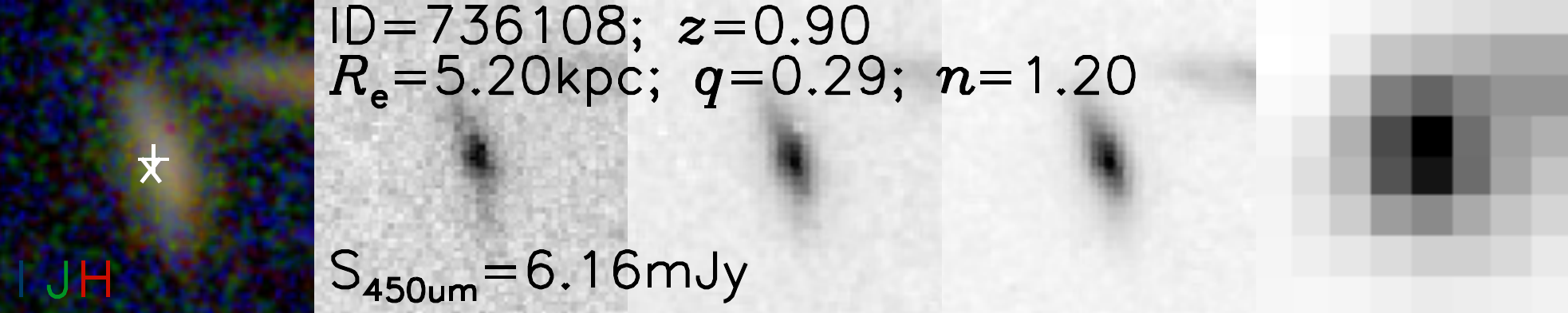}
\includegraphics[width=0.95\columnwidth]{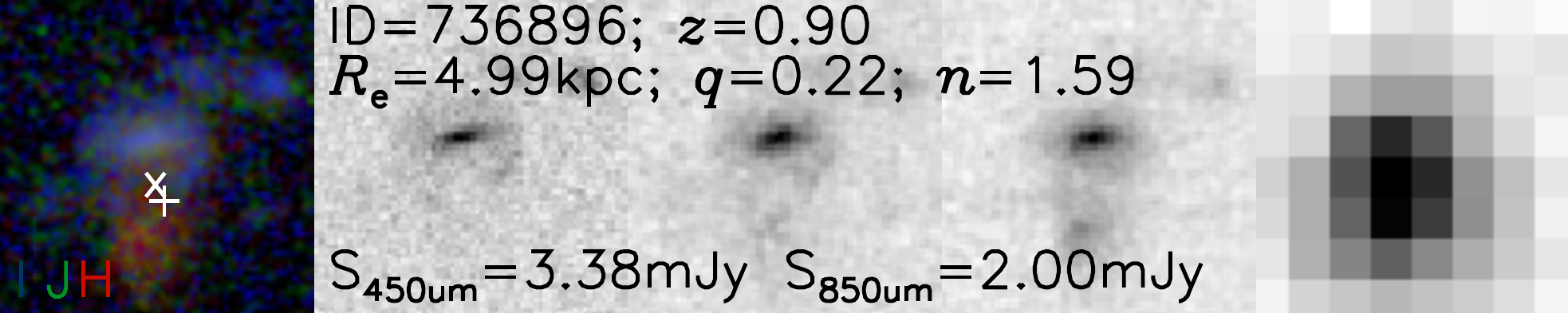}
\includegraphics[width=0.95\columnwidth]{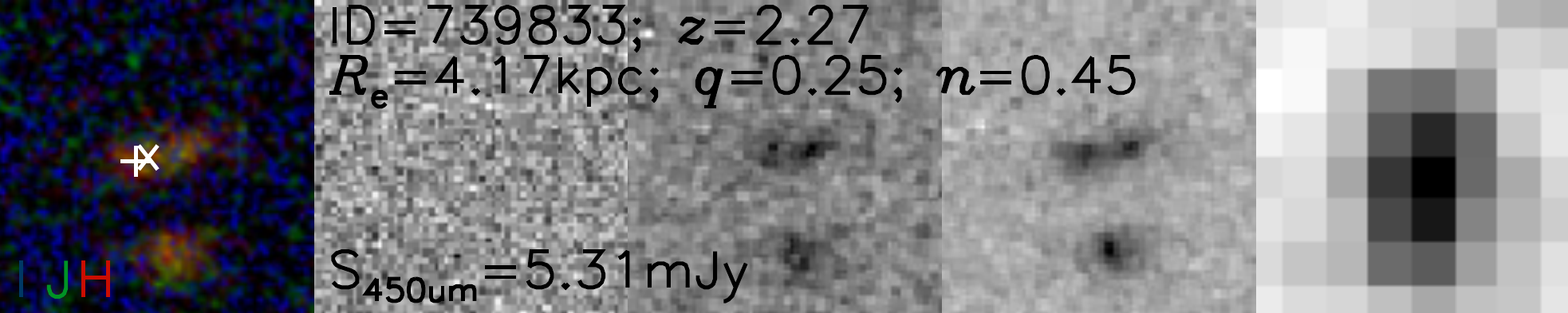}
\includegraphics[width=0.95\columnwidth]{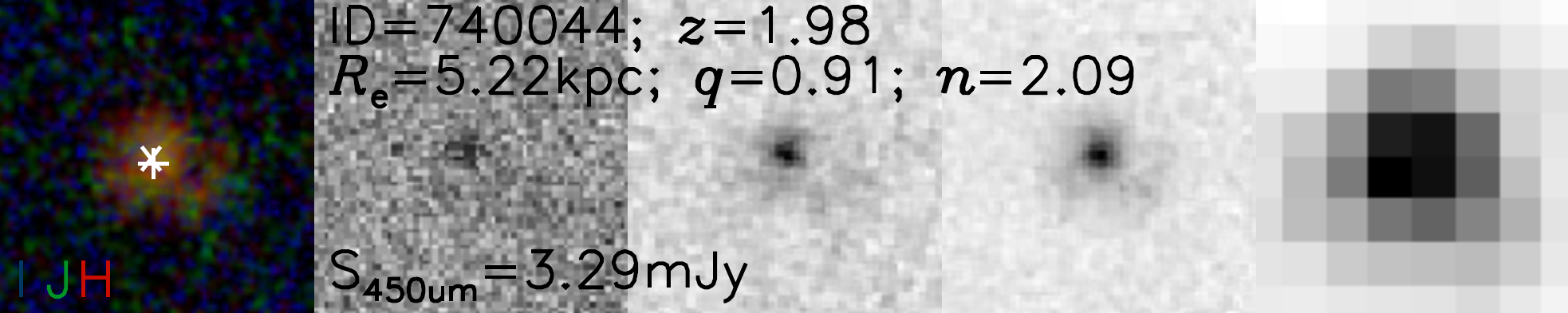}
\includegraphics[width=0.95\columnwidth]{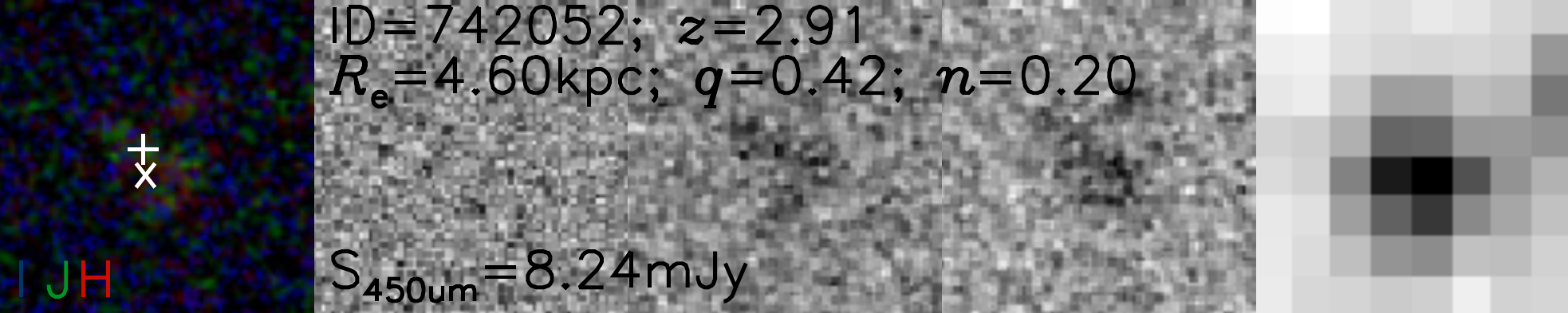}
\includegraphics[width=0.95\columnwidth]{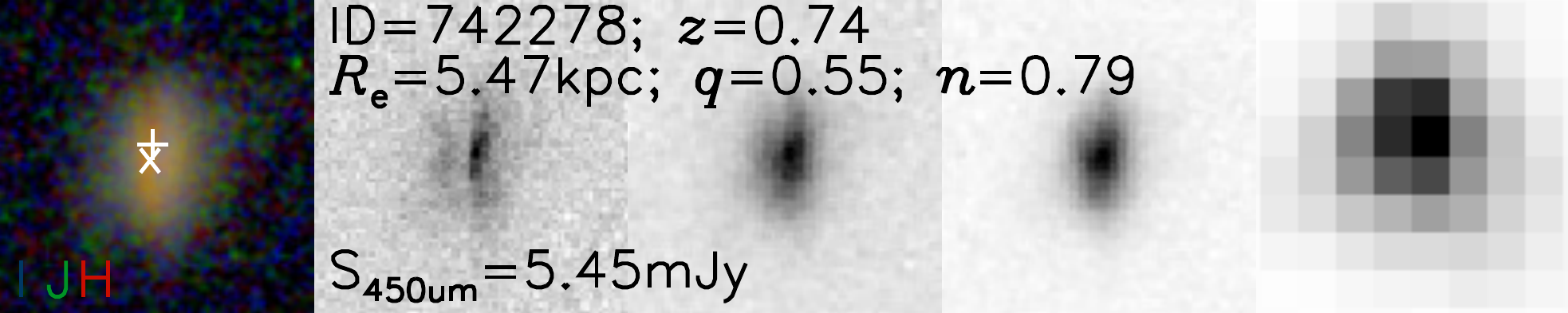}
\includegraphics[width=0.95\columnwidth]{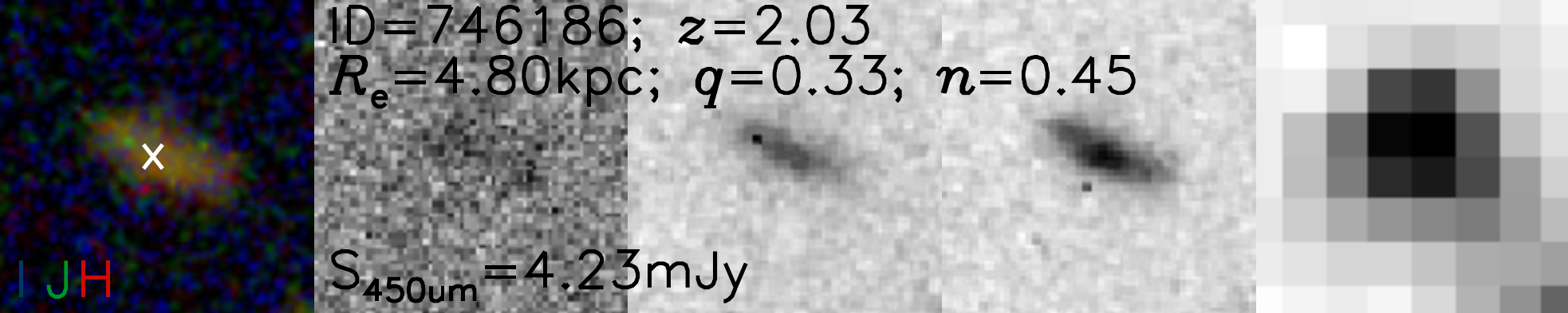}
\includegraphics[width=0.95\columnwidth]{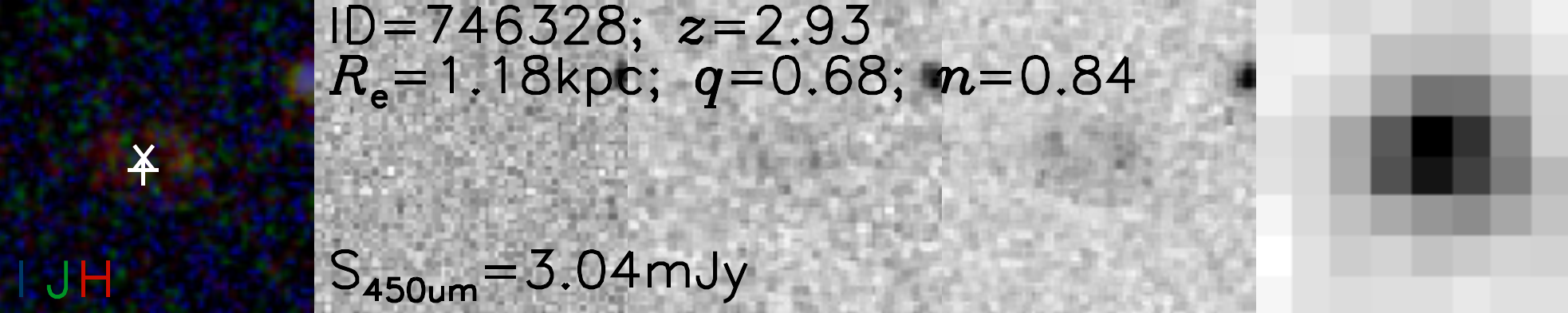}
\includegraphics[width=0.95\columnwidth]{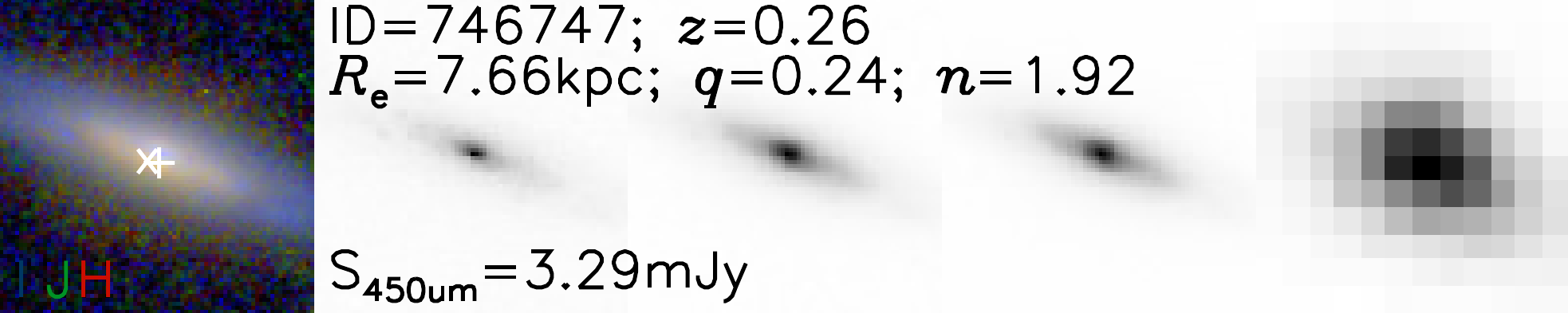}
\includegraphics[width=0.95\columnwidth]{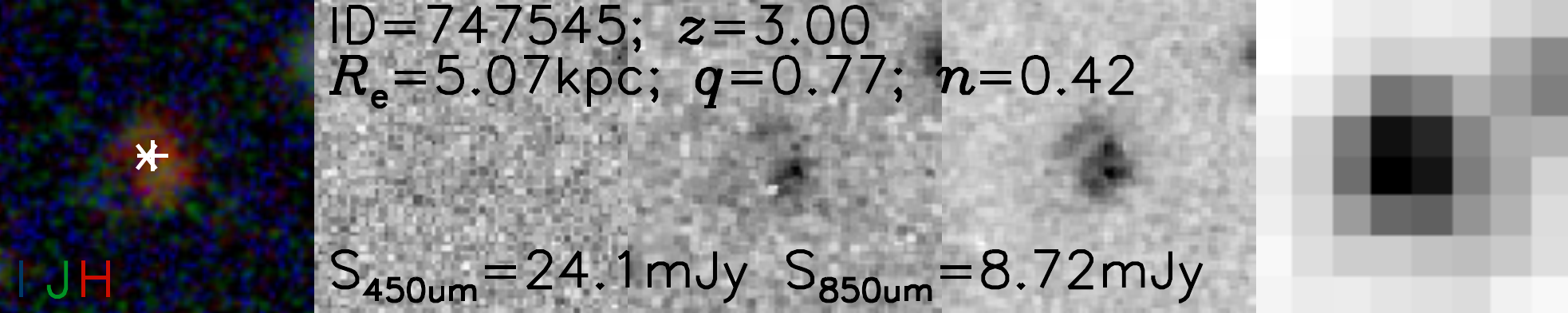}
\includegraphics[width=0.95\columnwidth]{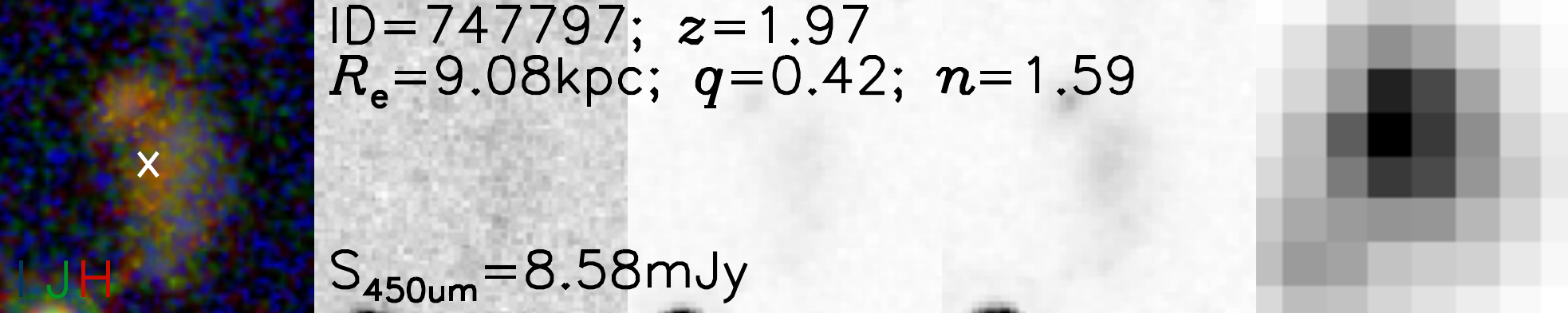}
\includegraphics[width=0.95\columnwidth]{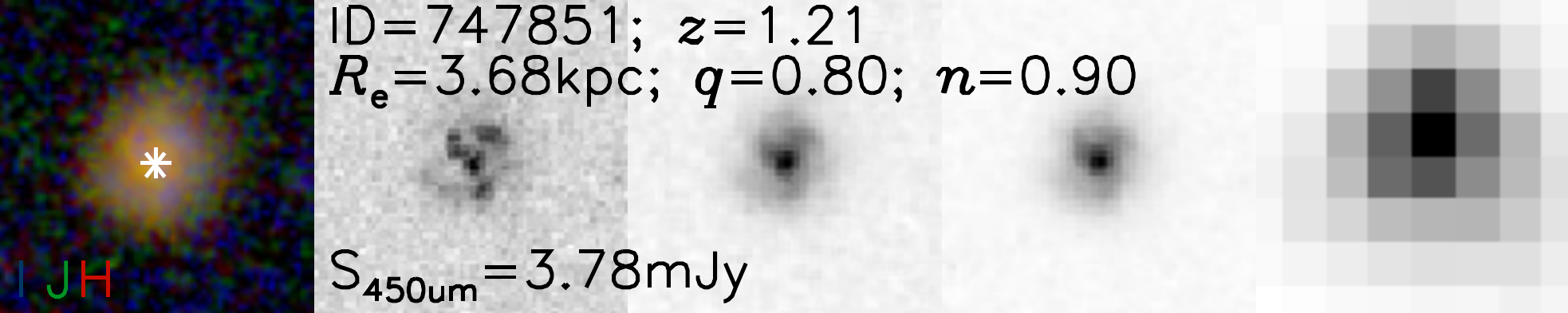}
\end{figure}
\begin{figure}
\centering
\includegraphics[width=0.95\columnwidth]{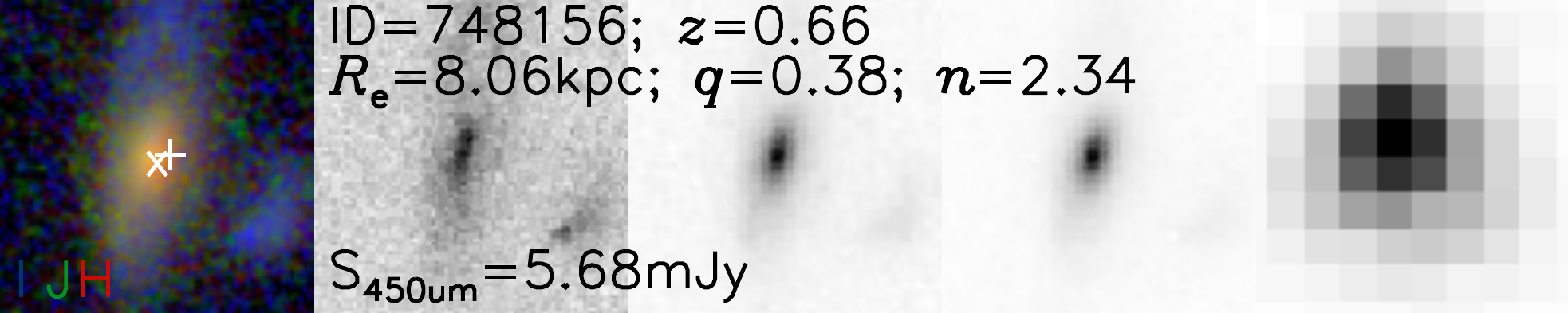}
\includegraphics[width=0.95\columnwidth]{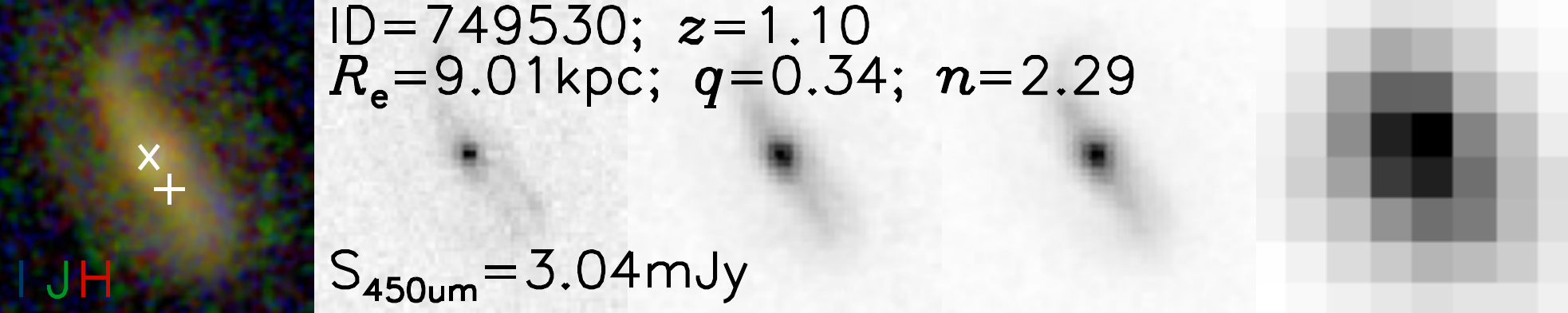}
\includegraphics[width=0.95\columnwidth]{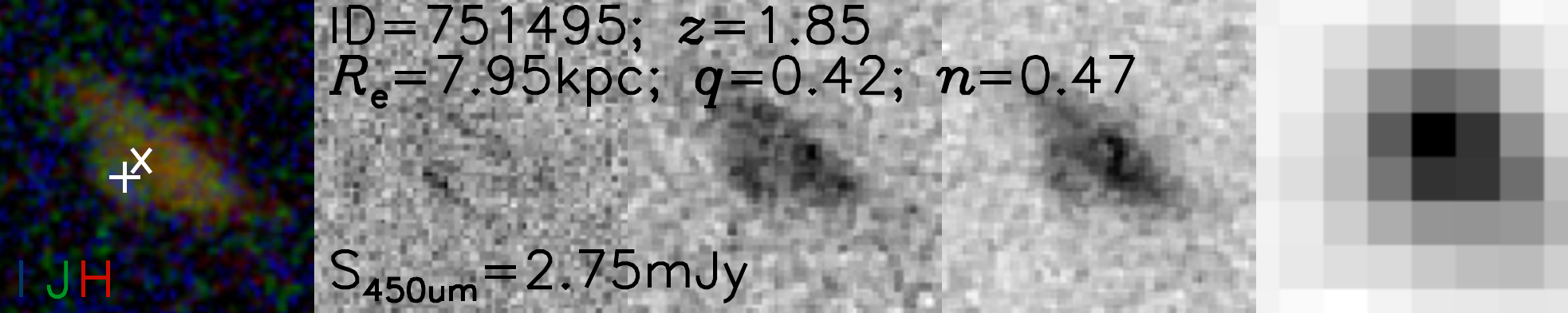}
\includegraphics[width=0.95\columnwidth]{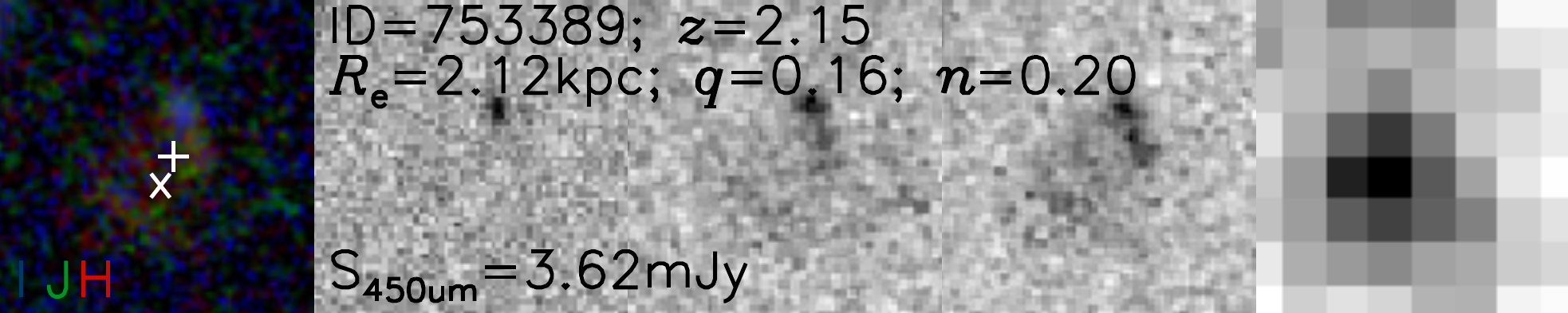}
\includegraphics[width=0.95\columnwidth]{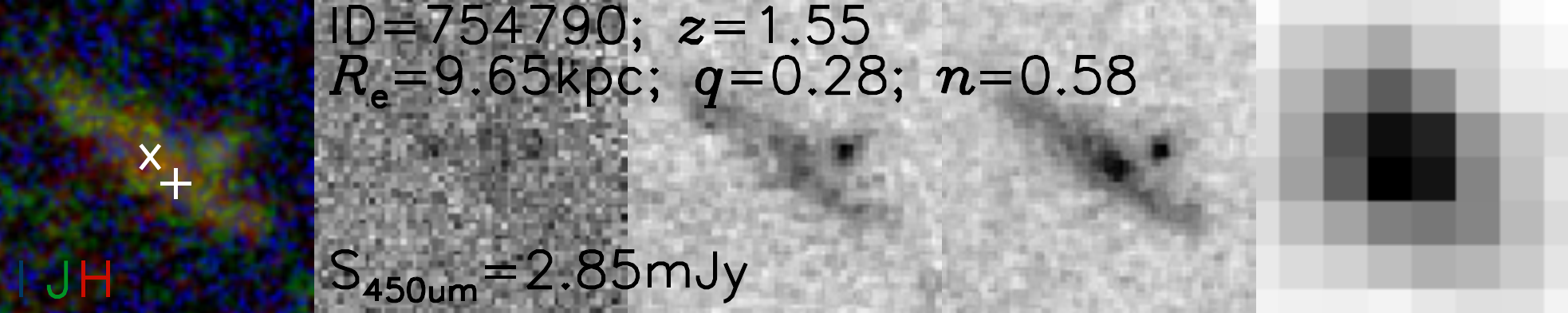}
\includegraphics[width=0.95\columnwidth]{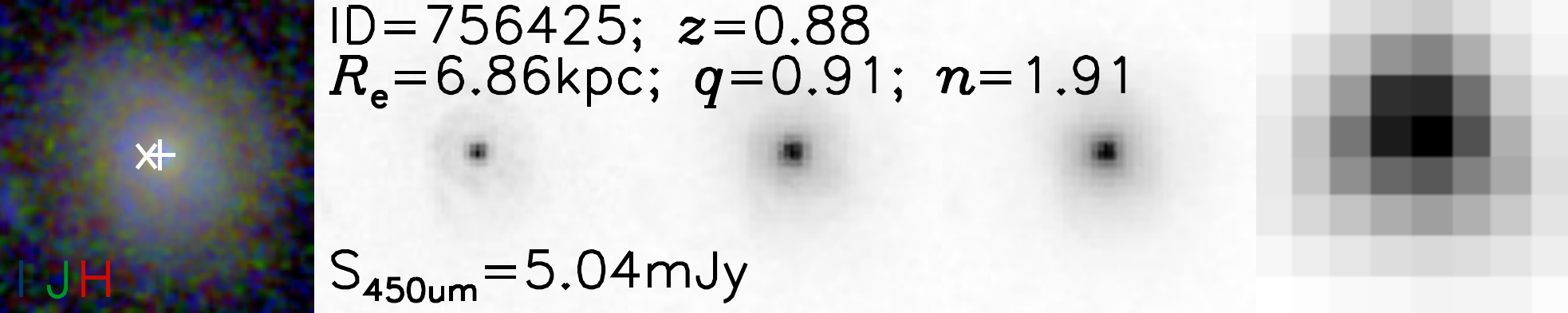}
\includegraphics[width=0.95\columnwidth]{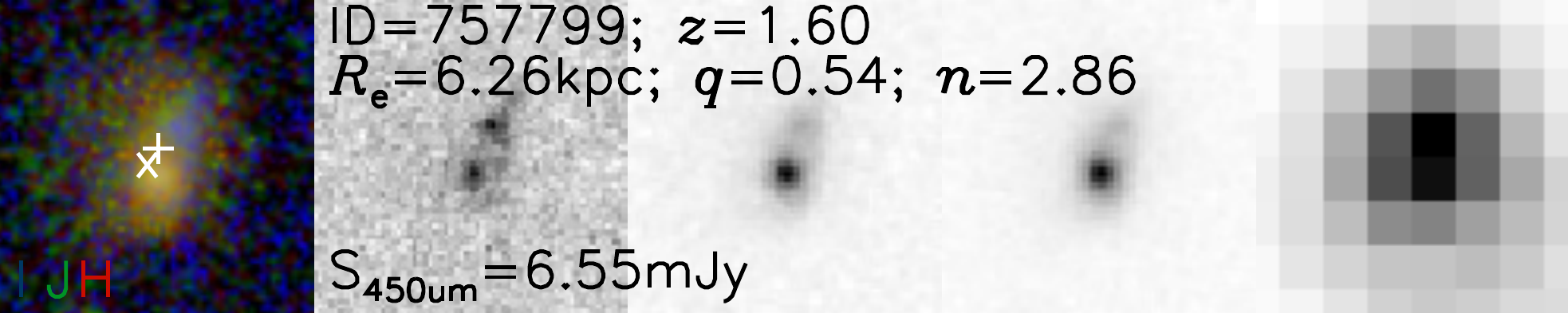}
\includegraphics[width=0.95\columnwidth]{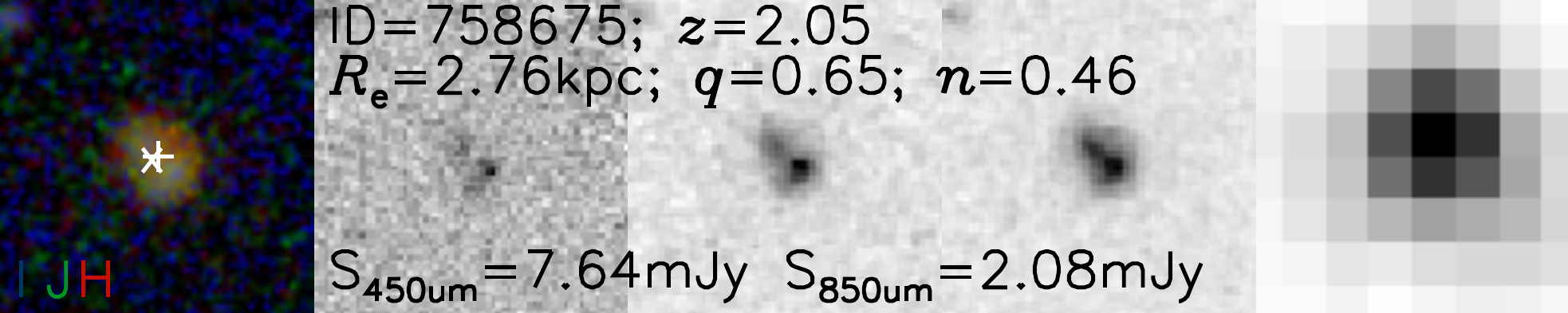}
\includegraphics[width=0.95\columnwidth]{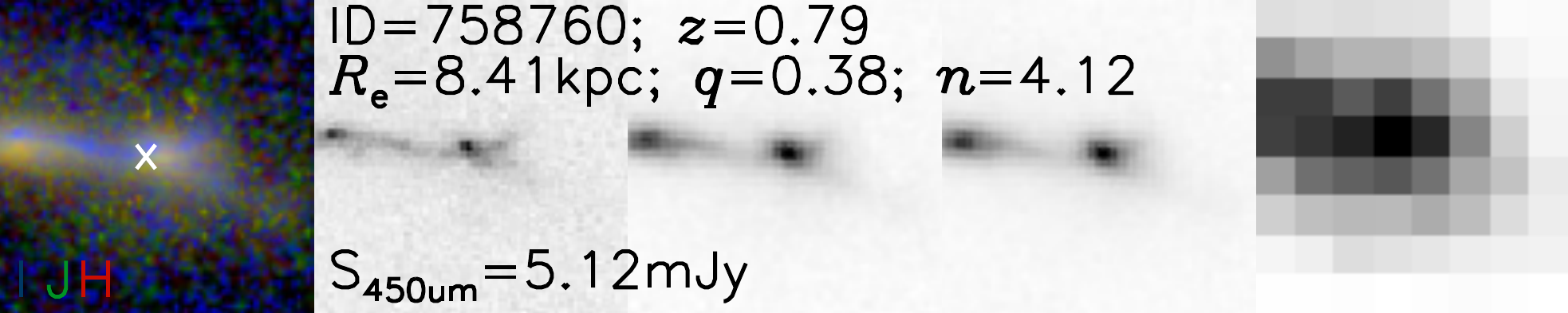}
\includegraphics[width=0.95\columnwidth]{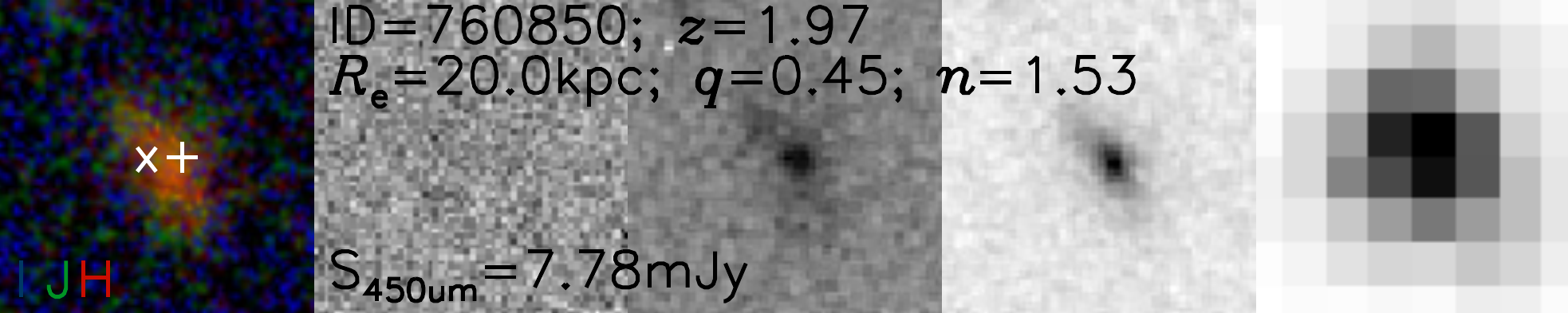}
\includegraphics[width=0.95\columnwidth]{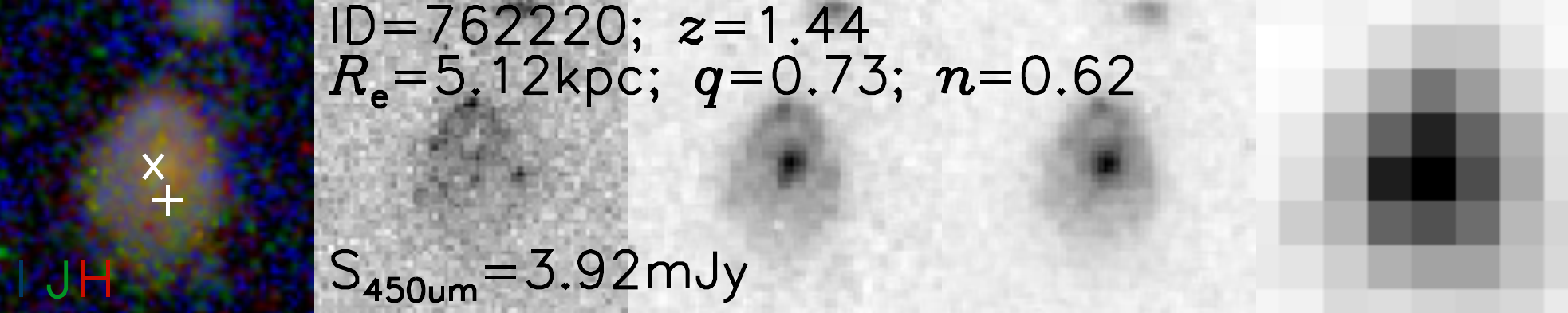}
\includegraphics[width=0.95\columnwidth]{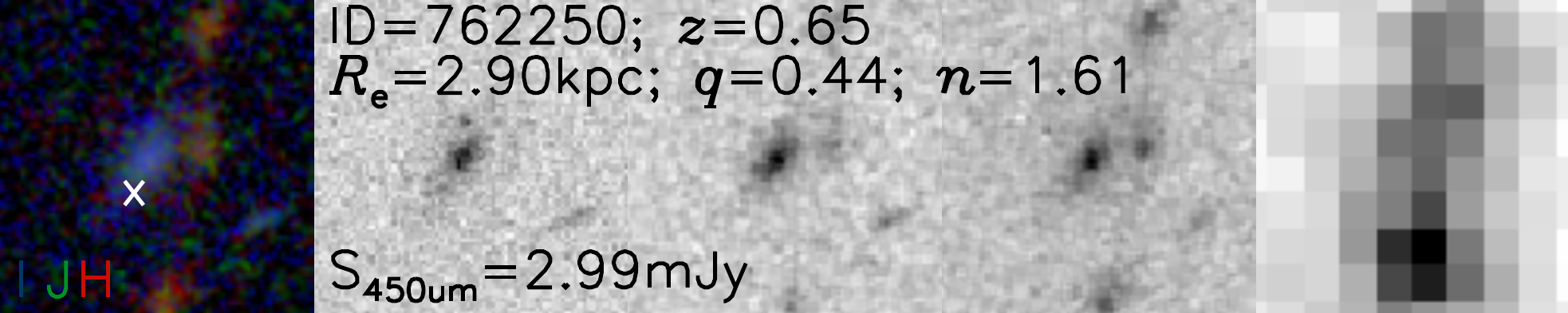}
\includegraphics[width=0.95\columnwidth]{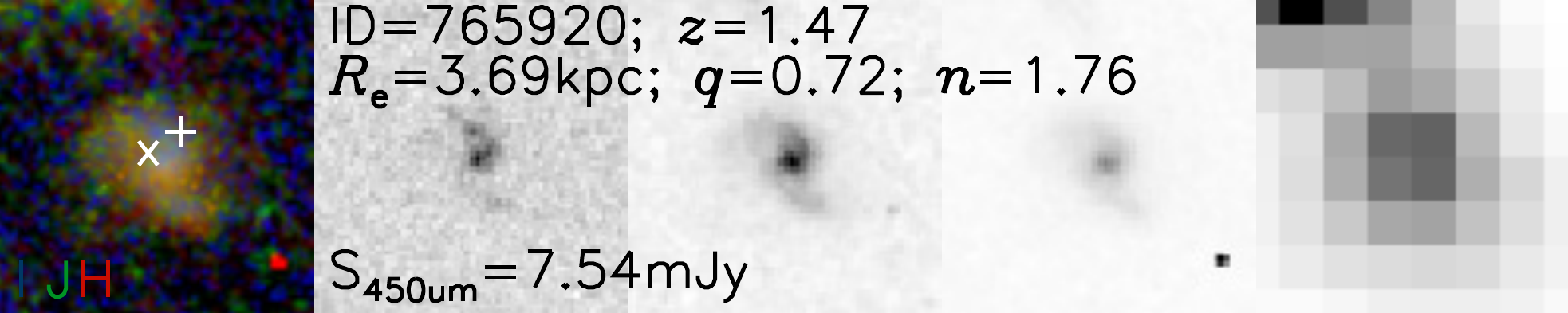}
\includegraphics[width=0.95\columnwidth]{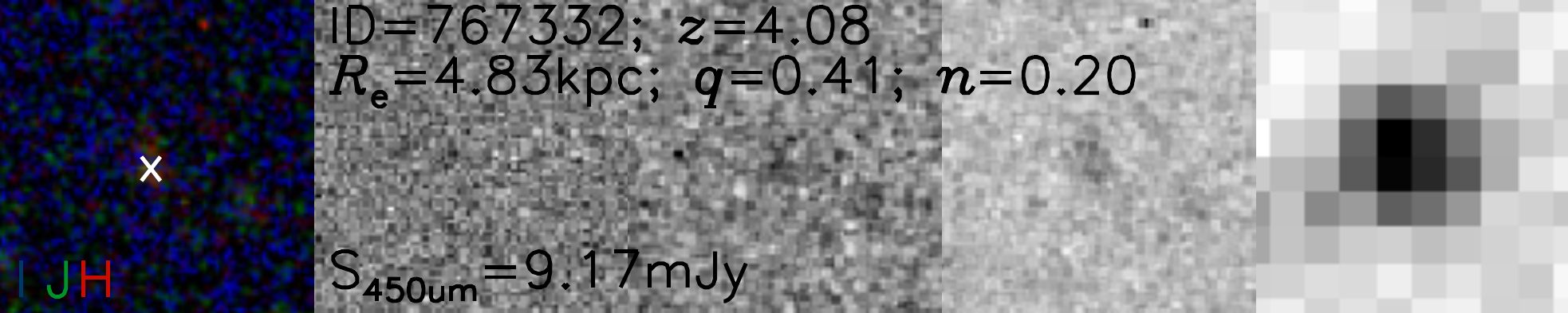}
\includegraphics[width=0.95\columnwidth]{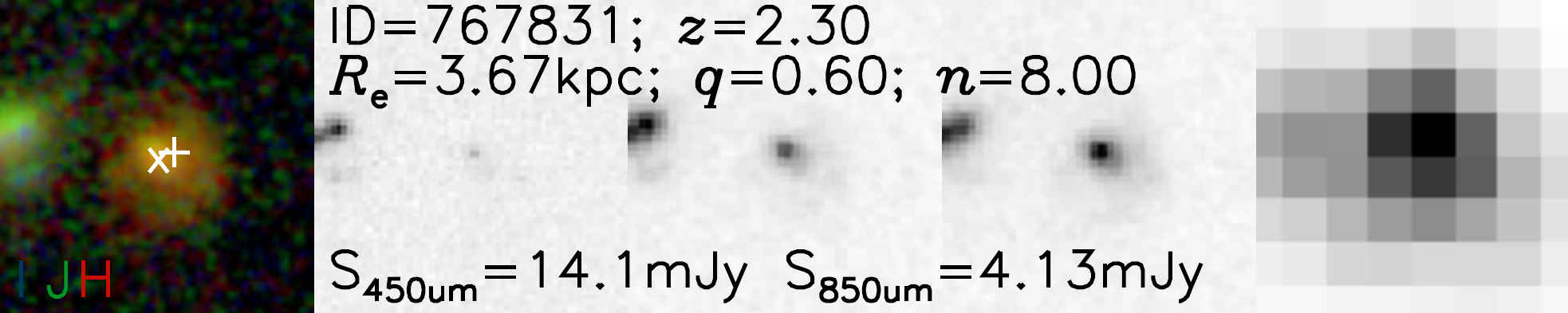}
\end{figure}
\begin{figure}
\centering
\includegraphics[width=0.95\columnwidth]{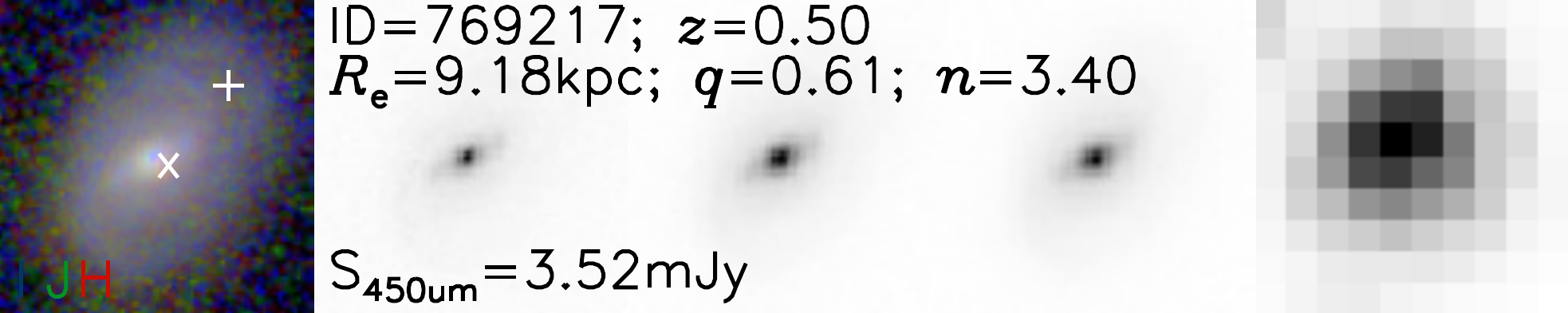}
\includegraphics[width=0.95\columnwidth]{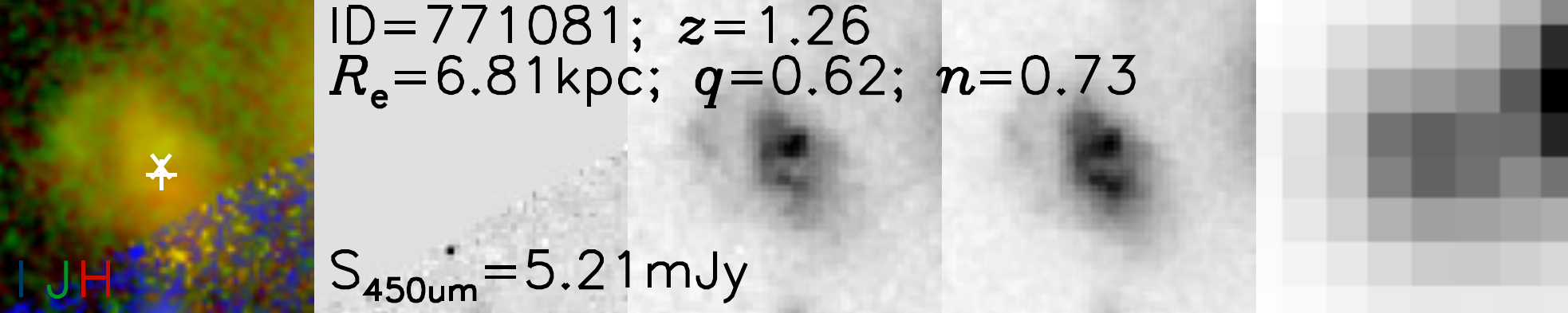}
\includegraphics[width=0.95\columnwidth]{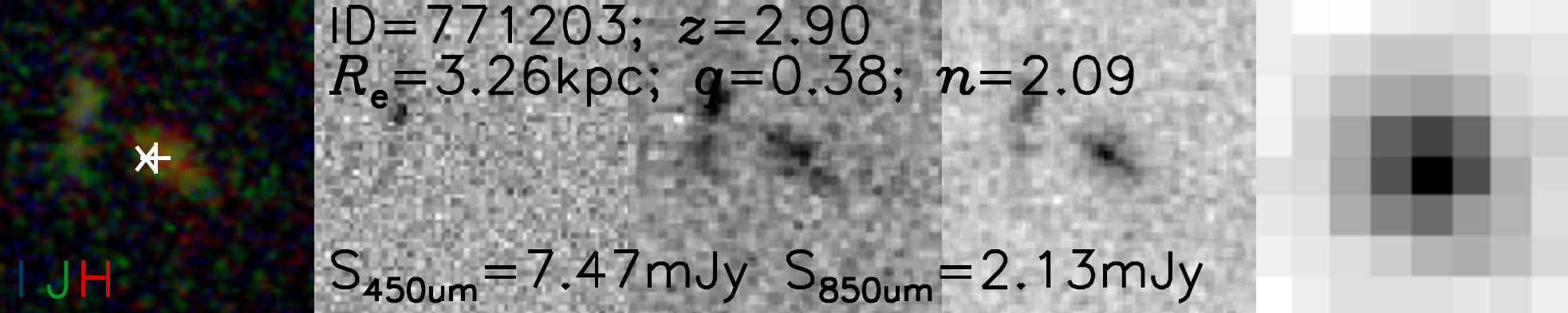}
\includegraphics[width=0.95\columnwidth]{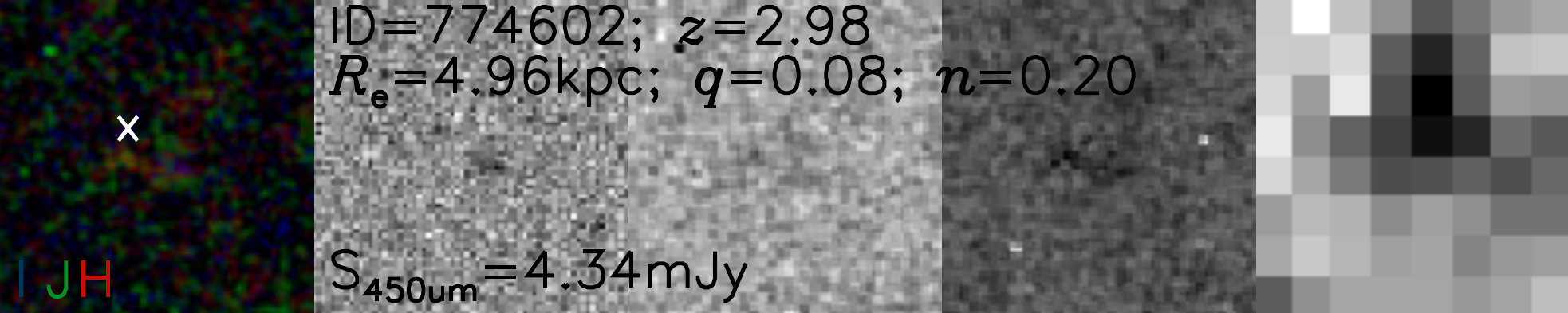}
\includegraphics[width=0.95\columnwidth]{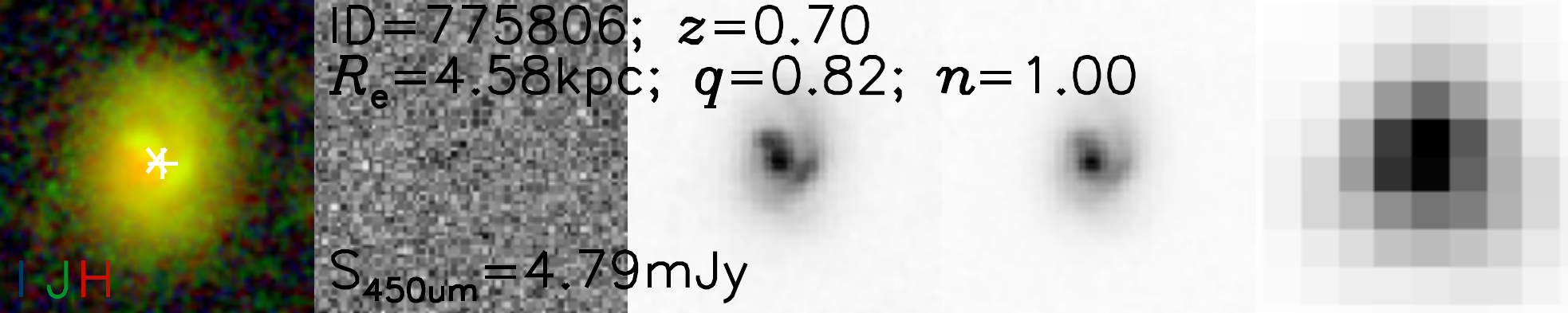}
\includegraphics[width=0.95\columnwidth]{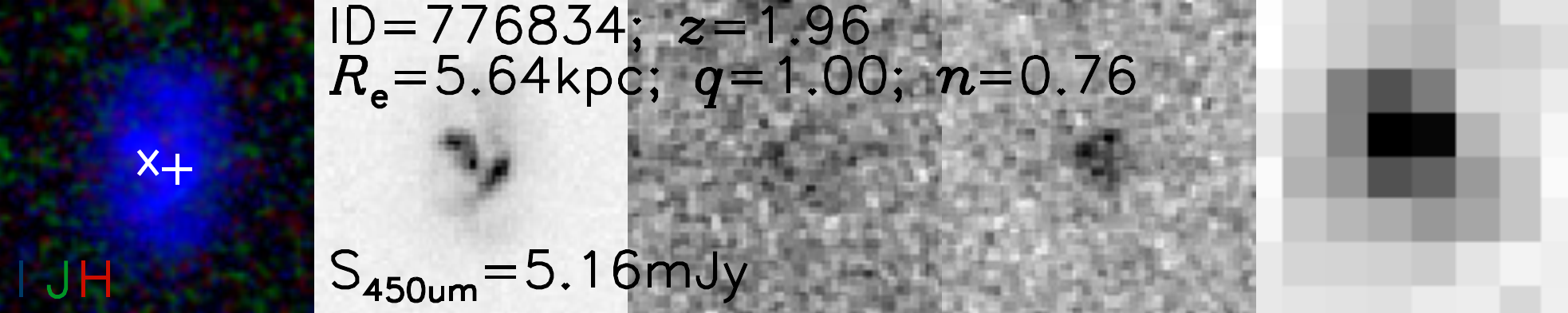}
\includegraphics[width=0.95\columnwidth]{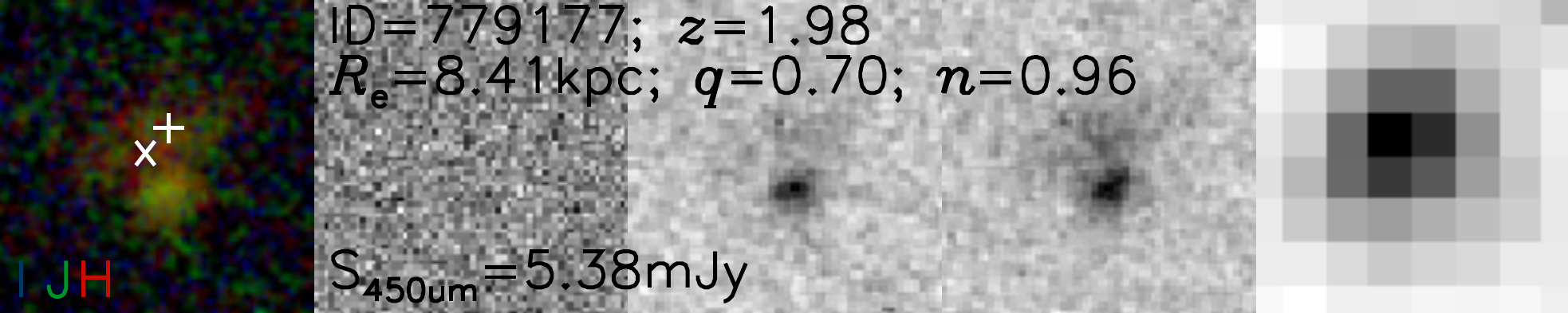}
\includegraphics[width=0.95\columnwidth]{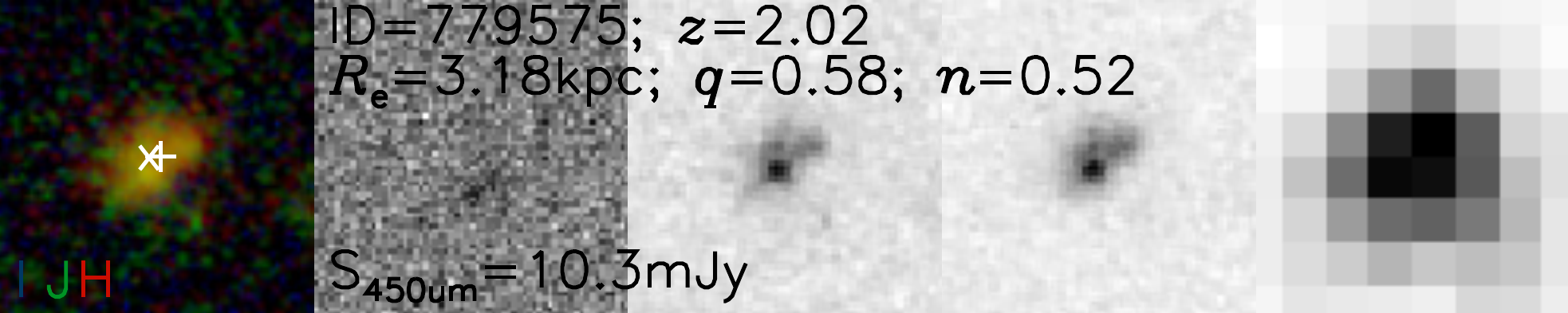}
\includegraphics[width=0.95\columnwidth]{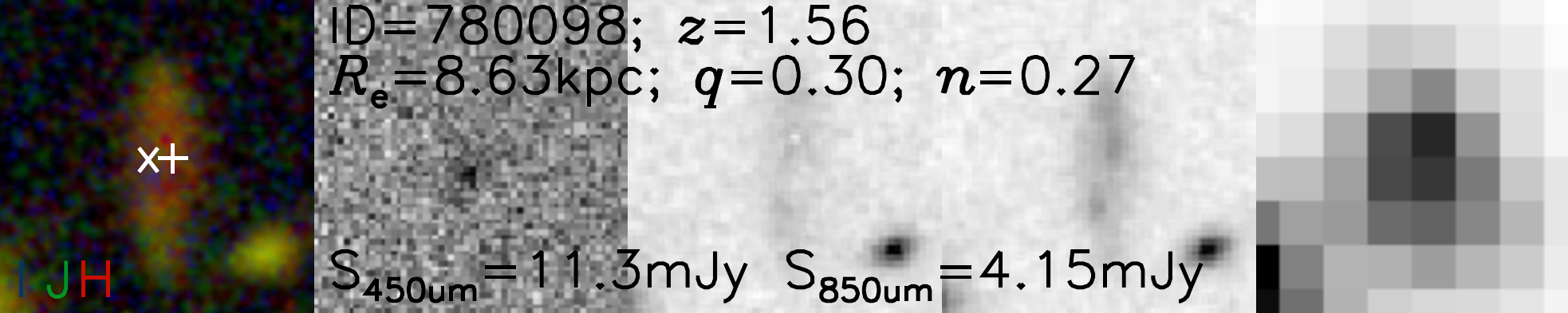}
\includegraphics[width=0.95\columnwidth]{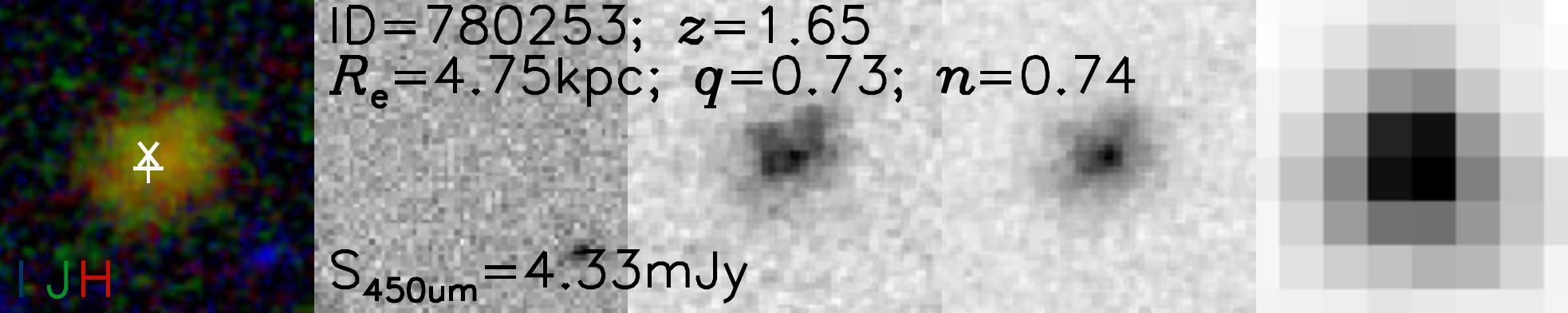}
\includegraphics[width=0.95\columnwidth]{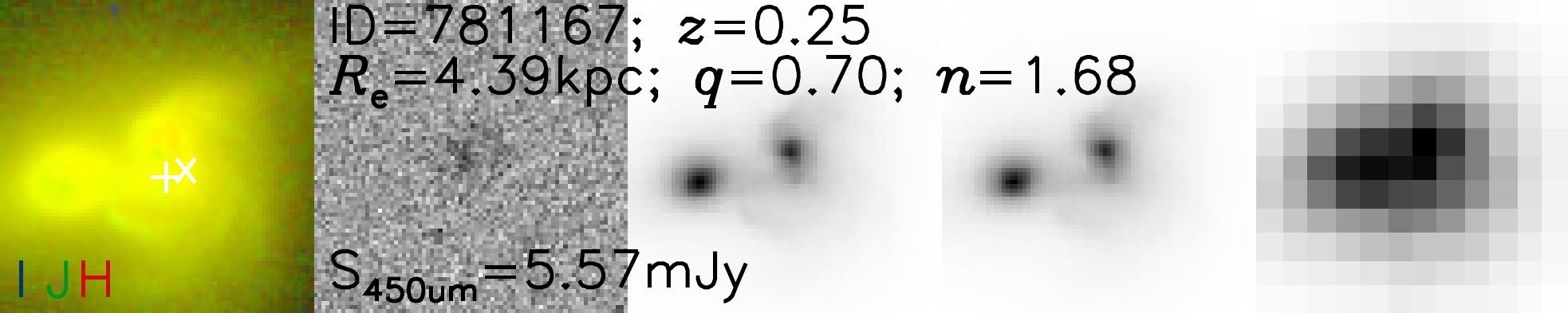}
\includegraphics[width=0.95\columnwidth]{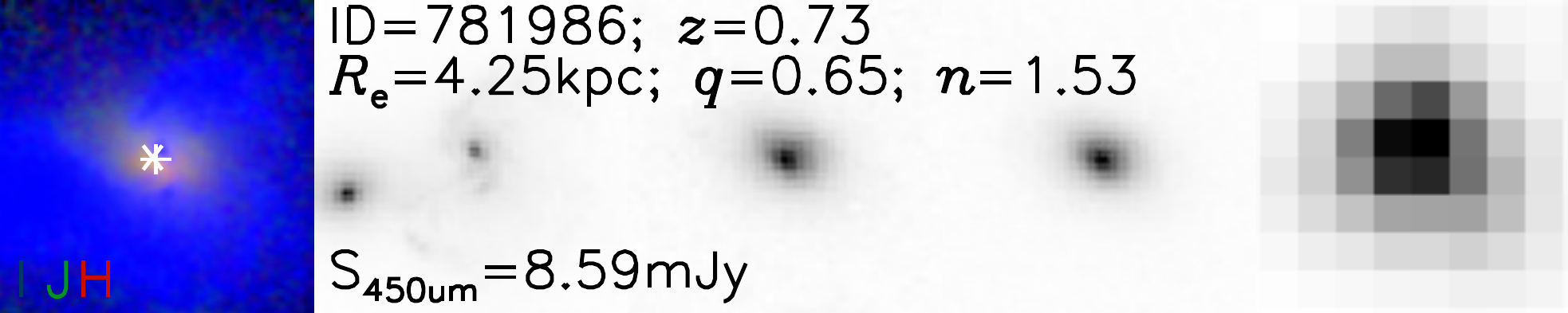}
\includegraphics[width=0.95\columnwidth]{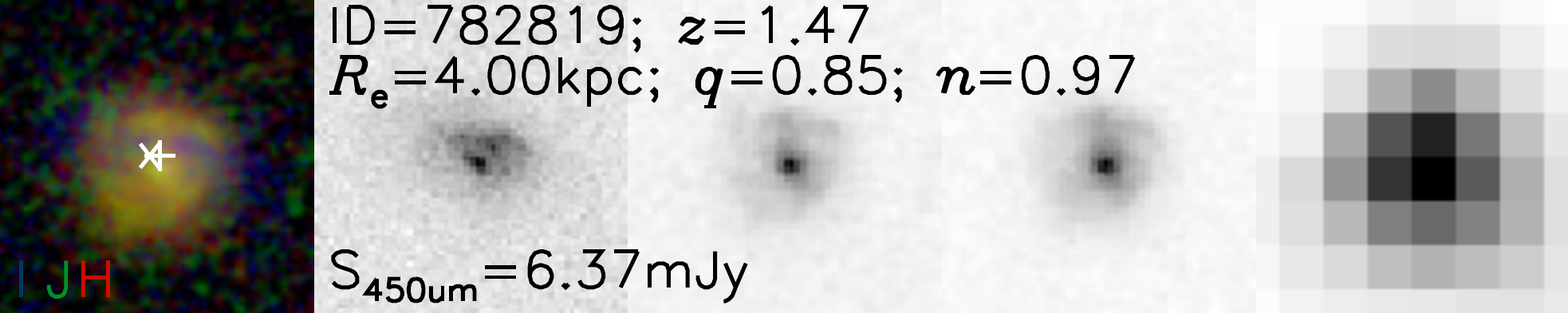}
\includegraphics[width=0.95\columnwidth]{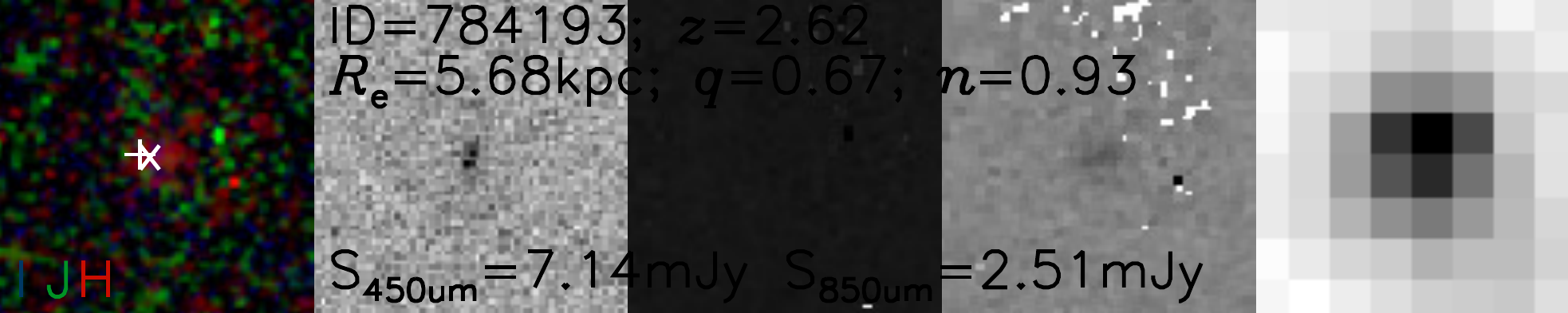}
\includegraphics[width=0.95\columnwidth]{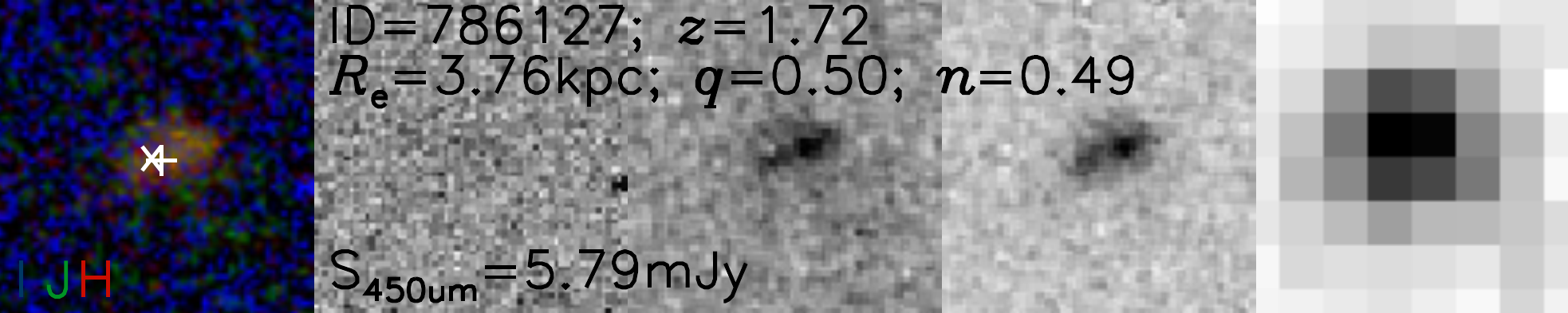}
\end{figure}
\begin{figure}
\centering
\includegraphics[width=0.95\columnwidth]{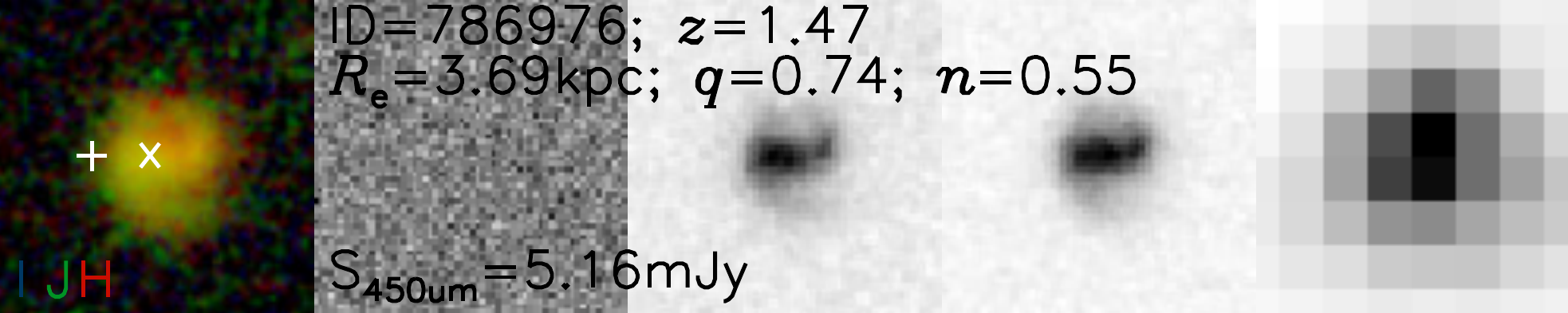}
\includegraphics[width=0.95\columnwidth]{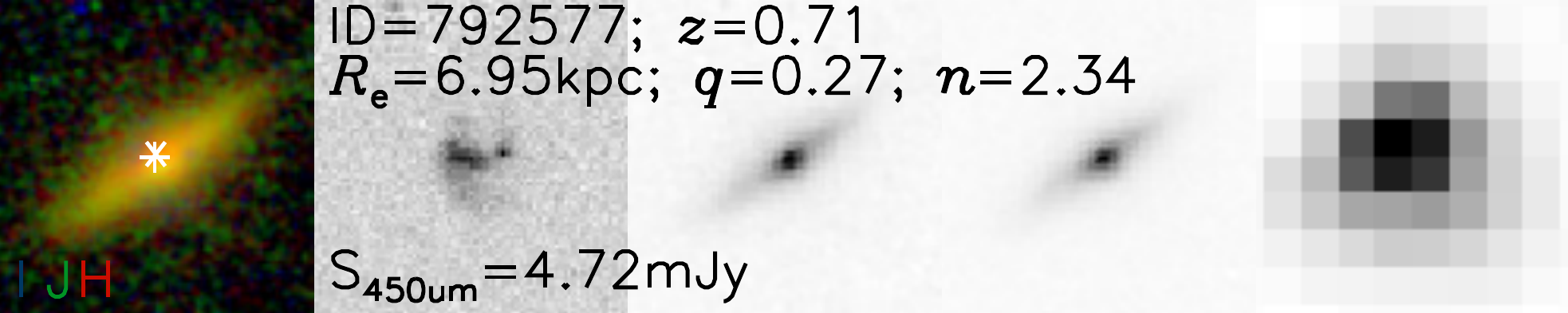}
\includegraphics[width=0.95\columnwidth]{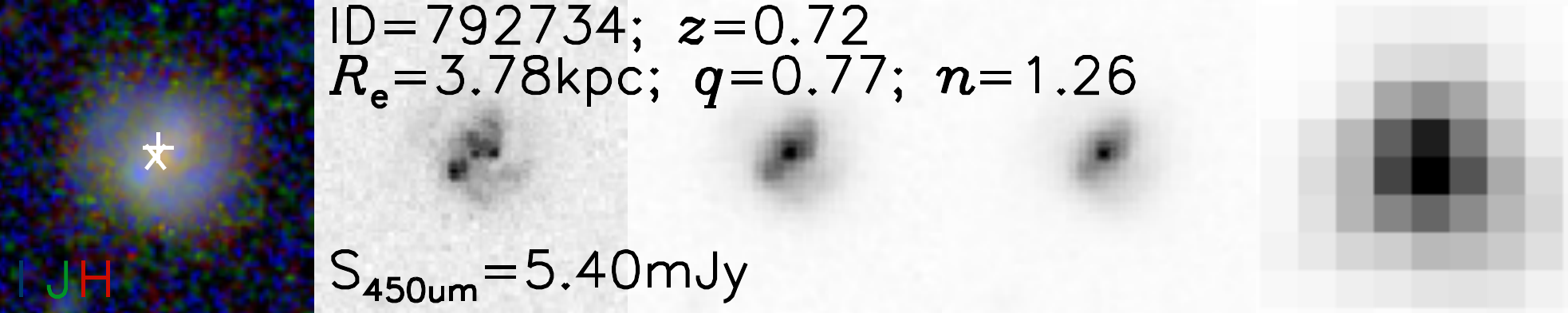}
\includegraphics[width=0.95\columnwidth]{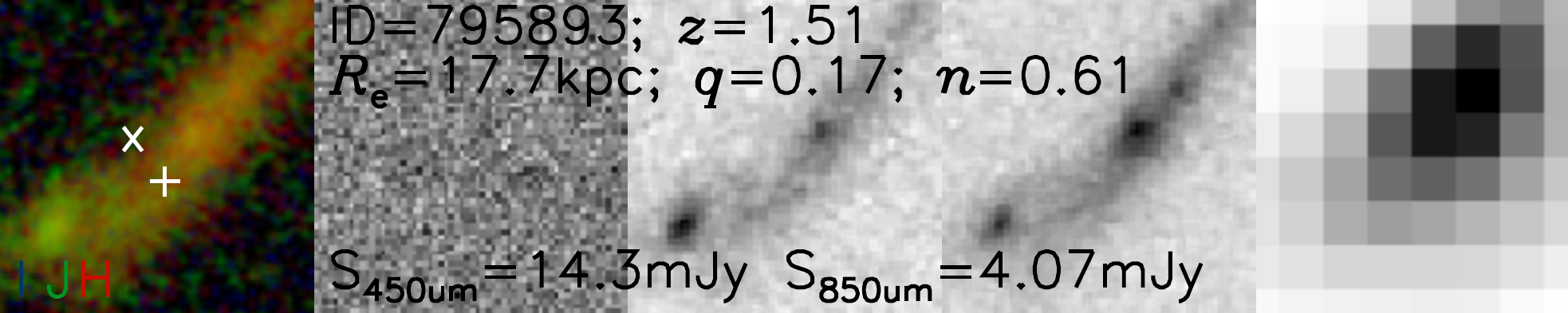}
\includegraphics[width=0.95\columnwidth]{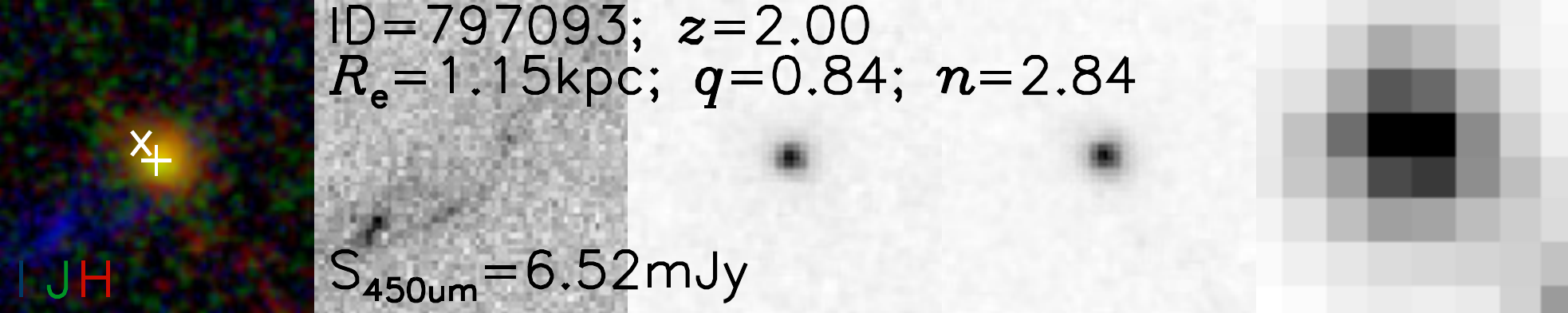}
\includegraphics[width=0.95\columnwidth]{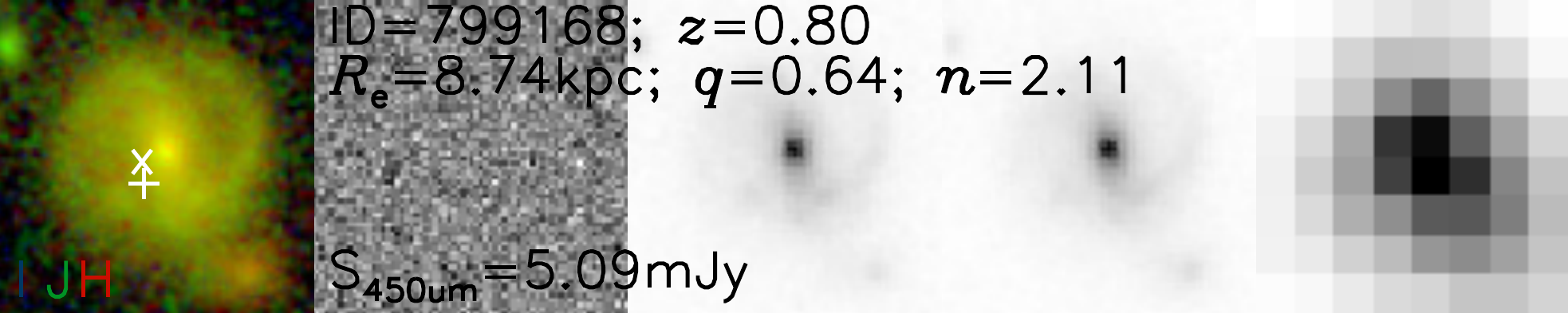}
\includegraphics[width=0.95\columnwidth]{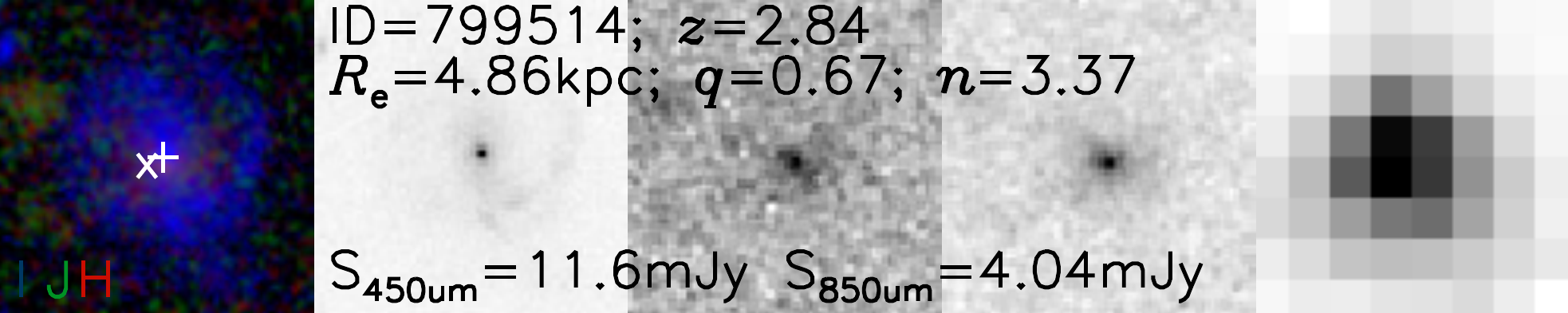}
\includegraphics[width=0.95\columnwidth]{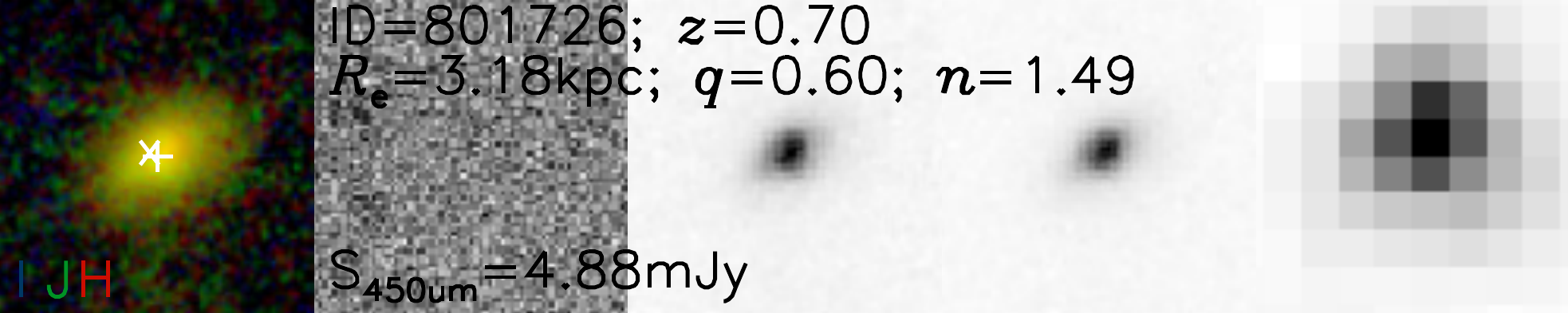}
\includegraphics[width=0.95\columnwidth]{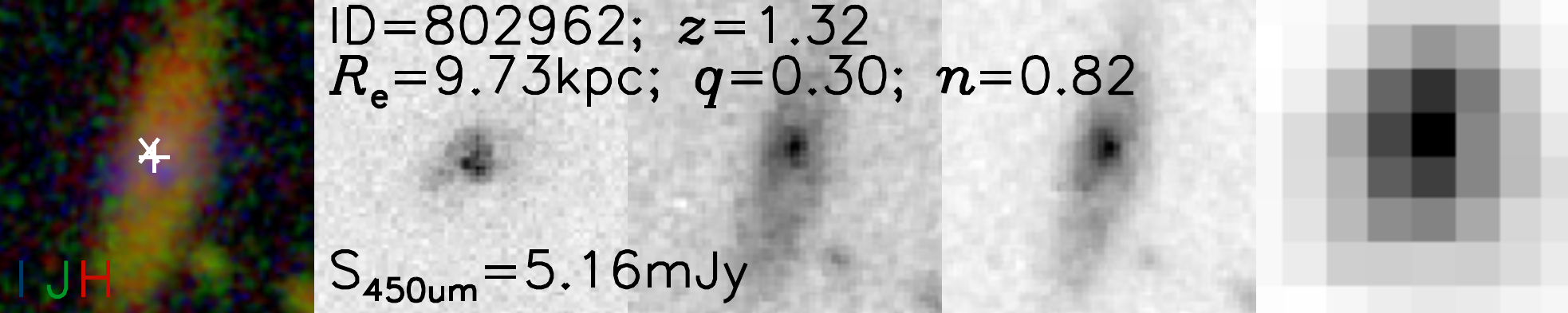}
\includegraphics[width=0.95\columnwidth]{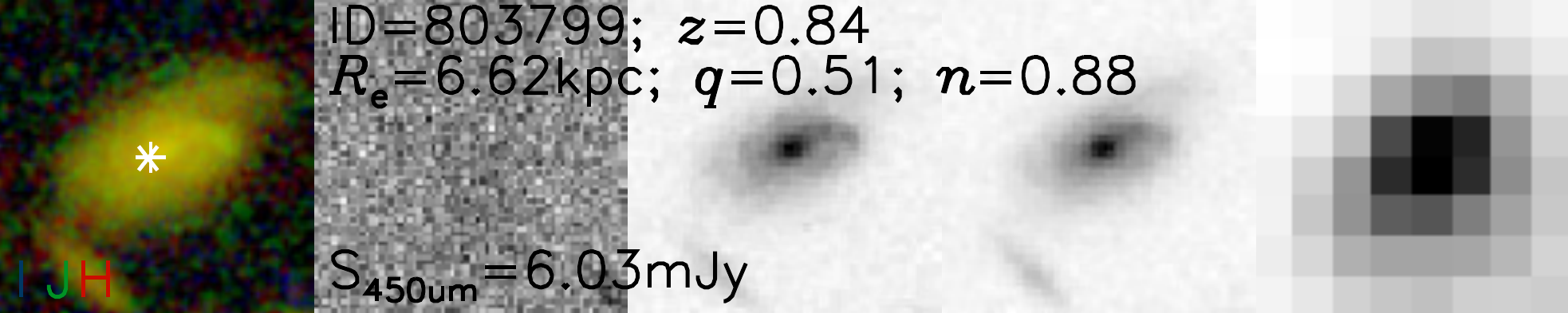}
\includegraphics[width=0.95\columnwidth]{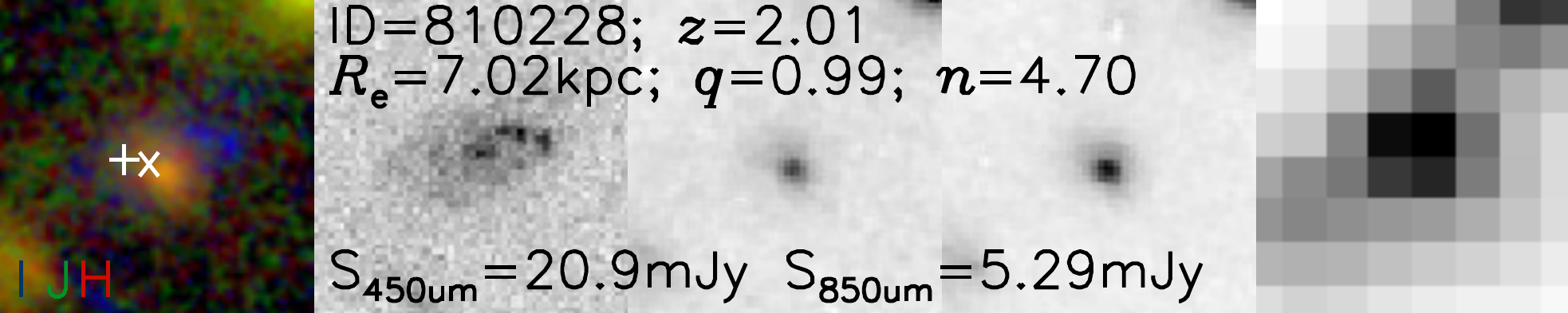}
\includegraphics[width=0.95\columnwidth]{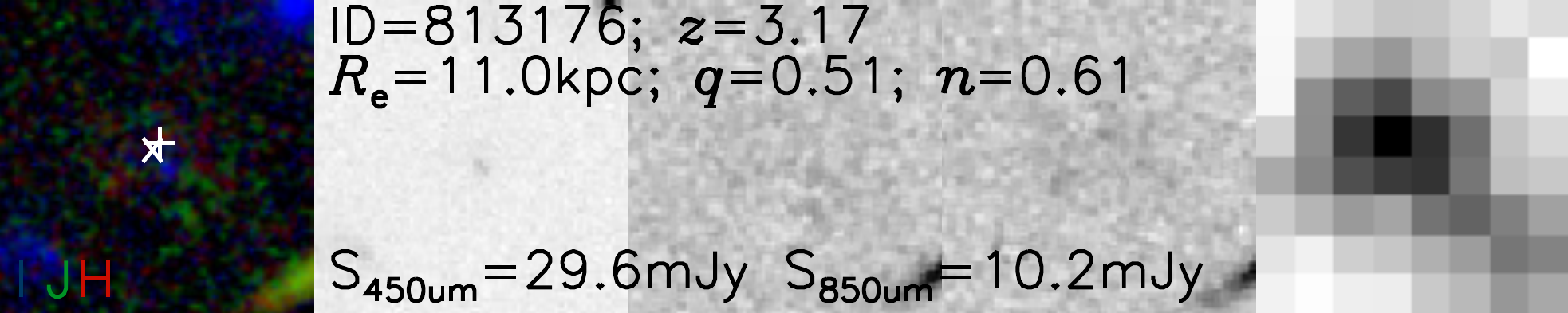}
\includegraphics[width=0.95\columnwidth]{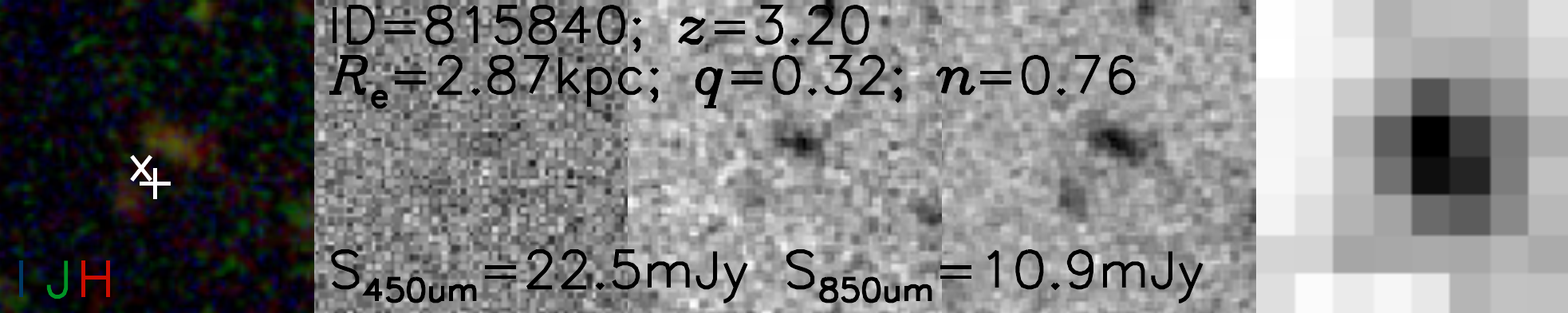}
\includegraphics[width=0.95\columnwidth]{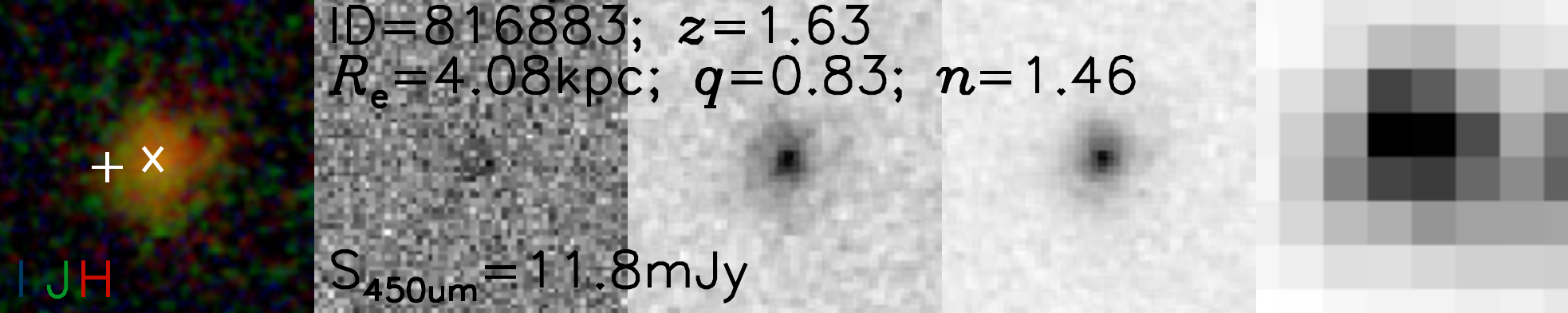}
\label{studies_fig}
\end{figure}

\centering
\bibliographystyle{aasjournal}



\end{document}